# CHEMICAL PROPERTIES OF THE LOCAL GALACTIC DISK AND HALO. I. FUNDAMENTAL PROPERTIES OF 1,544 NEARBY, HIGH PROPER-MOTION M DWARFS AND SUBDWARFS

Neda Hejazi,[1] Sébastien Lépine,[1] Derek Homeier,[2] R. Michael Rich,[3] and Michael M. Shara[4]

[1]*Department of Physics and Astronomy, Georgia State University, Atlanta, GA 30302, USA*

[2]*Zentrum für Astronomie der Universität Heidelberg, Landessternwarte, Königstuhl 12, D-69117 Heidelberg, Germany*

[3]*Department of Astrophysics, University of California at Los Angeles, Los Angeles, CA 90095, USA*

[4]*Department of Astrophysics, American Museum of Natural History, Central Park West at 79th Street, New York, NY 10024, USA*

## ABSTRACT

Due to their ubiquity and very long main sequence lifetimes, M dwarfs provide an excellent tool to study the formation and chemical enrichment history of our Galaxy. However, owing to their intrinsic faintness, the acquisition of high-resolution, high signal-to-noise spectra of low-mass stars has been limited to small numbers of very nearby stars, mostly from the Galactic disk population. On the other hand, large numbers of low-to-medium resolution spectra of M-type dwarf stars from both the local Galactic disk and halo are available from various surveys. In order to fully exploit these data, we develop a template-fit method using a set of empirically assembled M dwarf/subdwarf classification templates, based on the measurements of the TiO and CaH molecular bands near 7000 Å, which are used to classify M dwarfs/subdwarfs by spectral type and metallicity class. We further present a pipeline to automatically determine the effective temperature $T_{eff}$, metallicity [M/H], $\alpha$-element to iron abundance ratio [$\alpha$/Fe], and surface gravity log $g$ of M dwarfs/subdwarfs using the latest version of BT-Settl model atmospheres. We apply these methods to a set of low-to-medium resolution spectra of 1,544 high proper-motion ($\mu \geqslant 0.4''\mathrm{yr}^{-1}$) M dwarfs/subdwarfs, collected at the MDM observatory, Lick Observatory, KPNO, and CTIO. Our metallicity estimates appear to be consistent with the expected color-magnitude variation of stars relative to atmospheric composition, as our sample shows a clear stratification with respect to metallicity in the Hertzsprung-Russel diagram constructed from their Gaia DR2 parallaxes and optical magnitudes. Furthermore, the measured chemical parameters of the two components in 48 binary systems are in a good agreement with each other, which suggest a precision of $\pm0.22$ dex in [M/H], $\pm0.08$ dex in [$\alpha$/Fe], and $\pm0.16$ dex in the combined index [$\alpha$/Fe]+[M/H]. We find that the relationship between color and spectral subtype depends on metallicity class, as the color $G_{\mathrm{BP}}$ - $G_{\mathrm{RP}}$ is more sensitive to subtype for metal-rich M dwarfs, in comparison to metal-poor M subdwarfs. We also demonstrate that effective temperature as a function of spectral subtype has a steeper slope for metal-rich M dwarfs than metal-poor M subdwarfs. There is also a good consistency between "metallicity class", obtained from the empirical classification templates, and the index [$\alpha$/Fe]+[M/H] ($\sim$ [$\alpha$/H]), obtained from BT-Settl model-fitting, which means that the more easily measured "metallicity class" can be used as a relatively reliable indicator of absolute alpha-element abundance, [$\alpha$/H], in low-mass stars. Finally, we examine the distribution of our stars in the [$\alpha$/Fe] vs. [M/H] diagram, which shows evidence of clustering in chemical abundance makeup, suggestive of discrete populations among the local disk and halo stars. We predict that analyses of larger samples of spectra of nearby M-type stars will uncover a complex structure of our Galaxy.

*Keywords:* M dwarfs — M subdwarfs — Classification—Atmosphere models — Stellar Fundamental Properties—Chemical Parameters—Galactic Disk—Galactic Halo

## 1. INTRODUCTION

Although M dwarfs are amongst the faintest stars in the Galaxy, they have increasingly become attractive targets for various studies. These low-mass main sequence stars are by far the most common type of star in the Universe, making up $\sim 70\%$ of all stars by number in our Galaxy, thus dominating the stellar content of the Milky Way (Reid & Gizis 1997; Bochanski et al. 2010). Moreover, their main-sequence lifetimes are much longer than the current age of the Universe and many of them have completed a large number of Galactic orbits (Reid & Hawley 2000). M dwarfs can therefore be excellent tracers of the structure and kinematics of the Galactic populations (e.g., Reid et al. 2002; Lépine et al. 2003a, 2003b, and 2013; Lépine & Shara 2005; Lépine & Gaidos 2011; Bochanski et al. 2010, 2011 and 2013; West et al. 2011) as well as fossils to Galactic chemical (e.g., Woolf & West 2012; Hejazi et al. 2015) and dynamical evolution (e.g., West et al. 2006). Furthermore, M dwarfs have been under scrutiny as planet host candidates, because widely used methods in detecting exoplanets such as the radial velocity and transit techniques are more sensitive to detecting planets orbiting low-mass stars. Statistics from Kepler light curves and radial velocity surveys show that M dwarfs are the most abundant planet hosts in the Milky Way, having a high frequency of Earth-size planets in their habitable zones (Dressing & Charbonneau 2013; Petigura et al. 2013; Kopparapu 2013; Mulders et al. 2015; Gaidos et al. 2016). The determination of the fundamental parameters of M dwarfs has thus become a cornerstone for both Galactic and exoplanet astronomy.

The spectra of M dwarfs are mainly governed by molecular opacities, a characteristic of low-temperature atmospheres which allow simple molecules to form. While $H_2$ is the dominant molecular species in such atmospheres, a handful of other molecules - essentially TiO, CaH and VO in the optical and $H_2O$ in the NIR - make significant contributions to the opacity. The enormous number of molecular transitions crucially affects the spectral energy distributions of M dwarfs which bear little resemblance to a Planck function (Jones 1968, Allard et al. 2000a, 2000b; Krawchuk et al. 2000, Tennyson et al. 2007). These molecular bands also depress the true continuum, leaving only a detectable pseudo-continuum whose level is generally sensitive to chemical abundances (Veyette et al. 2016). Detailed studies of molecular bands are therefore critical for identifying and characterizing M dwarfs.

Essentially, the dependence of the strength of the TiO, VO and CaH bands to the physical parameters such as effective temperature and chemical composition has provided a useful tool to assign both spectral types (an indicator of effective temperature) and metallicity class (an indicator of overall metal abundances) for M dwarfs. In this system, the so-called metal-poor "M subdwarfs" are metal-deficient ([M/H] $\lesssim$ -1) stars whose lower molecular opacities result in smaller radii as compared to more metal-rich stars with the same mass (Kesseli et al. 2019). In order to develop an accurate classification system, Reid et al. (1995) introduced a set of molecular indices defined by the ratio of the flux within a spectral band of interest (e.g. in the deeper part of a molecular band) and the flux within a nearby spectral band defining a local pseudo-continuum. Their original spectral type assignment was based on a relationship between spectral subtype and a molecular band index associated with the most prominent TiO band around 7130 Å.

However, the strong dependency of TiO molecular bands on metallicity, and their obvious weakness in local metal-poor M subdwarfs implied that the strength of the TiO bands could not generally be used as an independent measure of a star's effective temperature. CaH molecular bands, on the other hand, are found to be just as strong in metal-rich M dwarfs and metal-poor M subdwarfs, and were therefore adopted as the primary spectral classification features in papers by Gizis 1997; Lépine et al. (2003a); Lépine et al. (2007); Lépine et al. (2013). It should be noted that the CaH molecular bands saturate in late-type M dwarfs (M7-M9) while TiO and VO bands remain prominent in late-type metal-rich M dwarfs and can still be used as the secondary spectral estimators (Kirkpatrick et al. 1995; Lépine et al. 2003a; Lépine et al. 2013).

The M subdwarfs are typically found to be kinematically associated with the local Galactic halo populations (Lepine et al. 2003a). These metal-poor, main sequence stars were formed out of primordial matter before heavy elements were synthesized in significant amounts by stellar nuclear reactions, leading to a unique chemistry as compared to the more recently formed M dwarfs, which are kinematically associated with the local Galactic disk populations. Due to the deficiency of metal content, M subdwarfs are TiO depleted, but maintain relatively strong hydrides such as CaH, leading to a peculiarly small bandstrength ratio of TiO to CaH. This is the basis of the metallicity classification system suggested by Gizis (1997), in which the separation between M dwarfs (dM), (moderately metal-poor) M subdwarfs (sdM) and extreme M subdwarfs (esdM) was determined by the relative strength of CaH bands with respect to the strength of TiO bands. With more metal-poor M subdwarfs being identified in spectroscopic surveys (Lépine et al. 2007), a fourth metallicity class (ultra M subdwarf, or usdM) was added, and the four classes (dM/sdM/esdM/usdM) were characterized by a ratio of spectral indices, dubbed $\zeta_{\mathrm{TiO/CaH}}$ (hereafter $\zeta$), which measures the variation of the TiO band strength relative to the CaH band strength, and has been calibrated against independent measurements of metallicity in a number of M subdwarfs (Woolf, Lépine , & Wallerstein 2009).

Using the spectra of 88 K and M subdwarfs, Jao et al. (2008) suggested an alternative method for assigning spectral type to cool subdwarfs on the basis of spectral morphology over the range 6000-8000Å. It was pointed out that the CaH and TiO band indices are affected by a combination of physical properties such as effective temperature, metallicity and gravity, and esdM/usdM stars defined by these indices may be explained by differences in surface gravity rather than by differences in metallicity. Jao et al. (2008) therefore characterized cool subdwarfs by comparing their optical spectra with model grids and evaluating the trends in all physical parameters, and then offered an alternative classification scheme.

In order to automate the procedure of spectral typing, Covey et al. (2007) provided a custom IDL-based package, named "the Hammer", to assign spectral types based on combined measurements of several spectral band indices. To improve the quality of their sample, West et al. (2011) visually inspected the spectra of over 70,000 M dwarfs from the Sloan Digital Sky Survey (SDSS) spectroscopic database to identify and correct any possible errors in the spectral types originally determined by the Hammer code. Although this approach improved the reliability of the spectral typing, the method proved both time-consuming and human-intensive, and a more efficient technique was therefore required for analyzing large samples. To this end, Zhong et al. (2015) developed an automated template-fit method, as an alternative to the direct measurements of band indices. They assigned spectral type and metallicity class to their M dwarf/subdwarf sample based on a comparison with empirical classification templates built from spectra of previously classified stars. This method is revisited in the present work (Section 3).

Beyond simple spectral classification, the accurate determination of M dwarf physical parameters has proven challenging. Given the difficulty of continuum identification and spectral rectification in their spectra, high-resolution, high signal-to-noise ratio (hereafter S/N) spectra are needed to accurately analyze the atmospheres of M dwarfs. However, due to the intrinsically low luminosity of M dwarfs, which for M subdwarfs is compounded by their relative rarity in the Solar Neighborhood, the acquisition of high-resolution, high S/N spectra has required long exposure times using large telescopes, achievable only for a limited number of very nearby stars. On the other hand, large samples of low-to-medium resolution spectra of M dwarfs/subdwarfs from both the local Galactic disk and halo populations are available from various surveys. Alternative techniques have therefore been developed to infer stellar parameters based on moderate-resolution spectra.

A number of M dwarfs/subdwarfs have, for example, been analyzed using high-resolution spectra in the optical (e.g., Woolf & Wallerstein 2005; Pineda et al. 2013; Neves et al. 2014; Rajpurohit et al. 2014; Maldonado et al. 2015; Kuznetsov 2019), in the near infrared (NRI) (e.g., Önehag et al. 2012; Lindgren et al. 2016, 2017; Schmidt et al. 2016; Souto et al. 2017), and in both the optical and NIR (Rajpurohit et al. 2018). Using samples of M+FGK binaries (consisting of an M dwarf as the secondary and an F, G, or K dwarf as the primary), the metallicity of M dwarfs has been calibrated on the basis of medium-resolution spectra (Rojas-Ayala et al. 2010, 2012; Terrien et al. 2012; Mann et al. 2013; Newton et al. 2014). Veyette et al (2017) empirically calibrated synthetic spectra using a calibration sample of M+FGK binaries to determine the physical parameters of M dwarfs. We refer the reader to the introduction (Section 1) of Veyette et al (2017) and Mann et al. (2019) for a comprehensive description of recent studies on the determination of M dwarf fundamental properties.

Despite of all these efforts, the stellar chemical parameters reported in the literature have not gained an acceptable level of consistency, as they are often found to differ in independent studies for some of the very same stars; this suggests that current estimates likely suffer from systematic errors. Authors focus on different spectral regions, lines, or features, and use different techniques with a wide range of spectral resolution and accuracy. Studies also assume different sets of elements to measure chemical parameters such as [M/H] and [α/Fe]. More importantly, metal-poor M subdwarfs have been overlooked in most studies, and only a handful of M subdwarf samples have been analyzed using high-resolution optical spectra (Woolf & Wallerstein 2005; Rajpurohit et al. 2014), and medium-resolution ultraviolet/optical/NIR spectra (Lodieu et al. 2019). Most previous calibrations are thus only valid for [Fe/H] $\gtrsim$ -0.7, and no reliable empirical relations have been established for M subdwarfs yet. Since these stars are associated with old Galactic populations, most of them are not accompanied by an opportune massive companion that might be used as an independent metallicity calibrator. Therefore, accurate metallicities have only been determined for a few selected M subdwarfs that by chance are the companions of metal-poor K subdwarfs in wide binary systems (e.g. Pavlenko et al. 2015). For this reason, it has become necessary to develop a means to extract physical parameters directly using synthetic spectra calculated from theoretical models of M dwarf/-subdwarf atmospheres.

Current direct techniques (without using a calibration sample) can be divided into two main groups. One method relies on comparing observed spectra with synthetic models and find the best fit by varying the fitting parameters. This can be done by fitting over a wide range of wavelengths (e.g., Önehag et al. 2012; Rajpurohit et al. 2014, 2016 and 2017) or by atomic line fitting (e.g., Lindgren et al. 2016, 2017). In the other method, a line by line analysis is performed by measuring the strength or the equivalent width (EWs) of several spectral lines, which are then converted into individual abundances using atmospheric models (e.g., Woolf & Wallerstein 2005). Alternatively, the measured EWs can be used to generate the observed curve-of-growth (CG) which is then compared with theoretical CGs to obtain chemical abundances (e.g., Tsuji et al. 2014, 2015 and 2016).

The accurate measurement of spectral line strengths/EWs or atomic line fitting normally requires high-S/N, high-resolution spectra, and model fitting over a wide spectral region thus remains the best approach using low-to-medium spectra. In this paper, we elaborate on this method to obtain effective temperature, gravity, metallicity, and alpha abundance measurements for a large number of M dwarfs and M subdwarfs by fitting their low-to-medium resolution spectra to a wide grid of synthetic spectra generated from the BT-Settl theoretical model of cool star atmospheres.

Section 2 describes the spectroscopic survey program from which the spectra were assembled, and Section 3 performs a generic spectral and metallicity classification of these stars, based on empirical classification templates of M dwarfs and M subdwarfs. In section 4, the distinct properties of stars with different metallicity classifications is shown in photometric and kinematic diagrams. In Section 5, we outline a more general method for directly extracting stellar parameters through comparison of observed spectra with BT Settl synthetic models. We also report the resulting parameters and their corresponding uncertainties in this section. We discuss the correlation between the classification and physical parameters of our stars in Section 6. In Section 7, we show again that stars of different inferred chemical abundances have very distinct properties in color-magnitude diagrams and in their kinematic distributions. Section

8 compares the measured chemical parameters of primaries and their companions for a set of common proper-motion pairs in our sample. In Section 9, we briefly discuss the relationship between chemical parameters, and outline future plans for using this relationship to uncover the chemo-dynamical evolution of local Galactic populations. Finally, we summarize our study in Section 10.

## 2. SPECTROSCOPIC OBSERVATIONS AND DATA REDUCTION

### 2.1. *Target selection*

Stars with proper motions $\mu > 40$ mas yr$^{-1}$ identified in the SUPERBLINK survey, and published in the LSPM-north catalog of Lépine & Shara (2005) were selected for follow-up observations as part of a long-term, stellar spectroscopy program. The initial list of targets consisted of 2,991 stars spanning the entire northern sky. Most of the brighter objects were known main-sequence stars of spectral subtype G or earlier, and these were not observed. Faint stars with relatively blue colors were targeted and observed as part of the survey for nearby white dwarf by Limoges et al. (2013). Our search for M dwarf and M subdwarf stars has mainly focused on the subset of 2,290 stars with red optical to infrared colors ($V$-$J > 2.5$). These were observed in multiple observing runs at the MDM observatory, the Kitt-Peak National Observatory (KPNO), and the Lick Observatory. In the course of the program, a few high proper-motion stars of interest were identified in the southern sky were included in the survey, and observed in two observing runs at the Cerro-Tololo Interamerican Observatory (CTIO). Stars that were found to be of spectral type earlier than K5 were excluded from the more detailed spectral analysis. Below we describe only the subset of 1,746 stars that are formally identified as late-K and M dwarfs/subdwarfs. A significant fraction of these stars had their spectra used and summarily discussed in Lépine et al. (2007), but the spectra have not been presented in a comprehensive catalog until now, and never analyzed with the minute level of detail we present in Sections 3 and 5 below.

The description of the observations is summarized in Tables 1 and 2. Table 1 lists the observatories, the number of stars observed at each observatory, the number of observing runs, years of observation, the telescopes and their respective spectrographs, slit modes, and slit widths. Table 2 presents the grating(s) used in each spectrograph, the resolving power corresponding to each grating, the exposure times, and the calibrations.

### 2.2. *Data reduction*

| Observatory | Number of Stars | Number of Runs | Years of Observation | Telescope | Spectrograph | Slit Mode | Slit Width ($''$) |
|---|---|---|---|---|---|---|---|
| MDM | 1,575 | 22 | 2002-2012 | McGraw-Hill 1.3 m Hiltner 2.4 m | MkIII | Spectroscopy | 1.0-1.5 |
| KPNO CTIO | 52 | 2 | 2002-2009 | Mayall 4 m Blanco 4m | RCspec | Long-slit | 1.0 |
| Lick | 119 | 5 | 2002-2004 | Shane 3 m | KAST | Long-slit | 1.0 |

**Table 1.** Observation description

| Observatory | Grating (lines/mm) | R | Exposure[a] Time (sec) | Calibration[b] |
|---|---|---|---|---|
| MDM | 300 | $\approx 2000$ | 150-600 | Internal flats: CCD response calibration |
|  | 600 | $\approx 4000$ |  | Arc lamp spectra of Ne, Ar, and Xe: wavelength calibration |
|  |  |  |  | Standard stars Feige 110, 66, and 34: spectrophotometric calibration |
| KPNO CTIO | 316 | $\approx 2000$ | 600-1800 | Internal flats: CCD response calibration |
|  | 600 | $\approx 4000$ |  | Arc lamp spectra of He, Ne, and Ar: wavelength calibration |
|  |  |  |  | Standard stars Feige 110, 66, and 34: spectrophotometric calibration |
| Lick | 600 | $\approx 4000$ | 120-900 | Dome and sky flats: CCD response calibration[c] |
|  |  |  |  | Arc lamp spectra of Ne and Ar: wavelength calibration |
|  |  |  |  | Standard stars Feige 110, 66, and 34: spectrophotometric calibration |

[a] Total integration times varied depending on seeing, telescope aperture, and target brightness.

[b] Arc lamp spectra were obtained for every pointing of the telescope to account for flexure in the system.

[c] An additional complication was that the CCD camera suffered significant amounts of fringing in the red, and dome flats had to be acquired at every pointing of the telescope; due to flexure in the instrument, however, fringing artifacts remained in some spectra after reduction, in regions redward of 7500.

**Table 2.** Observation description

Reduction of all spectra was performed using the CCDPROC and DOSLIT packages in IRAF. Reduction included bias and flat-field correction, removal of the sky background, aperture extraction, and wavelength calibration. The spectra were also extinction-corrected and flux-calibrated based on the measurements obtained from the spectrophotometric standards. We did not attempt to remove telluric absorption lines from the spectra, as some spectra were collected on humid nights or with light cirrus cover, which resulted in variable telluric features. However, telluric features generally do not affect standard spectral classification or the measurement of spectral band indices, since all the spectral indices and primary classification features of M dwarfs/subdwarfs avoid regions with telluric absorption. In the present work, we use a wide range of wavelengths to classify and infer the physical parameters of stars, and using model spectra, we identify three different spectral regions which are contaminated by telluric absorption bands, and thus excluded from our analysis (Sections 3 and 5).

A more common problem at the MDM observatory was slit loss from atmospheric differential refraction. Although this problem could have been avoided by the use of a wider slit, the concomitant loss of spectral resolution was deemed more detrimental to our science goals. Instead, stars were observed as close to the meridian as observational constraints allowed. In some cases, however, stars were observed up to $\pm 2$hr from the meridian, resulting in noticeable slit losses. Fluctuations in the seeing, which often exceeded the slit width, played a role as well. As a result, the spectrophotometric calibration was not always reliable, since the standards were only observed once per night to maximize survey efficiency. Flux recalibration was therefore performed using the procedure outlined below (Section 3).

## 3. SPECTRAL AND METALLICITY CLASSIFICATION

### 3.1. *Classification Templates*

Zhong et al. (2015) assembled their classification templates using the spectra of relatively bright late-K and M dwarfs drawn from SDSS, mostly from the catalog of M dwarf stars by West et al. (2011), complemented by additional identifications of M subdwarfs (Savchena et al. 2014). Instead of using the Hammer software, they re-classified all stars in their sample based on the methods described by Lépine et al. (2007, 2013). The templates were generated through three passes as follows. In the first pass, the SDSS spectra were classified by measuring the band indices, and the spectra of stars with the same classification were then co-added to produce a set of tentative templates. The radial velocity shift of atomic lines were measured, and the templates were subsequently shifted back to the stellar rest frame. In the second pass, the SDSS spectra were cross-correlated with their matching templates to obtain the corresponding radial velocity shifts for each star, which were used for shifting the spectra of all the stars to the rest frame. The radial velocity corrected spectra were then re-classified by re-calculating the molecular band indices, and a new set of templates were generated. In the third pass, the second pass was repeated using the updated templates.

The final co-adds from the third pass were recorded as the formal templates for spectral and metallicity classification of late K and M dwarfs. The spectral types of these templates ranged from K7.0 to M8.5 including every half-subtype while the metallicity classes covered the original sequence of dM/sdM/esdM/usdM. Zhong et al. (2015) increased the number of metallicity classes from 4 to 12 by adding two more subclasses to each class, a more metal-rich and a more metal-poor subclass, expanding each standard class to three subclasses labeled by "r" for the metal-rich, "s" for the standard and "p" for the metal-poor subclass. The extra templates were synthesized by linearly interpolating/extrapolating the gird one third of the way from each standard metallicity class, establishing 12 metallicity subclasses: $dM_r$, $dM_s$, $dM_p$, $sdM_r$, $sdM_s$, $sdM_p$, $esdM_r$, $esdM_s$, $esdM_p$, $usdM_r$, $usdM_s$, $usdM_p$.

Zhong et al. (2015) then applied the template-fit method to 83,500 spectra collected from the Large sky Area Multi-object fiber Spectroscopic Telescope (LAMOST) commissioning data (Cui et al. 2012), and around 2600 stars were confirmed as M dwarfs/subdwafs. To validate the quality of the method and the resulting classification, a visual inspection was performed for a number of spectra spanning a broad range of spectral types. Since the accuracy of both spectral and metallicity classification strongly depends on the S/N of spectra, the spectra with different ranges of S/N were examined separately. The comparison of each target spectrum with its best-fit template as well as the neighboring templates indicated that the classification is accurate to within the nearest half-subtype ($\pm0.5$ subtype) and the nearest metallicity class ($\pm1$ metallicity class) for high-to-medium S/N (S/N>5). However, the classification of the spectra with low S/N (S/N<5) were less reliable, but still accurate to within $\pm1.0$ subtype and $\pm2$ metallicity class.

### 3.2. *Spectral Types and Metallicity Classes from the Template-Fitting Method*

We adopt the classification templates from Zhong et al. (2015) for assigning spectral type and metallicity class to our 1746 observed spectra. To provide a more convenient designation system, we have simply numbered the 12 subclasses mentioned above from 1 for the most metal-rich M dwarfs ("$dM_r$") to 12 for the most metal-poor M subdwarfs ("$usdM_p$").

Our fitting spectral range is limited to 6300-8200 Å, and excludes the problematic regions 6850-6885, 7585-7670 and 8130-8175 Å which are contaminated by atmospheric absorption bands. The $H_\alpha$ emission line of hydrogen, generally attributed to stellar chromospheric activity, is detected in some of our stars, and the region 6550-6580 Å is thus also removed from our analysis. While many of our spectra have wavelength coverage blueward of 6,300Å, some do not, and in order to retain consistency, we limited the fitting range to regions redward of 6,300 Å. The spectral regions 8200-8430 Å and 9000-10000 Å are dominated by strong telluric absorptions (Rajpurohit et al. 2014; Kesseli et al. 2018), and are therefore excluded as well. Furthermore, we have noted that systematic errors in the classification of some stars occur if the region 8450-9000 Å is included in the fit; this is most likely due to flux calibration issues in this part of the spectrum with either our own spectra or the SDSS spectra used to generate the templates (or even both), leading to significant offsets in the spectral and metallicity classification. It is important to mention that the selected fitting spectral range 6300-8200 Å perfectly covers the most prominent TiO and CaH molecular bands which have been identified as the primary indicators for M dwarf and M subdwarf classification (Section 1).

We use a least squares minimization method by comparing the observed spectra and the templates to determine the spectral type and metallicity class of our stars.

As a general rule, the sum of the squared residuals is calculated by

$$\chi^2 = \sum_i (F_{obs}(\lambda_i) - F_{tem}(\lambda_i(1 + v_r c^{-1}))^2 \qquad (1)$$

where $v_r$ is the radial velocity of the star, c is the speed of light, $\lambda_i$ is the $i$th observed wavelength, $F_{obs}(\lambda_i)$ is the observed flux at the wavelength $\lambda_i$, and $F_{tem}(\lambda_i(1 + v_r c^{-1}))$ is the template flux at the shifted wavelength $\lambda_i + \lambda_i v_r c^{-1}$. The template and $v_r$ value which together minimize the $\chi^2$ is then assigned as the best fit.

Our selected templates correspond to a two-dimensional grid of metallicity class and spectral subtype covering all the 12 metallicity classes and the spectral subtypes from K7.5 to M7.0 (in steps of 0.5 subtype); these two dimensions define the primary free parameters in the fitting procedure. If a parameter has a best-fit value at either the upper or the lower limit of either the subtype or metallicity class range, this best fit may not necessarily signify that the respective template is the best representation of the spectrum because the parameter values beyond that limit have not been tested in the fitting process. The best way around this problem is to exclude from the analysis any star that is best fitted by a template at the edge of the parameter grid, keeping objects that have best fit parameters at least one step removed from the edge. This assures that the best fit does indeed minimize the $\chi^2$. As mentioned in Section 1, the spectral type assignment of the templates was based on CaH band indices which saturate towards late-type M dwarfs, and as a result, our templates for spectral types of M7.0 and later may not be reliable. For this reason, we select the spectral type M7.0 as the late-type edge of our grid, and any star classified as M7.0 is simply removed from our sample, making the subtype M6.5 as the latest-type stars in the present analysis. Similarly, the spectral type K7.5 is the earliest subtype for which a reliable separation of metallicity classes is possible, and we define this as the early-subtype edge of our grid, and all stars that have a best-fit K7.5 are hence excluded, leaving the subtype M0.0 as the earliest type in the final sample. This edge-rejection procedure is however only applied for spectral subtypes, and not for metallicity classes. As it happens, the full range metallicity subclasses - from 1 to 12 - are needed to classify all of our stars, and if the best fit is found at a subclass of 1 or 12, then this subclass is used for the classification. This means that stars with extreme high metallicity values will be simply classified as subclass 1, and stars with extreme low metallicity values will be simply classified as subclass 12.

According to Eq. 1, the template spectra $F_{tem}$ are linearly interpolated at wavelengths shifted by a factor of $\lambda_i v_r c^{-1}$. Since accurate radial velocities are not available for many of our stars, the value of $v_r$ is used as a secondary parameter (versus above mentioned primary parameters) in the minimization; we allow $v_r$ to vary from -500 to +500 km/s with a 10 km/s step. These large minimum and maximum radial velocities ensure that the spectra of very high-velocity stars in our sample, which typically belong to the local Galactic halo population, are properly corrected. Large radial velocities will typically shift part of the star's spectrum outside of the nominal template fit range (6300-8200 A). To avoid potential edge issues, we therefore adjust the initial wavelength region of all stars to a new standard range, 6311-8186 Å.

Prior to running the least squares minimization, special care has to be made to correct for any possible flux calibration issues in the observed spectra. The spectral energy distribution from a star is generally influenced by a number of matter-radiation interactions between the star and the Earth, before being recorded as a spectrum in a detector. Among the most important of these interactions are interstellar reddening, seeing fluctuations, slit loss combined with atmospheric differential refraction, and the throughput of the telescope and instrument in general. These processes may significantly affect the slope of the spectral energy distribution, which in turn may impact the stellar classification. While some effects can potentially be calibrated out (throughput) others may not (interstellar reddening). The usual approach for flux recalibration in earlier type stars is to "rectify" the spectrum, i.e., to normalize the continuum to unity, however, M dwarfs have such prevailing molecular bands that the continuum level cannot be easily rectified or corrected, even at high spectral resolution. We therefore correct the observed spectra by following the algorithm outlined below.

The flux of an observed spectrum ($F_{obs}(\lambda_i)$) is divided by the flux of a given template interpolated at the corresponding shifted wavelengths ($F_{tem}(\lambda_i(1 + v_r c^{-1}))$), and a polynomial is then fitted to the quotient. In the best of cases - perfect spectral calibration - this quotient would be a uniform (flat) function of wavelength, assuming the template is also a perfect match to the intrinsic spectrum of the star. However, owing to flux calibration errors, the variation of this ratio with wavelength can be relatively complicated. In both Lépine et al. (2013) and Zhong al. (2015), a first-order or second-order polynomial was used to fit the spectrum/template quotient for flux recalibration. However, these low-order polynomials are often not sufficient to properly rectify observed spectra over the full spectral range needed for classification. For this purpose, we settled on the use of polynomials of order n=6, 8, and 10, and defined these as another set of secondary fitting parameters. These poly-fit order values have been selected based on close visual inspection of a number of spectra spanning a wide range of noise levels. Although intermediate orders, i.e., n=7 and 9, may yield a better polynomial fit in some cases, the entire process can be computationally intensive if these numbers are also included. After some experimentation, we determined that the quality of the fit was not overly sensitive to variations of polynomial order by $\Delta n = \pm 1$, and we thus excluded these odd order values from the fit, keeping only even values.

Here is how we conduct the recalibration. We first perform a polynomial fit of the spectrum/template quotient, which is effectively a tabulated function of the wavelength $\lambda_i$. We then calculate the root mean square error (RMSE) between the quotient and the fit as the standard deviation, and eliminate the spectral outliers which fall outside ten standard deviations ($10\sigma$). A new polynomial is again fitted over the remaining spectral data, and the outliers outside $8\sigma$ are rejected, subsequently followed by the same procedure for $6\sigma$ outliers. In general, most outliers which are due to instrumental artifacts (e.g., strong cosmic rays or sky lines) are removed by these three rejection passes.

Once we have the final polynomial best-fit, we evaluate this polynomial at each wavelength of the observed spectrum. A flux-corrected spectrum is then calculated from the ratio of the original spectrum and this evaluated polynomial. The $\chi^2$ value between this corrected spectrum and the classification template (Eq. 1) is then the basis of the minimization process. The combination of primary (metallicity class and subtype) and secondary (radial velocity and polynomial order) parameters which minimizes the $\chi^2$ is assigned as the best template fit to the observed spectrum. It bears mentioning that the radial velocities determined by this minimization method do not generally represent the absolute radial velocity of stars because the observed spectra are not calibrated with the precision normally needed for accurate radial velocity measurements. A significant number of our stars still require relatively large $v_r$ values in any case, because they do have intrinsically large radial velocities, being kinematic members of the local Galactic halo.

### 3.3. Classification Results

#### 3.3.1. Observed Versus Template Spectra

The flux-corrected spectra (red) and best-fit templates (blue) of 28 selected stars are shown in Figures 1-3 as examples. We have selected stars from seven groups cor-responding to the spectral subtypes M0.0-M0.5, M1.0-M1.5, M2.0-M2.5, M3.0-M3.5, M4.0-M4.5, M5.0-M5.5, and M6.0-M6.5. These spectra are plotted at unshifted wavelengths. However, the templates present the flux values which have been interpolated at the shifted wavelengths associated with the best-fit radial velocity of each star. Spectral regions dominated by telluric absorption bands and the $H_\alpha$ emission line of hydrogen, which are not use in the fit, are left blank. Moreover, the small fitting spectral region 8175-8186 Å is not shown in these figures.

In general, and for all spectral subtypes, we find the observed spectra of both metal-rich dM and metal-poor sdM stars to be in excellent agreement with their respective best-fit templates. However, there are discrepancies between the observed and template spectra for some very metal-poor esdM and usdM stars. Part of this may be due to the esdM/usdM stars being typically noisier. However in some cases we do observe significant discrepancies in line strengths, which suggests additional physical parameters, besides temperature and metallicity, may be affecting the spectra.

We first used the simpler spectral fitting method outlined in Zhong et al. (2015), to identify 1,746 probable late-K and M dwarfs/subdwarfs out of our initial spectroscopic catalog of 2,991 stars with proper motions $\mu \geqslant 0.4''\text{yr}^{-1}$. We then applied our more elaborate pipeline to reclassify these stars, and after rejecting 22 stars having a subtype of K7.5 or M7.0 in the pipeline (i.e., at the upper/lower edge of the subtype range), and 23 stars which are clearly misclassified due to their noisy spectra, in addition to eliminating one star whose Gaia proper motion ($\simeq 0.14''\text{yr}^{-1}$) is significantly less than our lower limit of proper motion, we were left in the end with exactly 1,700 spectra of stars with spectral subtypes spanning M0.0-M6.5, and with metallicity classes from 1 to 12, including 1380 dM ($1 < \text{MC} < 3$), 139 sdM ($4 < \text{MC} < 6$), 140 esdM ($7 < \text{MC} < 9$), and 41 usdM ($10 < \text{MC} < 12$) stars.

Table 3 presents a complete list of these 1,700 stars. The table includes the observing date, the location of observation, and the spectrograph used (columns 2-4). The right ascension, declination, and their respective proper motions along with the parallax from the Gaia DR2 are also recorded in the columns 5-9. We find 1,636 stars cross-matched with the Gaia DR2, whose Gaia parallax and proper motion are listed in Table 3, while the proper motion of the remaining 64 stars are extracted from the SUPERBLINK catalog. Photometric and classification information is collected in Table 4, where we tabulate the spectral type and metallicity class in columns 2 and 3. The Gaia photometry $G$, $G_{\text{BP}}$, and $G_{\rm RP}$ as well as the 2MASS photometry $J$, $H$, and $K$ can be found in the columns 4-9 of this table.

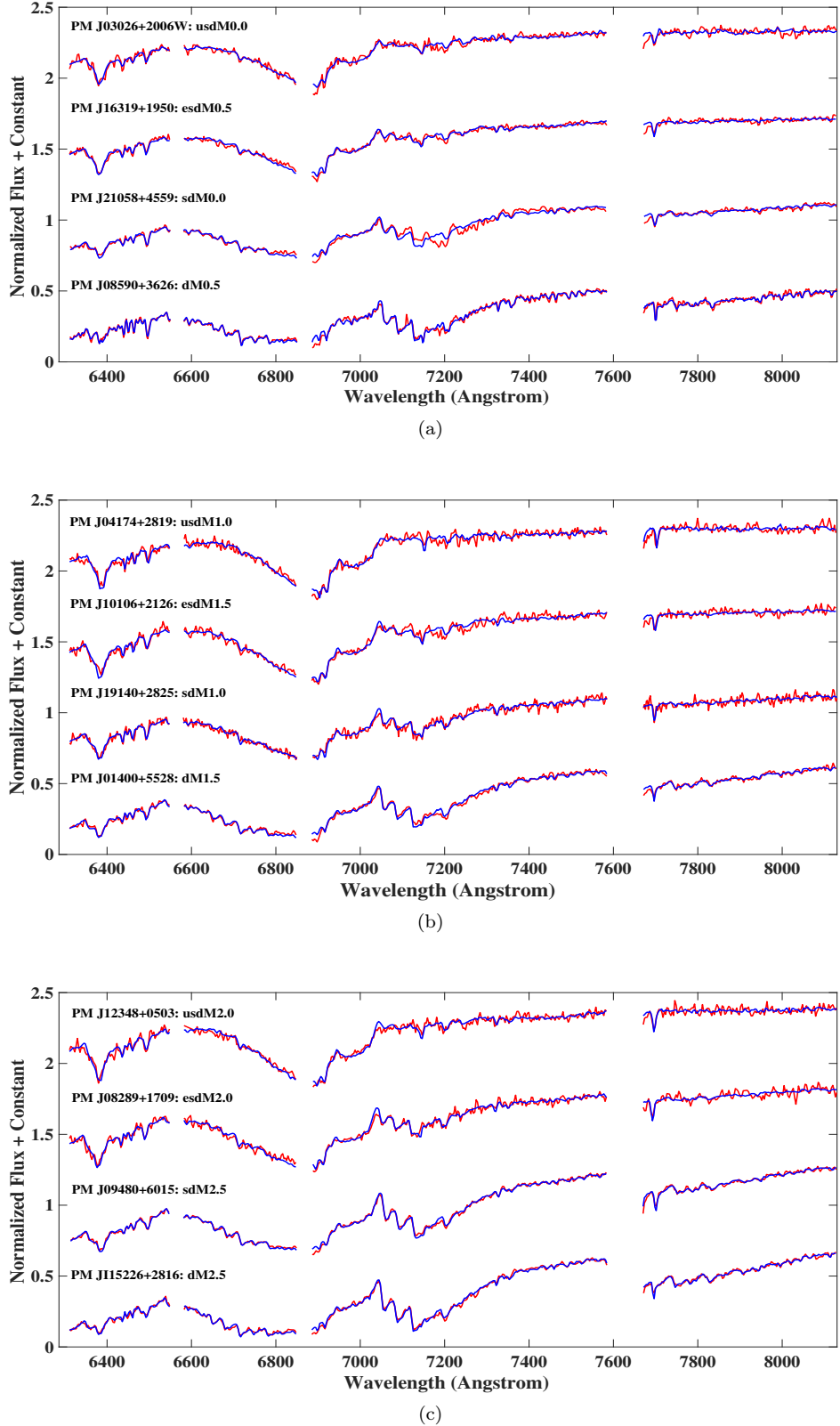

**Figure 1.** Comparison between the flux-corrected spectra of selected stars (red) and their respective best-fit templates (blue). **(a)** Stars with spectral subtypes M0.0-M0.5; the usdM star PM J03026+2006W: MC = 10, the esdM star PM J16319+1950: MC=8, the sdM star PM J21058+4559: MC=4, and the dM star PM J08590+3626: MC=1. **(b)** Stars with spectral subtypes M1.0-M1.5; the usdM star PM J04174+2819: MC=12, the esdM star PM J10106+2126: MC=9, the sdM star PM J19140+2825: MC=6, and the dM star PM J01400+5528: MC=3. **(c)** Stars with spectral subtypes M2.0-M2.5; the usdM star PM J12348+0503: MC=11, the esdM star PM J08289+1709: MC=7, the sdM star PM J09480+6015: MC=4, and the dM star PM J15226+2816: MC=2.

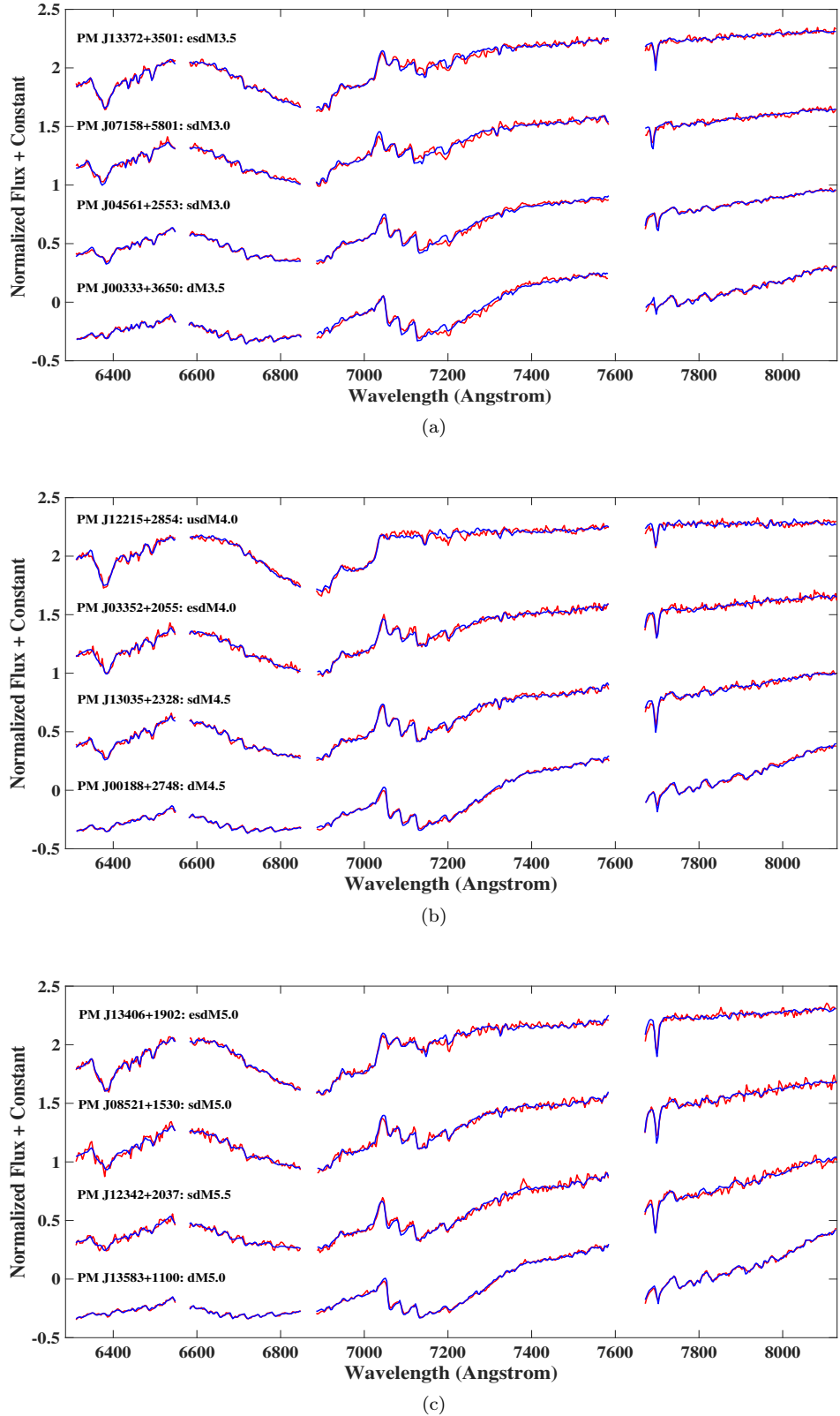

**Figure 2.** Comparison between the flux-corrected spectra of selected stars (red) and their respective best-fit templates (blue). **(a)** Stars with spectral subtypes M3.0-M3.5; the esdM star PM J13372+3501: MC=8, the sdM star PM J07158+5801: MC=6, the sdM star PM J04561+2553: MC=4, and the dM star PM J00333+3650: MC=1. **(b)** Stars with spectral subtypes M4.0-M4.5; the usdM star PM J12215+2854: MC=12, the esdM star PM J03352+2055: MC=7, the sdM star PM J13035+2328: MC=5, and the dM star PM J00188+2748: MC=2. **(c)** Stars with spectral subtypes M5.0-M5.5; the esdM star PM J13406+1902: MC=9, the sdM star PM J08521+1530: MC=6, the sdM star PM J12342+2037: MC=4, and the dM star PM J13583+1100: CM=1.

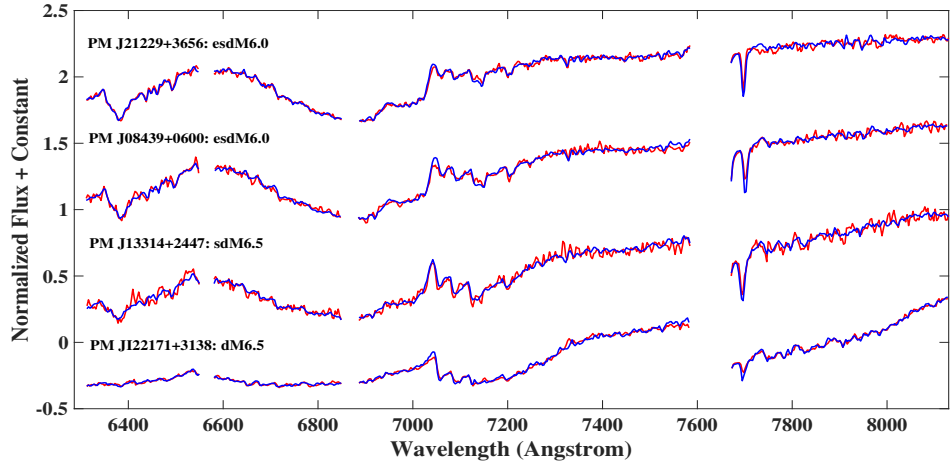

**Figure 3.** Comparison between the flux-corrected spectra of selected stars (red) and their respective best-fit templates (blue) for stars with spectral subtypes M6.0-M6.5; the esdM star PM J21229+3656: MC=8, the esdM star PM J08439+0600: MC=7, the sdM star PM J13314+2447: MC=6, and the dM star PM J22171+3138: MC=2.

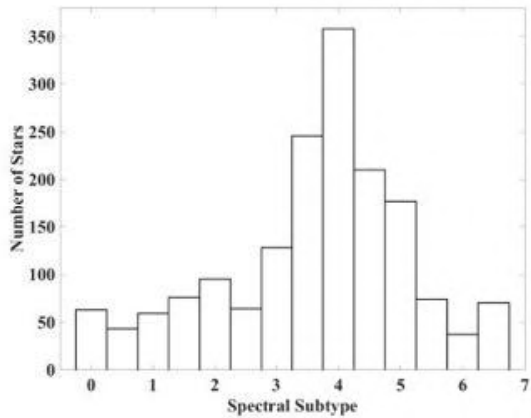

**Figure 4.** Spectral type distribution of 1,700 M dwarfs and M subdwarfs in our high proper-motion sample, with the spectral types ranging from M0.0 to M6.5, which are shown by numbers from 0 to 6.5 with the step size of 0.5.

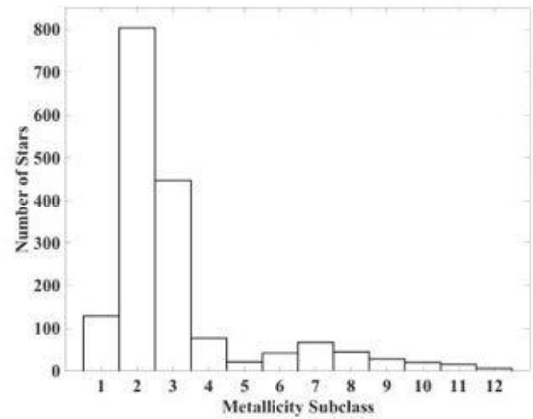

**Figure 5.** Metallicity class distribution of 1,700 M dwarfs and M subdwarfs in our high proper-motion sample, with the metallicity classes shown by numbers from 1 (for the most metal-rich stars) to 12 (for the most metal-poor stars), with the step size of 1. M dwarfs correspond to classes 1-3, and M subdwarfs to classes 4-12.

### 3.3.2. *Classification Distributions*

The spectral type distribution of the above described 1,700 stars is presented in Figure 4. The distribution peaks at the spectral type M4.0, and decreases relatively sharply towards the earlier and later subtypes. Figure 5 shows the metallicity class distribution of the same stars, which indicates that the sample is dominated by the metal-rich M dwarfs with metallicity classes of 2 and 3, which belong to the local disk population. The more metal poor M subdwarfs, likely associated with the local thick disk and halo populations, span a wider range of metallicity class. Despite their relative rarity in the

Solar Neighborhood, we manage to catch a remarkable number of these stars in our spectroscopic sample to scrutinize the trends of their parameters in comparison to metal-rich M dwarfs (Sections 4, 6, 7, and 8).

### 3.4. *Comparison of Spectral Types with Other Catalogs*

We have cross-matched our sample with two spectroscopic catalogs, the Palomar/Michigan State University (PMSU) Survey (Reid et al. 1995; Hawley et al. 1996), and the more recent catalog of nearby low-mass stars by Newton et al. 2014. We found 117 stars in common with our own subset, mostly stars which belong to the Galactic disk. Figure 6 (a) compares the spectral type of these

**Table 3.** Spectroscopic Catalog of the 1,700 Stars in Our Survey: Observational Facilities, Astrometry, and Kinematics

| Object Designation | Obs.Date. | Observatory | Spectrograph | R.A. (Deg) | Decl (Deg) | $\mu_{\mathrm{R.A.}}$ ($''\mathrm{yr}^{-1}$) | $\mu_{\mathrm{Decl}}$ ($''\mathrm{yr}^{-1}$) | prlx ($''$) |
|---|---|---|---|---|---|---|---|---|
| PM J00012+0659 | 2003-09-19 | MDM | MkIII+Wilbur | 0.31403 | 6.99284 | -0.4367 | -0.0835 | 0.04275 |
| PM J00031+0616 | 2003-09-19 | MDM | MkIII+Wilbur | 0.78058 | 6.27537 | 0.2395 | -0.513 | 0.03426 |
| PM J00051+4547 | 2009-09-21 | MDM | MkIII+Nellie | 1.30075 | 45.78592 | 0.8708 | -0.1513 | 0.08696 |
| PM J00101+1327 | 2004-08-11 | Lick | KAST | 2.54052 | 13.45341 | -0.2954 | -0.4168 | 0.02279 |
| PM J00110+0420 | 2003-12-07 | MDM | MkIII+Wilbur | 2.75391 | 4.33806 | 0.1477 | -0.5241 | 0.0126 |
| PM J00119+3303 | 2006-10-03 | MDM | MkIII+Wilbur | 2.9823 | 33.05302 | -0.5565 | -0.3963 | 0.04776 |
| PM J00132+6919N | 2011-10-04 | MDM | CCDS | 3.32508 | 69.32573 | 0.7614 | -0.3322 | — |
| PM J00132+6919S | 2011-09-08 | MDM | CCDS | 3.32508 | 69.32573 | 0.7756 | -0.2772 | — |
| PM J00133+3908 | 2003-09-16 | MDM | MkIII+Wilbur | 3.33314 | 39.14574 | -0.4503 | -0.188 | 0.02714 |
| PM J00138+3537 | 2003-09-17 | MDM | MkIII+Wilbur | 3.45507 | 35.61685 | -0.1917 | -0.4274 | 0.03687 |
| PM J00146+6546 | 2001-07-24 | Lick | KAST | 3.67664 | 65.78068 | 0.888 | 0.368 | 0.01999 |
| PM J00147-2038 | 2009-09-08 | CTIO-4m | RCspec | 3.68832 | -20.64294 | 0.5177 | -0.1668 | 0.03145 |
| PM J00162+1951E | 2011-10-12 | MDM | MkIII+Wilbur | 4.07055 | 19.8608 | 0.7081 | -0.7487 | 0.06525 |
| PM J00162+1951W | 2011-10-12 | MDM | MkIII+Wilbur | 4.06422 | 19.85715 | 0.7145 | -0.7614 | 0.06572 |
| PM J00178+0006 | 2003-09-17 | MDM | MkIII+Wilbur | 4.45785 | 0.11263 | 0.2742 | -0.4031 | 0.01848 |
| PM J00179+2057W | 2008-11-20 | MDM | MkIII+Nellie | 4.49307 | 20.95368 | -0.2532 | -0.3645 | 0.03596 |
| PM J00183+4401 | 2009-09-21 | MDM | MkIII+Nellie | 4.61267 | 44.02473 | 2.8915 | 0.4119 | 0.28069 |
| PM J00184+4401 | 2009-09-21 | MDM | MkIII+Nellie | 4.62475 | 44.0287 | 2.8633 | 0.3365 | 0.28079 |
| PM J00188+2748 | 2007-11-10 | MDM | MkIII+Wilbur | 4.72523 | 27.81337 | 0.3945 | -0.1118 | 0.05301 |
| PM J00190+0420 | 2003-09-17 | MDM | MkIII+Wilbur | 4.76823 | 4.34715 | -0.3933 | -0.2191 | 0.04199 |

Note—This table is available in its entirety in a machine-readable form in the online journal. A portion is shown here for guidance regarding its form and content.

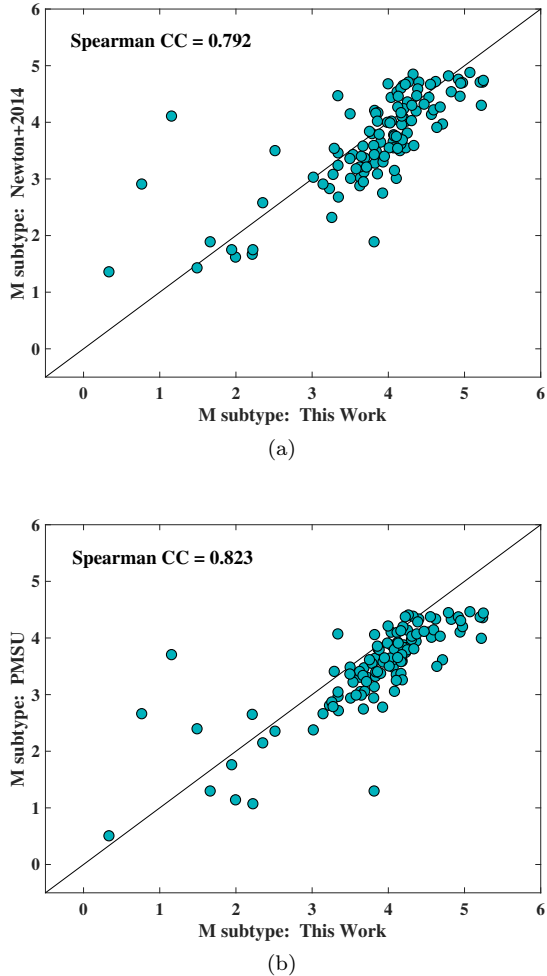

**Figure 6.** **(a)** Comparison between the spectral subtype of 117 disk M dwarfs from the present work and those from the catalog of Newton et al. 2014. **(b)** Comparison between the spectral subtype of the same stars from the present work and those from the PMSU survey. The subtypes from this study are randomized within $\pm 0.25$ subtype in both panels.

stars derived in the present work, which are randomized within $\pm 0.25$ subtype[1], with those calculated by the method described in Newton et al 2014 using NIR spectra. In general, the spectral subtypes from these two different techniques are in agreement, although our own subtypes tend to be slightly later compared with the Newton assignments. In addition, we identify three obvious outliers, the two stars far above the dividing line (the black line), PM J18353+4544 ($SP_{This\ work}$ = M1.0, $SP_{Newton}$ = M4.1, and $SP_{PMSU}$ = M3.7) and

PM J21000+4004E ($SP_{This\ work}$ = M1.0, $SP_{Newton}$ = M2.9, and $SP_{PMSU}$ = M2.7), which are classified as early-type M dwarfs using our pipeline while assigned a mid-subtype in the catalog of Newton et al. 2014, and the star rather far below the line, PM J00184+4401 ($SP_{This\ work}$ = M4.0, $SP_{Newton}$ = M1.9, and $SP_{PMSU}$ = M1.3), which we classify as a mid-subtype M dwarf, but an early-subtype is assigned to this star using the method of Newton et al. 2014. In the panel (b) of Figure 6, we compare the spectral subtype of the above mentioned stars obtained from our pipeline with those drawn from the PMSU survey which uses the strongest TiO feature in optical M dwarf spectra as the primary indicator of spectral type (in contrast to our method in which spectral types are mainly based on CaH molecular band indices). Apart from the same three outliers described above, the subtypes determined from these two methods are also consistent, although our own subtypes again tend to be marginally later compared with the PMSU assignments.

On close inspection, our spectra of the two outliers, PM J18353+4544 and PM J21000+4004E, appear to be equally well-fitted to their corresponding best-fit templates, the evident difference between our subtypes and those from Newton et al 2014 and the PMSU catalogs suggests a more careful revision of our classification templates for early-type M dwarfs. For the third outlier (PM J00184+4401), on the other hand, we are confident that our subtype of M4.0 is a correct representation of that star, and the respective subtypes from Newton et al 2014 and the PMSU catalogs (between M1.0 and M2.0) are mostly likely to be incorrect.

We calculate the Spearman's rank correlation coefficient[2] (hereafter, Spearman CC) of the subtypes from this study with respect to the subtypes from the catalog of Newton et al. 2014, Spearman CC = 0.792, and from the PMSU sample, Spearman CC = 0.823. Although our results are in a better agreement with the study of Newton et al. 2014, there is a slightly stronger correlation between our subtypes and those of the PMSU catalog.

Newton et al. 2014 demonstrated a systematic discrepancy between their subtypes and those from the PMSU as a function of metallicity for stars earlier than M5.0; the NIR subtype is, on average, half a subtype later than the PMSU spectral type for early- and mid-type M dwarfs, with more metal-poor stars showing the largest differences between the spectral types of these two catalogs. These systematic offsets in spectral subtypes can be attributed to TiO molecular bands, which are highly dependent on metallicity, and are employed as the main indicator of subtype in the PMSU survey. Our reliance on the optical CaH band, while PMSU uses primarily the optical TiO band and Newton et al. use NIR spectra, combined with variations due to metallicity, may explain the discrepancies in spectral type assignments between the three methods.

---

[1] Since our spectral subtypes are stated by grid numbers, to avoid a stripe-shaped distribution, we randomized these numbers within a small range, normally not larger than the step size. This randomization does not affect the results of our analysis.

[2] The Spearman correlation coefficient evaluates the strength of a monotonic relationship between two data sets. This coefficient is the nonparametric version of the Pearson correlation coefficient which measures the linear relationship between two variables.

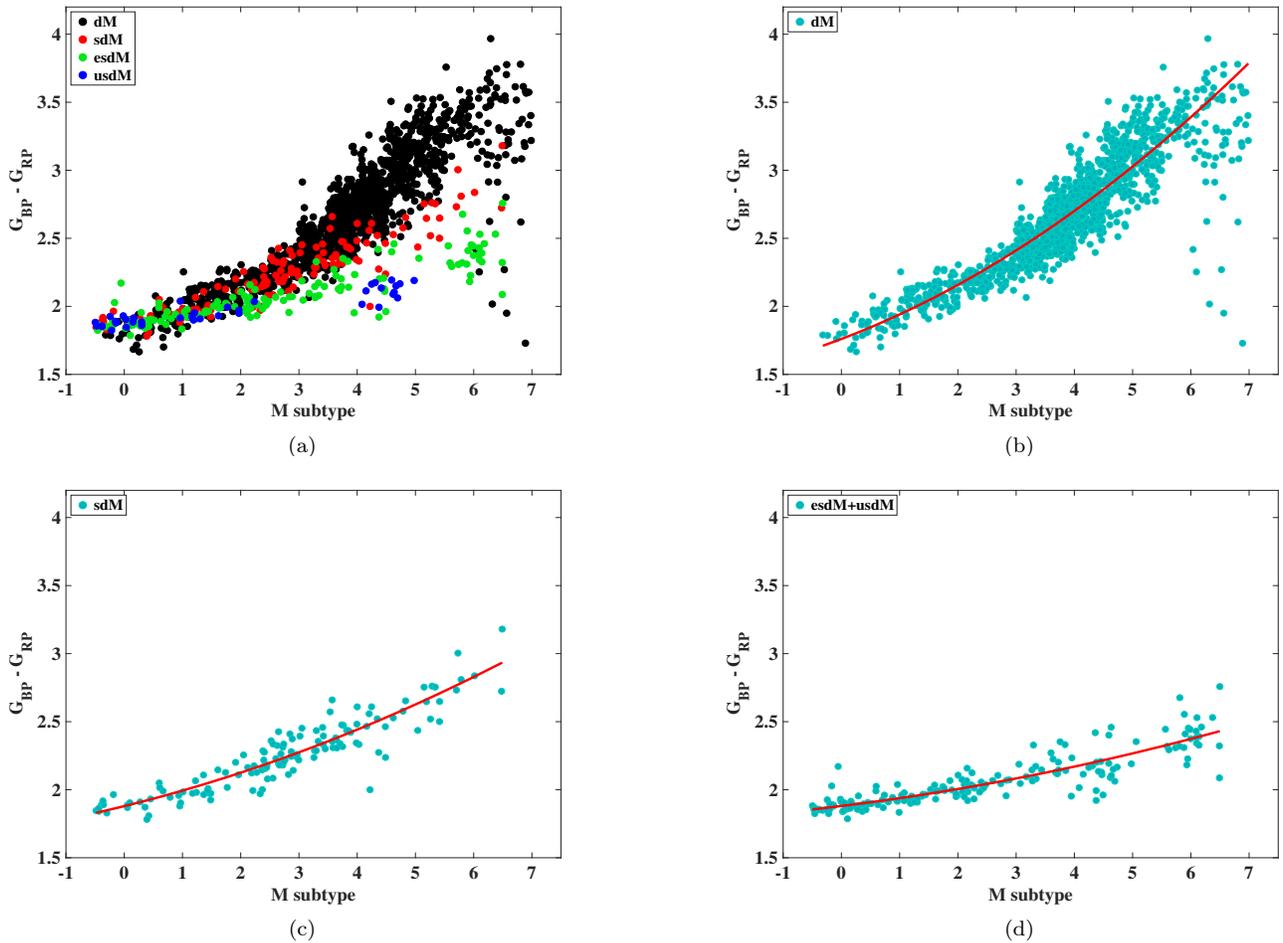

**Figure 7.** **(a)** The color $G_{BP}$ - $G_{RP}$ versus M subtype from M0.0 to M6.5, which are shown by numbers from 0 to 6.5 in steps of 0.5, for 1622, including 1313 dM (black), 133 sdM (red), 136 esdM (green), and 40 usdM (blue) stars. The same plot for **(b)** 1313 dM stars, **(c)** 133 sdM stars, and **(d)** 177 esdM and usdM stars. The red line in panels (b), (c), and (d) shows the best-fit of the second-order polynomial over the data points which are obtained after three passes of $3\sigma$ outlier rejection. The spectral subtypes are randomized within 0.5 subtype.

### 3.5. Color-Spectral Types Relationship

Figure 7 (a) shows the relationship between the Gaia color $G_{BP}$ - $G_{RP}$ (Gaia collaboration et al. 2018) and the spectral type (which are randomized within ±0.5 subtype) of our stars. We excluded 78 stars with uncertain or missing Gaia photometry, including several stars with Gaia colors far outside the M dwarf/subdwarf range ($1.5 \lesssim G_{BP}$ - $G_{RP} \lesssim 4$). The remaining sample consists of 1622 stars, consisting of 1313 dM (black), 133 sdM (red), 136 esdM (green), and 40 usdM (blue) stars.

We present the color-spectral type relationship for dM, sdM and a combined sample of esdM+usdM, in panels (b), (c) and (d) of Figure 7, respectively. There is a clear decrease in the slope of the color-subtype trend from metal-rich to metal-poor stars, which indicates that the color $G_{BP}$ - $G_{RP}$ term becomes less sensitive to subtype for low-metallicity stars, as compared to high-metallicity stars. We perform a second-order polynomial fit over the data points in these panels, in the form of

$$G_{BP} - G_{RP} = a\text{Sp}^2 + b\text{Sp} + c \qquad (2)$$

**Table 4.** Spectroscopic Catalog of of the 1,700 Stars in Our Survey: Classification and Photometry

| Object Designation | Spectral Type | Metallicity Class | $G$ | $G_{BP}$ | $G_{RP}$ | $J$ | $H$ | $K$ |
|---|---|---|---|---|---|---|---|---|
| PM J00012+0659 | dM6.0 | 2 | 14.69 | 16.8 | 13.34 | 11.29 | 10.74 | 10.42 |
| PM J00031+0616 | dM4.5 | 2 | 14.02 | 15.7 | 12.78 | 11.04 | 10.53 | 10.3 |
| PM J00051+4547 | dM1.5 | 2 | 9.08 | 10.2 | 8.06 | 6.7 | 6.1 | 5.85 |
| PM J00101+1327 | dM4.5 | 3 | 14.94 | 16.46 | 13.74 | 12.12 | 11.59 | 11.34 |
| PM J00110+0420 | esdM6.0 | 8 | 16.91 | 18.15 | 15.81 | 14.34 | 13.81 | 13.76 |
| PM J00119+3303 | dM4.0 | 2 | 11.84 | 13.29 | 10.67 | 9.07 | 8.4 | 8.16 |
| PM J00132+6919N | dM4.0 | 3 | 11.73 | — | — | 8.56 | 7.98 | 7.75 |
| PM J00132+6919S | dM4.0 | 3 | 11.73 | — | — | 8.56 | 7.98 | 7.75 |
| PM J00133+3908 | dM3.0 | 3 | 12.68 | 13.88 | 11.6 | 10.18 | 9.61 | 9.36 |
| PM J00138+3537 | dM6.5 | 2 | 13.69 | 15.36 | 12.45 | 10.68 | 10.12 | 9.86 |
| PM J00146+6546 | esdM4.5 | 7 | 15.65 | 16.98 | 14.52 | 12.85 | 12.41 | 12.16 |
| PM J00147-2038 | dM4.0 | 1 | 13.63 | 15.2 | 12.41 | 10.73 | 10.19 | 9.95 |
| PM J00162+1951E | dM4.0 | 1 | 11.89 | 13.53 | 10.65 | 8.89 | 8.34 | 8.1 |
| PM J00162+1951W | dM4.5 | 2 | 10.9 | 12.53 | 9.69 | 7.88 | 7.32 | 7.09 |
| PM J00178+0006 | dM4.0 | 2 | 14.83 | 16.32 | 13.63 | 11.98 | 11.48 | 11.24 |
| PM J00179+2057W | dM1.5 | 3 | 10.99 | 12.05 | 9.99 | 8.69 | 8.06 | 7.83 |
| PM J00183+4401 | dM1.5 | 2 | 7.22 | 8.36 | 6.18 | 5.25 | 4.48 | 4.02 |
| PM J00184+4401 | dM4.0 | 2 | 9.68 | 11.31 | 8.47 | 6.79 | 6.19 | 5.95 |
| PM J00188+2748 | dM4.5 | 2 | 12.52 | 14.15 | 11.28 | 9.53 | 8.93 | 8.65 |
| PM J00190+0420 | dM4.0 | 2 | 12.98 | 14.52 | 11.77 | 10.13 | 9.55 | 9.27 |

Note—This table is available in its entirety in a machine-readable form in the online journal. A portion is shown here for guidance regarding its form and content.

where Sp stands for spectral subtype, while $a$, $b$, and $c$ are the polynomial coefficients whose final values after three passes of $3\sigma$ outlier rejection are listed in Table 5. The best-fit polynomials are plotted as red lines in panels (b), (c), and (d).

**Table 5.** Second-Order Polynomial Coefficients of Color-Subtype Relations

| Metallicity Class | a | b | c |
|---|---|---|---|
| dM | 0.0183 | 0.1620 | 1.7605 |
| sdM | 0.0088 | 0.1054 | 1.8793 |
| esdM+usdM | 0.0049 | 0.0526 | 1.8804 |

Noticeably, there are some outliers that deviate from the general rising trends, and are found significantly above or below the fits. These outliers are explained in Appendix A.1

## 4. PHOTOMETRIC AND KINEMATIC VARIATIONS WITH METALLICITY CLASS

### 4.1. *Distribution in Color-Color Diagrams*

Shown in Figure 8 (a) is the *J-H* vs. *H-K* diagram of the stars in our sample; all values on the plot are randomized to within $\pm 0.004$ of their original values for easier interpretation. The diagram excludes 38 stars that are potentially too faint ($J > 15$) to have reliable 2MASS colors. In addition, 44 stars with obviously inaccurate 2MASS photometry[3], which most probably suffer from large instrumental errors, are also excluded. The final subset displayed in this panel thus contains 1,618 stars, which are divided into four main metallicity classes (dM/sdM/esdM/usdM). Panel (b) in Figure 8 presents the same *J-H* vs. *H-K* diagram, but dM stars are further divided into three subgroups with MC=1, 2, and 3. The color codes corresponding to these two panels are described in the caption of Figure 8.

Panel (c) in Figure 8 shows the *G-K* vs. *J-K* (the color *J-K* is randomized within 0.004) diagram of the subset above. However, this diagram excludes 30 more stars with inaccurate or missing Gaia magnitudes, or with extreme colors beyond the normal range of M dwarfs/subwarf, i.e., $2.2 \lesssim G\text{-}K \lesssim 5.1$. The final subsample presented in this panel then consists of 1588 stars, which are divided into four main metallicity classes. Panel (d) in Figure 8 demonstrates the same color-color diagram, in which dM stars are additionally separated into three subgroups with MC=1, 2, and 3. The respective color codes are also outlined in the caption of Figure 8.

Evidently, all the colors *J-H*, *H-K*, *J-K*, and *G-K* are sensitive to metallicity; from the most metal-rich dM to the most metal-poor usdM stars, these color terms become bluer on average with decreasing metallicity. The color-color diagrams potentially allow one to separate stars by metallicity class, with better separation apparently achieved in the *G-K* vs. *J-K* diagram in part due to the higher sensitivity of the *J-K* and *G-K* color term to metallicity and higher accuracy of Gaia magnitudes. The larger overlap in the *J-H* vs. *H-K* diagram may reflect classification error in some cases, but more likely is due to measurement errors, although another possibility is contaminated light from an unresolved fainter companion, which would have little effect on the optical spectrum but could affect the infrared color more significantly. But one cannot also rule out that optical and infrared colors are affected by physical parameters other than metallicity; for one thing they are clearly dependent on temperature/mass, but they may also be dependent on stellar age and on element-to-element variations in the chemical make-up of the star.

In particular, we identify a number of clear outliers, whose optical-infrared colors appear inconsistent with their metallicity class. In some cases the inconsistency appears to result from errors in spectral classification, but other cases are more perplexing. The most notable outliers are described in detail in Appendix A.2.

This analysis of color outliers demonstrates that accurate optical-infrared photometry is useful in verifying the metallicity classification of M dwarfs/subdwarfs, and can identify abnormal photometric measurements, possibly even pointing to the presence of unresolved companions.

### 4.2. *Distance-Transverse Velocity Diagram*

The transverse velocity, $V_{trans}$, of 1636 stars in our sample, which are matched with the Gaia DR2, is calculated by:

$$V_{trans} = 4.74\mu D \qquad (3)$$

where $\mu$ is the proper motion in units of arc second per year, and the D is distance in units of pc, yielding $V_{trans}$ in units of km/s.

We plot the transverse velocity versus distance of our stars in the logarithmic scale, as shown in Figure 9. This subsample is divided into four groups, as described in the caption. The lower limit of proper motion in our initial target selection ($\mu > 400$ mas/yr) creates a lower envelope in the diagram, including only stars having $V_{trans} \gtrsim 1.8D$. As can be seen from this figure, the metal-rich M dwarfs tend to be closer to the Sun, and have lower velocities, while the metal-poor M subdwarfs are typically farther away from the Sun, moving with higher velocities. These trends are even more obvious in the distance and transverse velocity histograms for dM, sdM, esdM and usdM stars separately, as shown in Figures 10 and 11. Clearly, the maximum of these distributions is shifted from low to high distances and velocities, as one moves from metal-rich dM stars to very metal-poor usdM stars. It is important to note that our high proper-motion sample includes nearby or/and high-velocity stars. Since M dwarfs typically move more slowly than M subdwarfs, the proximity of these metal-rich stars is due to the sample selection effect, as distant stars of low transverse velocities have low proper motions, and are thus not selected. The fact that subdwarfs are detected at much large distances however indicates two things: (1) the local density of M subdwarfs is much lower than that of M dwarfs, and (2) M subdwarfs have much larger average transverse motions, which is how they get selected at large distances.

---

[3] The 2MASS colors of these stars are far outside the range determined by dwarf stars, while their spectra represent them as M dwarfs/subdwarfs.

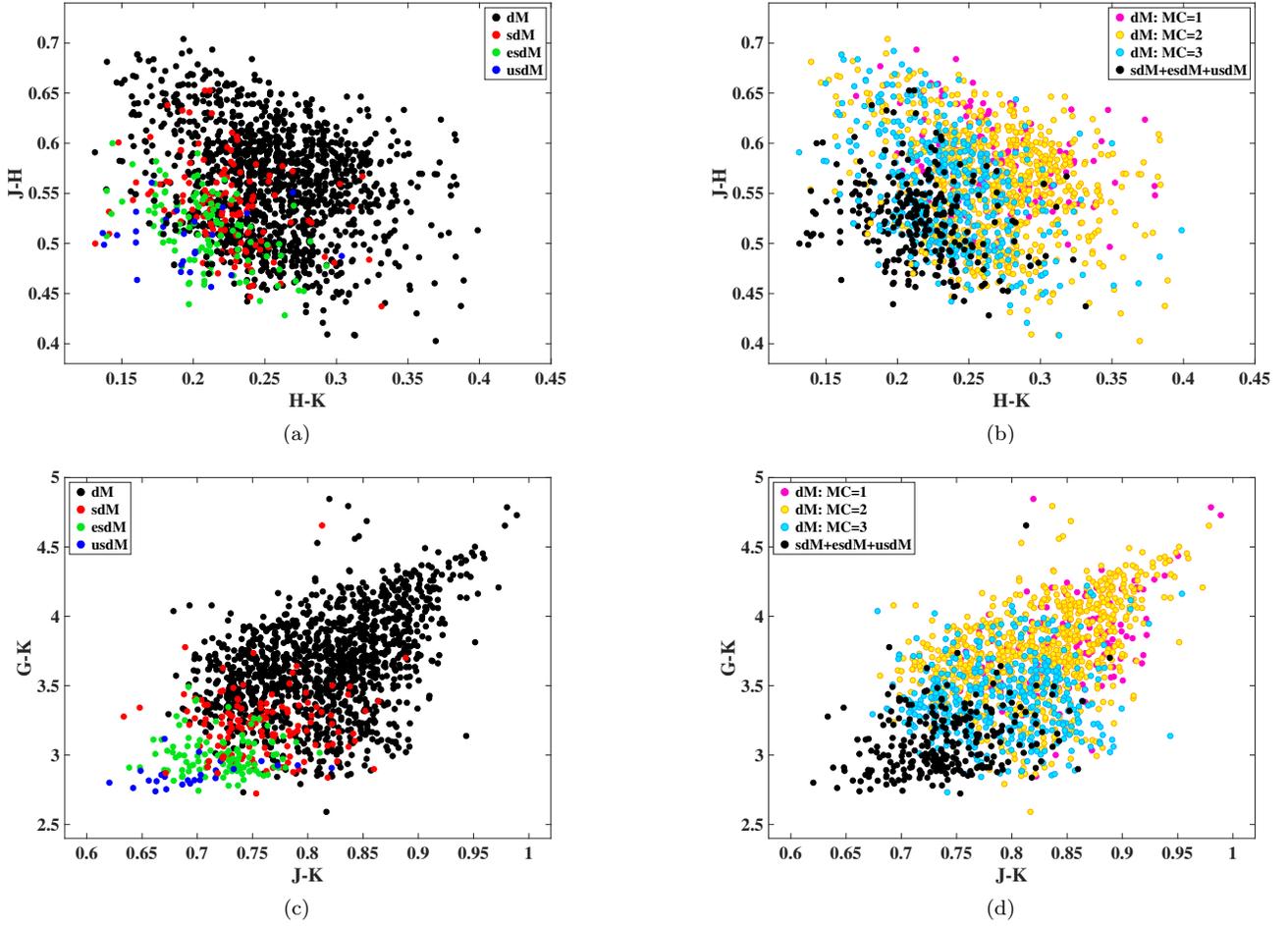

**Figure 8.** Distribution of color-color diagram for stars in our sample, with color-coding to emphasize the metal-poor (left panels) and metal-rich (right panels) stars. **(a)** Distribution of *J-H* vs. *H-K* for 1,618 stars, including 1,353 dM (black), 131 sdM (red), 108 esdM (green), and 26 usdM (blue) stars. **(b)** Distribution of *J-H* vs. *H-K* for 1,618 stars, including 126 stars with MC=1 (magenta), 788 stars with MC=2 (yellow), 439 stars with MC=3 (cyan), and 265 M subdwarfs with $4 \leqslant$ MC $\leqslant 12$ (black). **(c)** Distribution of *G-K* vs. *J-K* for 1588 stars, including 1,324 dM (black), 130 sdM (red), 108 esdM (green), and 26 usdM (blue) stars. **(d)** Distribution of *G-K* vs. *J-K* for 1588 stars, including 124 stars with MC=1 (magenta), 772 stars with MC=2 (yellow), 428 stars with MC=3 (cyan), 264 M subdwarfs with $4 \leqslant$ MC $\leqslant 12$ (black).

It should be kept in mind that the transverse velocity indicates only one component of the total velocity, and the accurate values of radial velocity are also required for a more accurate kinematical analysis of M dwarfs and M subdwarfs.

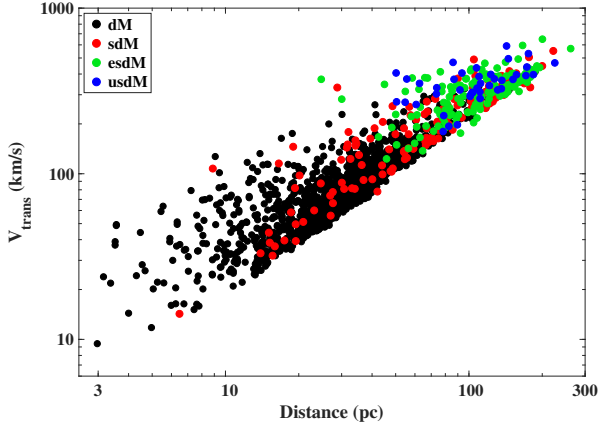

**Figure 9.** Transverse velocity (km/s) vs. distance (pc), in the logarithmic scale, of 1,636 stars, including 1,324 dM (black), 133 sdM (red), 138 esdM (green), and 41 usdM (blue) stars.

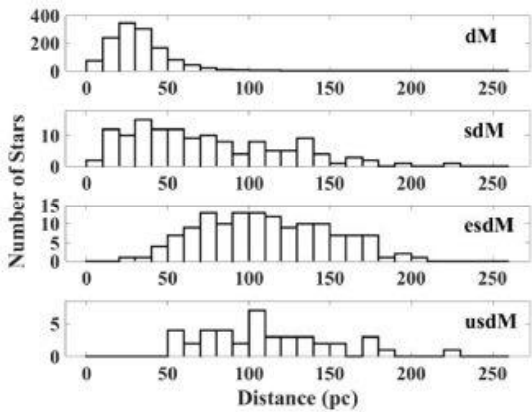

**Figure 10.** Distance distribution of the dM, sdM, esdM, and usdM subsamples described in Figure 9.

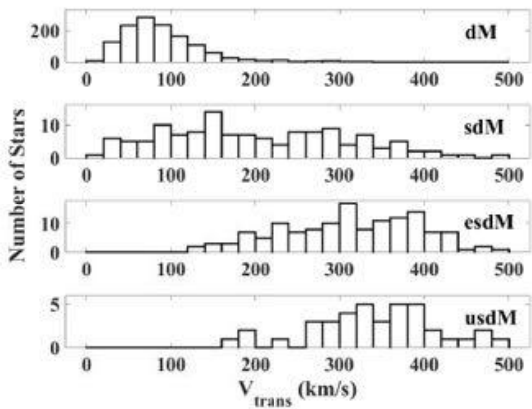

**Figure 11.** Transverse velocity distribution of the dM, sdM, esdM, and usdM subsamples described in Figure 9.

### 4.3. Hertzsprung-Russell Diagram

Figure 12 demonstrates the Hertzsprung-Russell diagram (hereafter H-R diagram) of absolute $G$ magnitude ($M_G$) versus $G_{BP}$ - $G_{RP}$ color, calculated using Gaia parallaxes and optical magnitudes, for 1622 stars in our sample after removing 14 stars with clearly inaccurate or missing Gaia magnitudes/parallaxes, or with colors outside the M dwarf range, i.e., $1.5 \lesssim G_{BP}$ - $G_{RP} \lesssim 4$. Panel (a) shows the H-R diagram of this subsample which are divided into four main metallicity classes, as outlined in the caption. While different metallicity classes fall into separate loci, which suggest a stratification with metallicity, there are some overlaps between metallicity classes, especially between early-subtype dM and sdM stars. These overlaps can be explained in part by uncertainties in the classification; as pointed out in Lépine et al. (2007), at early subtypes (which may not be completely distinguished from late-type K dwarfs), metallicity class separators converge in [CaH2+CaH3, TiO5] plane, and the metallicity class assignment based on these molecular band indices becomes increasingly uncertain.

The H-R diagram of the same sample is shown again in Figure 12 panel (b), but this time the color scheme separates out the three subgroups of dM stars based on their metallicity class, which are explained in the caption. As seen from the metallicity class distribution in Figure 5, the dM population in our sample is dominated by stars with MC=2, but stars with MC=1, the most metal rich ones, make the smallest contribution to this population. There is an obvious stratification between the three metallicity subclasses, with the stars being increasingly shifted to the blue as one go from MC=1 (most metal rich) to MC=3 (least metal-rich). The distribution of stars with MC=1 is thus very distinct from that of stars with MC=3. However, stars with MC=2 are more spread out, overlapping with the sequence of stars with MC=1, which occupy the rightmost part of the diagram. There is also a significant overlap between the distributions of early-type stars with MC=2 and 3, which is likely due to the unreliability of metallicity classes at early-type M dwarfs. We list the most noticeable outliers or apparently misplaced objects in Figure 12 in Appendix A.3.

Despite some overlap in the loci of different metallicity subclasses, the Gaia H-R diagram appears to be an excellent tool for estimating the metallicity of an M-type dwarf - provided the Gaia photometry and parallax data is reliable. A check in the H-R diagram will at least flag stars whose spectral classification may be incorrect.

Panel (a) in Figure 13 presents the H-R diagram for the above sample again, but this time color-code is based on spectral subtype ranges. We divide the sample of 1622 stars into 7 subtype groups, as described in the caption. Apart from the outliers described in Section 3.5, and a few other outliers that may be misclassified, or have uncertain Gaia magnitudes/parallaxes, the different subtype groups are clearly stratified by color, with earlier stars to the blue and later stars to the red. The trend is most obvious for the dM stars, but can also be seen in the sdM/esdM/usdM stars to some extent. As mentioned in Section 3.5, the color $G_{BP}$ - $G_{RP}$ is less sensitive to subtype for metal-poor stars as compared to metal-rich dM stars.

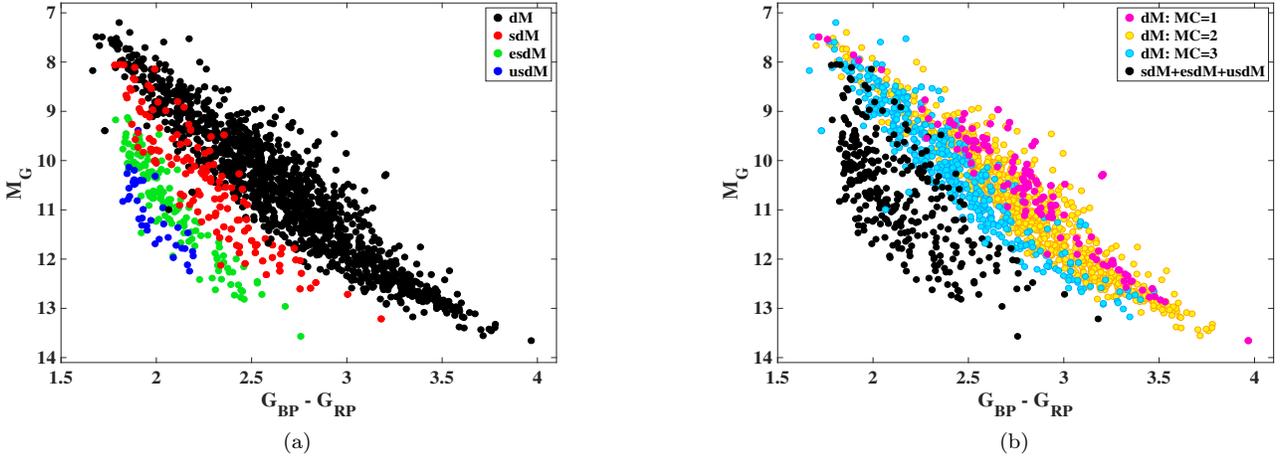

(a)                                                         (b)

**Figure 12.** H-R diagram built from $M_G$ vs. ($G_{BP}$ - $G_{RP}$) color of 1,622 stars, with color-coding to emphasize the metal-poor (left panel) and metal-rich (right panel) stars. **(a)** 1313 dM stars (black), 133 sdM stars (red), 136 esdM stars (green), and 40 usdM stars (blue). **(b)** 124 dM stars with metallicity MC=1 (magenta), 766 stars with MC=2 (yellow), 423 stars with MC=3 (cyan), and 309 M subdwarfs (black).

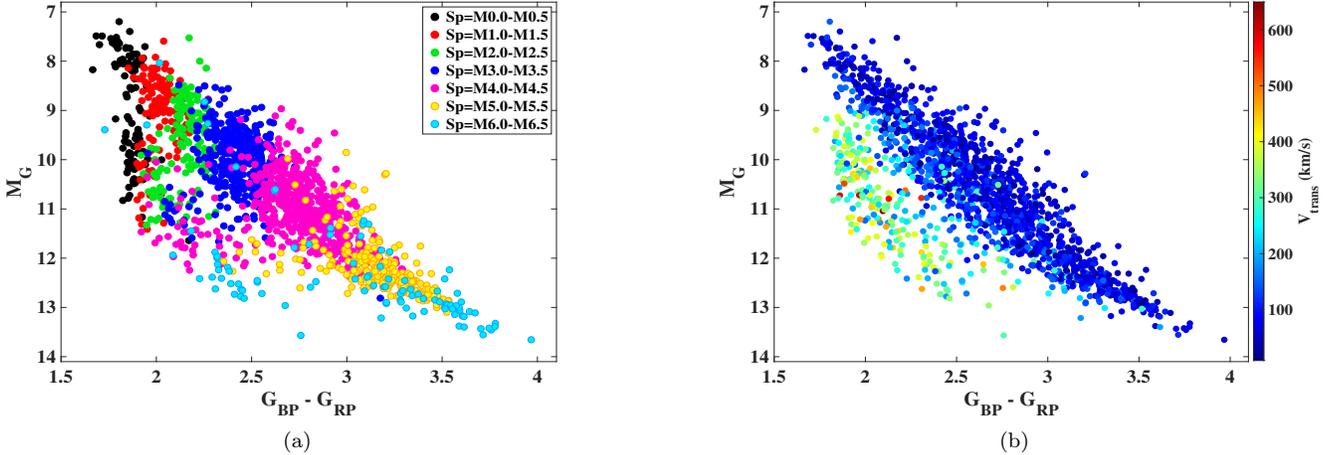

(a)                                                         (b)

**Figure 13.** **(a)** H-R diagram of the 1,623 stars in our spectroscopic sample, divided into 7 groups of different spectral subtype: 102 stars with M0.0-M0.5 (black), 129 stars with M1.0-M1.5 (red), 152 stars with M2.0-M2.5 (green), 359 stars with M3.0-M3.5 (blue), 535 stars M4.0-M4.5 (magenta), 245 stars M5.0-M5.5 (yellow), and 100 stars M6.0-M6.5 (cyan). **(b)** H-R diagram of the same sample, but color-coded based on transverse velocity, and showing a clear separation between the low-velocity disk stars, and the high-velocity halo stars.

Panel (b) in Figure 13 demonstrates the extent to which metallicity is locally a function of stellar kinematics, by color coding the stars in the H-R diagram with transverse velocity. In general, the transverse velocity increases from metal-rich dM stars to metal-poor M subdwarfs. However, there are some high-velocity stars that are found within the locus of the dM stars, and these may represent metal-rich stars in the local old thick disk population. There are also a few low-velocity stars in the M subdwarf domain, but since radial velocities are missing in this analysis, it is possible that these stars may have high total velocity values (see Section 6.3 or our future paper, Hejazi et al. in Prep, for more details).

In summary, we can consider two opposite, diagonal directions in a H-R diagram; one from the upper left to the lower right of the diagram, along which spectral subtype changes, and one from the upper right to the lower left, along which metallicity class and velocity change. These parameters could potentially be calibrated over the H-R diagram, if their accurate values can be determined for representative subsets of stars. Our sample appears to cover a relatively broad range of subtype-/metallicity values, and is thus well-suited for this purpose.

## 5. PHYSICAL PARAMETER DETERMINATION FROM MODEL FITS

### 5.1. *Model Atmospheres*

We employ the latest version of BT-Settl model atmospheres (Allard et al. 2012 a,b, Allard et al. 2013; Baraffe et al. 2015) which has been used in many recent M dwarf/subdwarf studies (e.g., Rajpurohit et al. 2014, 2016, and 2018; Veyette et al. 2016, 2017). These models are computed by a general-purpose state-of-the-art stellar and planetary atmosphere code, referred to as the PHOENIX version 15.5 (Hauschildt 1997; Allard et al. 2001). This model solves the radiative transfer equation in 1D (or more recently in 3D, Seelmann et al. 2010) spherical symmetry under the assumption of hydrostatic and chemical equilibrium, and with a detailed treatment of convection and line by line opacity sampling.

Compared to the previous BT-Settl models (Allard et al. 2001), the current version has been updated in different aspects, including improved molecular linelists, more accurate solar abundance estimates, and better treatment of convection (Rajpurohit et al. 2014; Baraffe et al. 2015). Notably, there has been significant improvement in the water-vapor linelist (Barber et al. 2006), which has an important effect on M dwarf spectra in the IR region. The linelists of metal hydrides such as CaH, FeH, CrH and TiH from Bernath (2006), metal oxides, in particular VO and TiO from Plez (1998), and $CO_2$ from Tashkun et al. (2004) have also been remarkably modified. Although the TiO linelist provided by Plez (1998) is not as complete as the AMES line list (Schwenke 1998) used in the previous BT-Settl models, the Plez linelist reproduces the strength of TiO molecular bands, and accordingly the optical colors, in better agreement with observations.

Stellar model atmospheres generally assume scaled solar abundances for all elements heavier than H and He, and the solar chemical composition is thus required for modeling synthetic spectra. The most recent models adopt the solar elemental abundances of Caffau et al. (2011), which present a considerable reduction of C, N, and O abundances compared to the abundances from Grevesse et al. (1993) previously used in the older version of models.

M dwarfs of the spectral type M3 and later are fully convective, with the convection zone extending all the way from the surface to the core. In stellar atmosphere modeling, convection is treated using the Mixing Length Theory (hereafter MLT, Kippenhahn & Weigert 1990) based on a fixed value of the mixing length $l_{mix}$, specified in terms of the pressure scale height $H_P$. The present BT-Settl model atmospheres use the $l_{mix}$ calibraton of Ludwig et al. (1999, 2002), which relies on the comparison between the 1D MLT models and the 2/3D radiation-hydrodynamical simulations, applicable to main-sequence (hereafter MS) and pre-MS stars down to the hydrogen-burning limit (Freytag et al. 2010, 2012).

Overall, the new models show substantial improvement over past generations. However, there are still discrepancies between synthetic spectra generated from these models and spectra obtained from observations. These likely reflect the incompleteness of opacities and the limitation of the MLT formalism used in modeling cool atmospheres. Further studies on model atmospheres are therefore needed to address these shortcomings and better reproduce observational constraints.

### 5.2. *Model Grid Selection*

The BT-Settl grid available from the CIFIST[4] project utilizes the solar abundances of Caffau et al. (2011). They comprise models calculated with effective temperatures from $T_{eff} = 400$ to 8000 K in steps of 100 K, overall metallicities[5] (i.e, the content of all elements heavier than H and He relative to their solar abundances, on a logarithmic scale) from [M/H] = -2.5 to +0.5 dex in steps of 0.5 dex, and surface gravities from log = 2.5 to 5.5 dex in steps of 0.5 dex. In this grid, $\alpha$-element enhancement $[\alpha/Fe]$[6], defined by the logarithmically-scaled ratio of the amount of $\alpha$-process elements to the amount of iron, compared to the ratio found in the Sun, is not a free parameter. Rather, it is set as a function of [M/H] in the following way: [α/Fe] = 0 (no enhancement) for [M/H] > 0, [α/Fe] = -0.4 × [M/H] for -1 ⩽ [M/H] ⩽ 0 dex, and [α/Fe] = +0.4 dex for [M/H] < -1 dex. This selection of [α/Fe] values is based on the rough estimates of α-element enrichment for the thin and thick disks, adopted by Gustaffsson et al. (2008) in a grid of the MARCS model atmospheres.

---

[4] https://phoenix.ens-lyon.fr/Grids/BT-Settl/CIFIST2011/

[5] [M/H] = $\log_{10}(N_M/N_H)$ - $\log_{10}(N_M/N_H)_\odot$ where $N_M$ is the number density of all heavy elements and $N_H$ is the number density of hydrogen.

[6] $[\alpha/Fe]$ = $\log_{10}(N_\alpha/N_{Fe})$ - $\log_{10}(N_\alpha/N_{Fe})_\odot$ where $N_\alpha$ is the number density of $\alpha$-process elements and $N_{Fe}$ is the number density of iron.

In this paper, we intend to examine the relationship between [M/H] and [α/Fe], and for this purpose, we relax the tight dependency of [α/Fe] on [M/H] mentioned above to a great extent, and treat [α/Fe] as an independent parameter in our model-fitting method. A new grid of BT-Settl models[7] has therefore been calculated, which covers $T_{eff}$ ranging from 2400 to 4000 K, in steps of 100 K, [M/H] spanning from -3.0 to +0.5 dex, in steps of 0.5 dex, and log $g$ = 4.5, 5.0 and 5.5 dex. In addition, this model grid varies [α/Fe] over a range of values, as partially illustrated in Figure 14 (blue grid), which is -0.4 ⩽ [α/Fe] ⩽ +0.4 dex for [M/H] ⩾ 0, -0.2 ⩽ [α/Fe] ⩽ +0.6 dex for [M/H] = -0.5 dex, and 0 ⩽ [α/Fe] ⩽ +0.8 dex for [M/H] ⩽ -1 dex , all in steps of 0.2 dex. These ranges roughly follow typical estimates of [α/Fe] for the Galactic thin disk, thick disk, and halo. As we will show in Section 6.3 below, observational measurements do indeed suggest that metal-poor metal-poor M subdwarfs have generally larger ("enhanced") values of [α/Fe], and conversely, metal-rich M dwarfs generally do not show significant enhancements of α-process elements.

We have interpolated this new grid at every step of 50 K in $T_{eff}$, 0.05 dex in [M/H], 0.025 dex in [α/Fe] and 0.1 in log $g$, using a 4D-cubic interpolation routine in MATLAB[8]. The flux of interpolated spectra varies more smoothly around the original grid points when using the cubic interpolation, as compared to the linear method employed in previous studies (e.g., Rajpurohit et al. 2014). We restrict our grid to $T_{eff}$ = 2600 to 4000 K, [M/H] = -2.5 to +0.5 dex, [α/Fe] = -0.2 to +0.4 dex for -0.45 ⩽ [M/H] ⩽ +0.5 dex, [α/Fe] = -0.2 to +0.6 dex for [M/H] = -0.5 dex , and [α/Fe] = 0 to +0.6 dex for -2.5 ⩽ [M/H] < -0.5 dex, and log $g$ = 4.8 to 5.2 dex (as expected for M dwarfs, except for latest types which are not included in our sample, Gizis 1996; Casagrande et al. 2008; Rajpurohit et al. 2016), yielding a total of 221,125 grid points in the 4D-parameter space. The direct calculation of this large number of models would be prohibitively time consuming, but these interpolated synthetic spectra provide a relatively good approximation of models. Figure 14 compares the distribution of [α/Fe] and [M/H] values in the original model grid (blue dots) and in the interpolated grid (red dots). For comparison, the black squares in this figure show the BT-Settl grid points previously used by the CIFIST, in which [α/Fe] is a function of [M/H], as mentioned above.

### 5.3. Synthetic Spectra Variation

The sensitivity of synthetic spectra to the stellar parameters is crucial in model fitting. We briefly address this point by showing some examples below. Due to the limited observed spectral resolution, we convolve the synthetic spectra using a Gaussian profile of full width at half maximum C ≃ 2.35σ, where σ matches the spectroscopic resolution of our observations.

In Figure 15, we show a selected set of model spectra, smoothed with the Gaussian kernel of C = 7.5 Å, and display how these spectra change with variations of atmospheric parameters. Panel (a) in this figure presents four model spectra with the same $T_{eff}$ = 3200 K, [α/Fe] = +0.2 dex, and log $g$ = 5.0 dex, but different values of [M/H] = +0.5, -0.5, -1.5, and -2.5 dex, showing how the shape and strength of spectral lines/features varies as one moves from very metal-rich to very metal-poor atmospheres. Clear changes in spectral features can be also perceived for the four model spectra with different values of [α/Fe] = 0, +0.2, +0.4, and +0.6 dex, but the same $T_{eff}$ = 3200 K, [M/H] = -1.0 dex, and log $g$ = 5.0 dex, as shown in panel (b). Panel (c) compares four synthetic spectra with a wide range of effective temperatures, $T_{eff}$ = 2800, 3200, 3600, and 4000 K, but equal [M/H] = -1.0 dex, [α/Fe] = +0.2 dex, and log $g$ = 5.0 dex. As can be seen from this panel, both the overall slope and the line/feature strength of the spectra are strongly dependent on the effective temperature.

On the other hand, we observe only very subtle changes in the spectral morphology over our adopted range of log $g$ values, with part of this is due to the limited spectral resolution. As demonstrated in panel (d), except for minor variations in atomic line strengths, there are no significant differences in the major molecular-band features between models with values of log $g$ = 4.8, 5.0, and 5.2 dex, but the same $T_{eff}$ = 3200 K, [M/H] = -1.0 dex, and [α/Fe] = +0.2 dex. Nevertheless, as mentioned later, this narrow range of log $g$ plays an important role in the fine adjustment of parameters in the following fitting pipeline.

By comparing panels (a) and (b) in Figure 15, similarities in the variation of synthetic spectra are noticeable when one either decreases [M/H] or increases [$\alpha$/Fe]. A consequence of this is that a variation due to a decrease in [M/H] can largely be compensated by a sufficient increase in [$\alpha$/Fe]. This is due to the fact that the spectral morphology in this wavelength range is strongly dependent on the $\alpha$ element abundance, which can be varied by changing either [M/H] (which includes all heavy elements) or [$\alpha$/Fe]. This degeneracy can be lifted by using high-resolution spectra in which more spectral details would distinguish apparently similar spectra with different values of [M/H] and [$\alpha$/Fe], but this cannot be achieved at such low resolution. Therefore, special attention should be given when developing a model-fit pipeline using low/medium-resolution spectra, in particular if both [M/H] and [$\alpha$/Fe] are allowed to vary simultaneously.

---

[7] These models have not yet become publicly available by the team.

[8] The syntax interpn is a more extensive version of interp in MATLAB which returns interpolated values of a n-variable function at specific query points, using the method of interest.

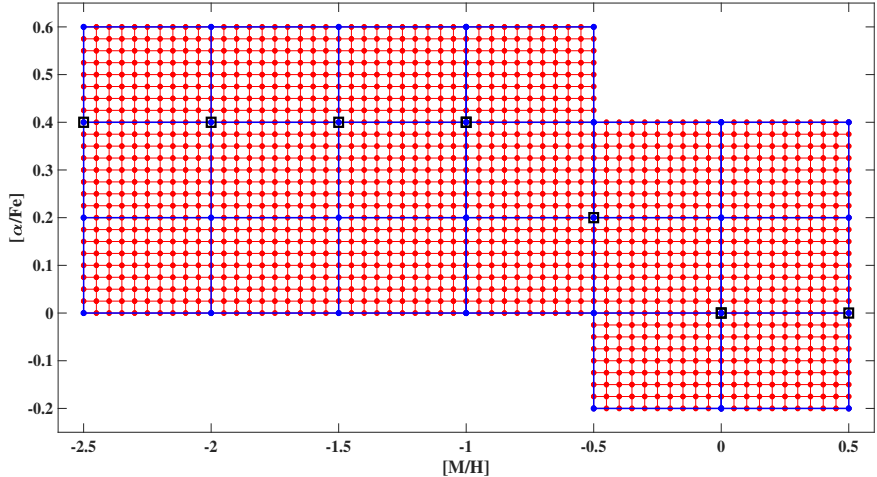

**Figure 14.** Adopted BT-Settl model grid points in metallicity [M/H] and $\alpha$-element enhancement [$\alpha$/Fe]. The original, calculated grid points are shown as blue dots; they follow a step size of 0.5 dex for [M/H] and 0.2 dex for [$\alpha$/Fe]. The interpolated grid points are shown as red dots; they follow a step size of 0.05 dex for [M/H] and 0.025 dex for [$\alpha$/Fe]. The black squares show the much coarser BT-Settl model grid previously adopted by the CIFIST.

We demonstrate the variation in morphology of metal-rich, synthetic spectra with different values of [$\alpha$/Fe] and log $g$, in panels (a) and (b) of Figure 16, respectively. Panel (a) shows the model spectra with equal $T_{eff}$ = 3200 K, [M/H] = +0.3 dex, log $g$ = 5.0 dex, and three values of [$\alpha$/Fe] = 0, +0.2, and +0.4 dex. Panel (b) presents three model spectra of different log $g$ = 4.8, 5.0, and 5.2 dex, but the same $T_{eff}$ = 3200 K, [M/H] = +0.3 dex, and [$\alpha$/Fe] = +0.2 dex.

A comparison between panel (a) in this figure and panel (b) in Figure 15 shows that the overall spectral shape is less sensitive to [$\alpha$/Fe] at high-metallicity values than low-metallicity ones. As shown in Section 5.6, this creates a rather large uncertainty in the estimated values of [$\alpha$/Fe], and is partly the reason for the relatively large scatter of estimated [$\alpha$/Fe] values for metal-rich stars noticeable in the [$\alpha$/Fe] vs. [M/H] diagram (Section 9). Although our very metal-poor spectra are generally noisier, the rather high sensitivity of the overall spectral profile to the [$\alpha$/Fe] ratio in the low-metallicity regime means that the estimated [$\alpha$/Fe] values obtained from model fitting are more reliable. Once more, the variation of log $g$ does not significantly change the appearance of synthetic spectra, though this parameter is important in fine-tuning final parameters.

### 5.4. *Physical Parameters from Synthetic-Model Fitting*

Our main goal for this study is to infer the physical parameters of our stars from their low/medium-resolution spectra, and determine the relationship between these parameters with reasonably high precision[9]. We used a procedure analogous to the spectral classification pipeline described in section 3 above, and use a least-squares minimization algorithm to match the observations with the best-fit BT-Settl synthetic spectra. Physical properties, i.e., $T_{eff}$, log $g$, [M/H] and [$\alpha$/Fe], are treated as primary free parameters, and radial velocity and polynomial order (for the flux correction/recalibration) as secondary free parameters. In this fitting process, we choose the same range of radial velocity shifts (from -500 to 500 km/s) as used in the spectral classification pipeline (section 3.2), but we include both even and odd numbers, i.e., 6, 7, 8, 9, and 10, in the set of polynomial orders. Prior to the minimization process, we smooth the synthetic spectra at the resolution of the observed spectra. Depending on stellar brightness, weather conditions at the time of observation, slit width and instrumental configuration, the spectral resolution differs from one star to another. For this reason, we introduce the FWHM of the Gaussian kernel, C, as a new secondary fitting parameter, with values ranging from 4.5 to 15.5 Å, which approximately reflects the effective resolution range of our sample.

---

[9] It is useful to remember that precision is independent of accuracy, by definition. Some measurements can be very precise but not accurate, or inversely, very accurate but imprecise.

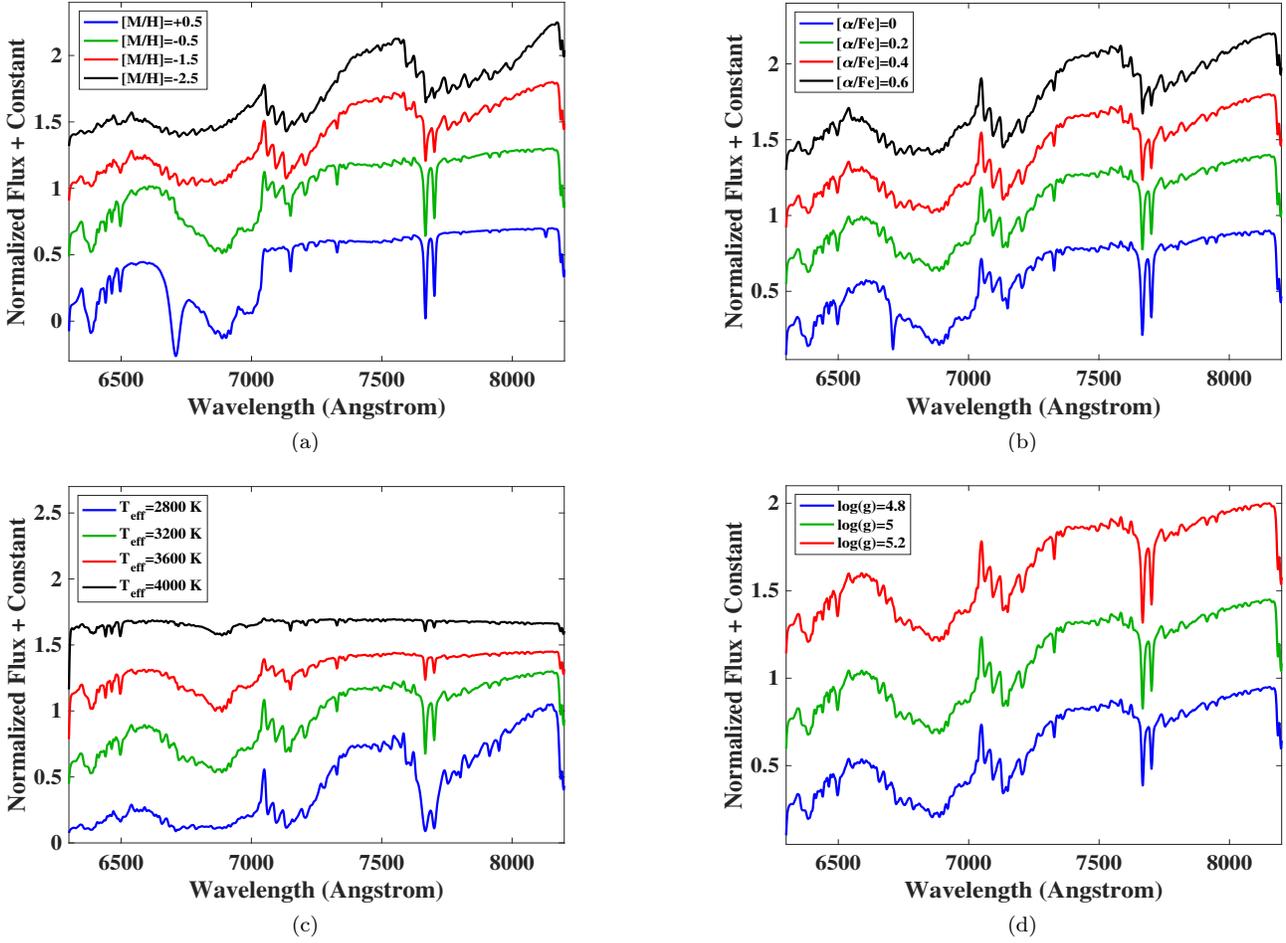

**Figure 15.** Comparison between synthetic models illustrating the effects of varying only one of the model parameters while keeping the others constant. **(a)**: Metallicity variation, $T_{eff}$ = 3200 K, $[\alpha/Fe]$ = +0.2 dex, log $g$ = 5.0 dex, and four different values of [M/H]. **(b)**: $\alpha$-element abundance variation, $T_{eff}$ = 3200 K, [M/H] = -1.0 dex, log $g$ = 5.0 dex, and four different values of $[\alpha/Fe]$. **(c)**: Effective temperature variation, [M/H] = -1.0 dex, $[\alpha/Fe]$ = +0.2 dex, log $g$ = 5.0 dex, and four different values of $T_{eff}$. **(d)**: Surface gravity variation, $T_{eff}$ = 3200 K, [M/H] = -1.0 dex, $[\alpha/Fe]$ = +0.2 dex, and three different values of log $g$.

In order to obtain estimates of the physical stellar parameters, we need to examine all possible combinations of the primary and secondary free parameters, and then find that combination which minimizes the $\chi^2$. Given the large number of the grid points in the 4D-parameter space (Section 5.2) and the large number of stars in our sample, this process can be exceedingly time consuming, and a computationally efficient method is therefore required to determine the stellar physical parameters in a reasonable time. As outlined below, we have developed an automated pipeline, comprising of five passes. In each pass, minimization processes are performed by varying a selected set of parameters while holding the other parameters fixed. The resulting best fits of varying parameters from one pass are then used as starting values in the next pass. Since the stellar physical parameters are highly correlated with each other, we have found that it is difficult to converge to the best solution if all these parameter are allowed to vary at the same time in any one of the passes. Therefore we require that at least one primary parameter should remain fixed in each pass.

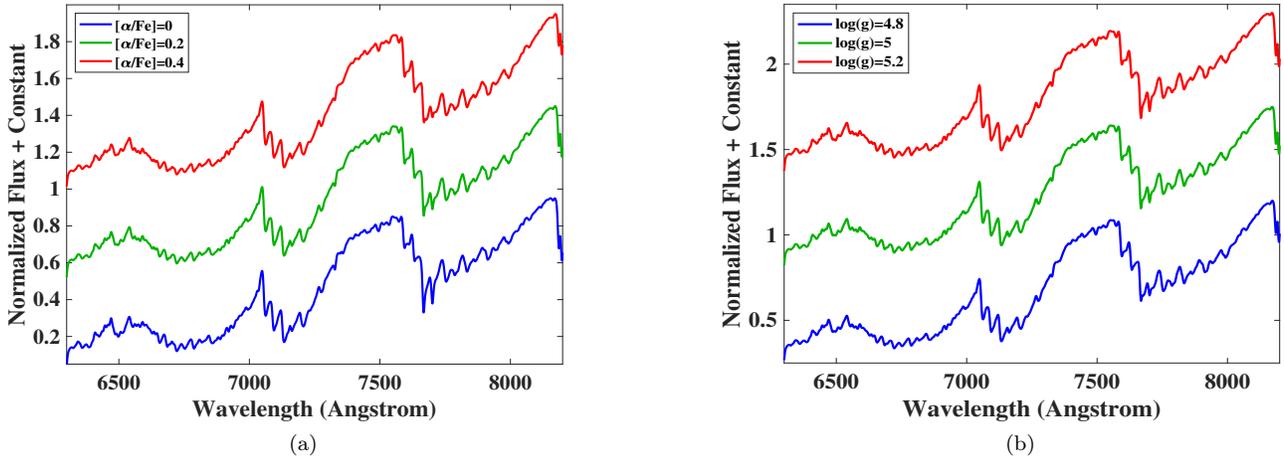

(a)

(b)

**Figure 16.** Comparison between synthetic models showing alpha-element abundance and surface gravity effects in relatively metal-rich stars. **(a):** $T_{eff}$ = 3200 K, [M/H] = +0.3 dex, log $g$ = 5.0 dex., and three different values of [$\alpha$/Fe]. **(b):** $T_{eff}$ = 3200 K, [M/H] = +0.3 dex, [$\alpha$/Fe] = +0.2 dex, and three different values of log $g$.

In general, the parameters $T_{eff}$ and [M/H] have greater effects on the spectral morphology, in both the overall slope of the spectrum and in the strength of spectral lines/features, as opposed to the parameters [$\alpha$/Fe] and log $g$, which have weaker effects. To find a rough estimate of the best-fit model, it is therefore easier to start by varying $T_{eff}$ and [M/H], while keeping [$\alpha$/Fe] and log $g$ fixed at some initial, pre-determined value, and then search for more accurate best-fit models by introducing [$\alpha$/Fe] and log $g$ as varying parameters in the next steps. As pointed out in Section 5.3, since model spectra are not remarkably sensitive to [$\alpha$/Fe] and log $g$ in the high-metallicity regime, we do not vary these two parameters together in any single pass, i.e., one of them is kept fixed as the other is allowed to vary. While progressing from the first to the fifth pass, the minimum value of the $\chi^2$ decreases, and the consistency between the observed and model spectra, and accordingly the obtained best fits progressively improves, and converges to the final set of best-fit parameter values.

The spectral fitting region used in all five passes is chosen to be the same as that used in the template-fit procedure of spectral classification described in Section 3.2. This region extends from 6300 Å to 8200 Å, excluding those sections contaminated by telluric bands and the $H_\alpha$ emission line. By utilizing the same spectral fitting region for both classification and physical parameter determination, we will be able to not only consistently determine the relationship between the resulting classification and physical parameters, but also assess the quality of the classification templates with respect to the synthetic spectra. All these five passes are explained at length in Appendix B. Although we need to perform different passes for each star in our fitting pipeline, the overall time required to complete the entire process is substantially shorter than the time it would take for running a least square minimization over all 221,125 synthetic spectra at once.

### 5.5. *Results*

#### 5.5.1. *Observed Versus Model Spectra*

We have applied our model-fit pipeline to the sample of 1700 M dwarfs/subdwarfs, and inferred their physical parameters, $T_{eff}$, log $g$, [M/H] and [$\alpha$/Fe]. We illustrate the quality of the model-fitting procedure in a series of plots that compare the observed spectra from some of our stars to their best-fit synthetic models. Figure 17 shows the flux-corrected spectrum of the solar-metallicity M dwarf PM J11285+5643N (red) versus its best-fit model (blue) with $T_{eff}$ = 3200 K, [M/H] = 0, [$\alpha$/Fe] = +0.15 dex, and log $g$ = 5.0 dex. Similar to Figures 1-3, the spectral ranges contaminated by telluric absorption bands and the $H_\alpha$ emission line of hydrogen are blanked, and the small fitting region 8175-8186 Å is not shown. The spectral ranges covered by prominent molecular lines and features[10] (see Anders &

---

[10] The electronic transition systems of TiO, $\gamma$ ($A^3\Phi$–$X^3\Delta$) and $\gamma'$ ($B^3\Pi$–$X^3\Delta$), are calculated by Plez 1998 and publicly available in the VALD database (http://vald.astro.uu.se). The list of VO lines are also calculated by Plez based on the technique used in Plez 1998 for calculating the list of TiO lines. A full description of VO electronic transition systems, in particular B-X system ($B^4\Pi^+$–$X^4\Sigma^+$), can be found in Anders & Grevesse 1989. The list of CaH lines, including electronic tran-

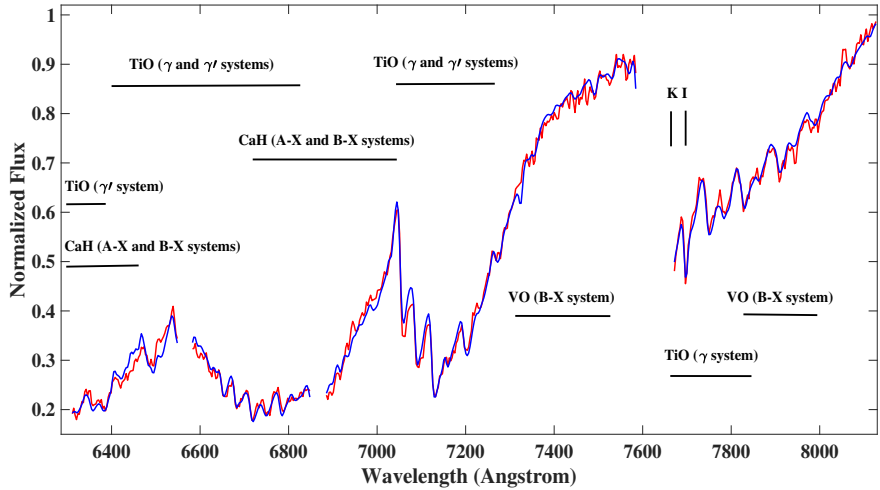

**Figure 17.** Comparison between the flux-corrected spectrum of the M dwarf PM J11285+5643N (red) and its best-fit model (blue) with $T_{eff}$ = 3200 K, [M/H] = 0, [$\alpha$/Fe] = +0.15 dex, and log $g$ = 5.0 dex. The spectral regions covered by the prominent lines and features of the molecules TiO, CaH, and VO are shown by the black horizontal lines. The position of the atomic lines of K I doublet are also depicted by the black vertical lines.

Grevesse 1989; Plez 1998; Weck et al. 2003; Pavlenko 2014; Pavlenko & Schmidt 2015 for more details) as well as the position of the atomic lines of K I doublet are also denoted. One observes that not only are the broad molecular bands comparable between the observations and the models, but so is their fine structure, and many narrow features that might have been perceived as "noise" in the data are revealed to be real.

In Figures 18-25, we further compare the flux-corrected spectra (red) of 16 M-type dwarfs, spanning a wide range of metallicity and $\alpha$-element enhancement values, against their respective best-fit models (blue). In order to better present the level of consistency between our observed and model spectra, the ratio of the flux-corrected spectrum to best-fit model for each star is also shown in the lower panel of each figure. It is important to note that the quality of fits, which can be assessed by the subtraction of flux-corrected spectra from their best-fit synthetic models (although not included in the figures), is strongly depends on spectral noise. On the other hand, the above ratios (while dependent on the local flux) can adequately indicate the discrepancies between observed and model spectra due to the insufficient evaluation of opacities in spectral modeling. These discrepancies differ from one star to another, depending on the values of physical parameters, as described below.

Each of Figures 18-24 shows two stars with nearly the same [M/H] (with a maximum difference of 0.1 dex) and nearly the same $T_{eff}$ (with a maximum difference of 50 K), but with different [$\alpha$/Fe] values. This allows one to examine the effect of $\alpha$-element enhancement on the morphology of spectra for different metallicities. Due to the small number of very metal-poor M subdwarfs in our sample, there are, however, no such two M subdwarfs with [M/H] $\lesssim$ -2.3 dex, having relative parameter values with respect to each other as above. Accordingly, Figure 25 demonstrates two M subdwarfs with similar physical parameters except for their effective temperatures[11].

We identify six distinct regions over which significant discrepancies are noted; these are listed in Table 6. The differences in some regions such as region 3 may be a result of either the insufficient opacities utilized in stellar atmosphere models, or the influence of telluric bands (as this region is adjacent to a spectral range contaminated by atmospheric absorptions), or even a combination of both. In addition, one notices a small difference between the observed spectrum and corresponding synthetic model of some stars, in sharper features, including atomic K I lines. This difference is, in part, due to inaccuracies in the best-fit spectral convolution factor estimated for these stars, which will create larger discrepancies in sharp spectral features. Given the step size of our selected convolution set, i.e., 1 Å (Section sition systems A-X ($A^2\Pi_r - X^2\Sigma^+$) and B-X ($B^2\Sigma^+ - X^2\Sigma^+$) is calculated by Weck et al. 2003, and more recently (2013) by Kurucz (http://kurucz.harvard.edu/ molecules/).

---

[11] In fact, there is an M subdwrf, PM J12044+1329, which has nearly the same temperature as another one, but its spectrum is too noisy to be useful for a proper comparison.

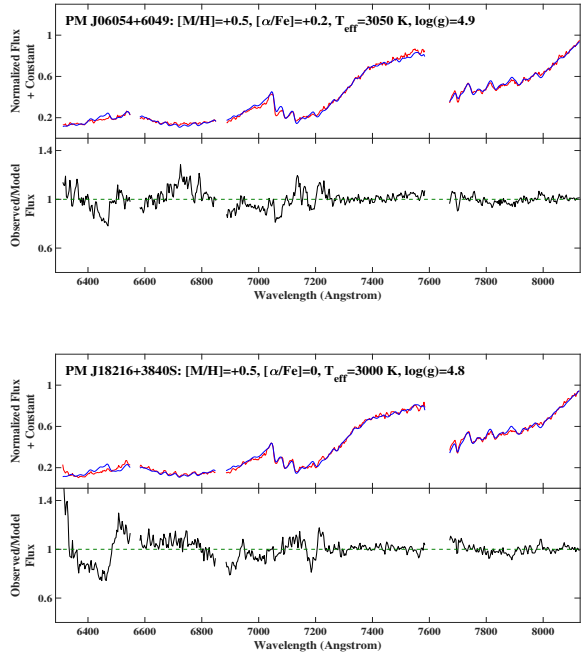

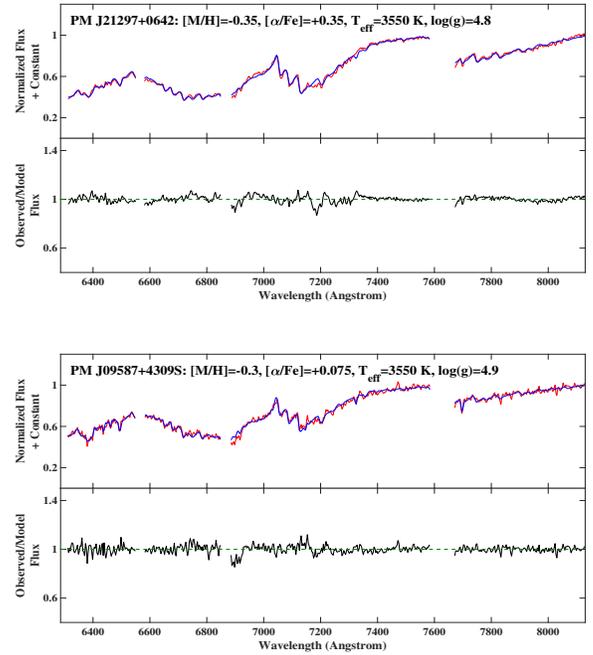

**Figure 18.** Comparison between the flux-corrected spectrum (red) and their respective best-fit model (blue), and the observed-to-model spectrum ratio (black) of two super metal-rich M dwarfs: PM J06054+6049 and PM J18216+3840S.

**Figure 20.** Comparison between the flux-corrected spectrum (red) and their respective best-fit model (blue), and the observed-to-model spectrum ratio (black) of two slightly metal-poor M dwarfs: PM J21297+0642 and PM J09587+4309S.

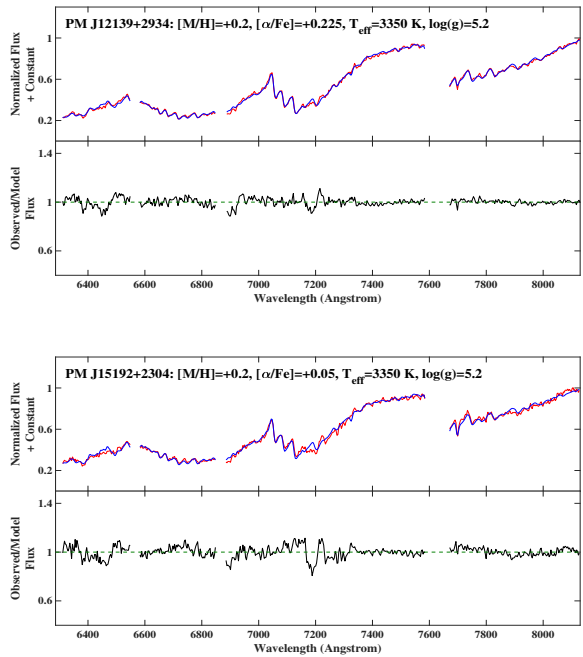

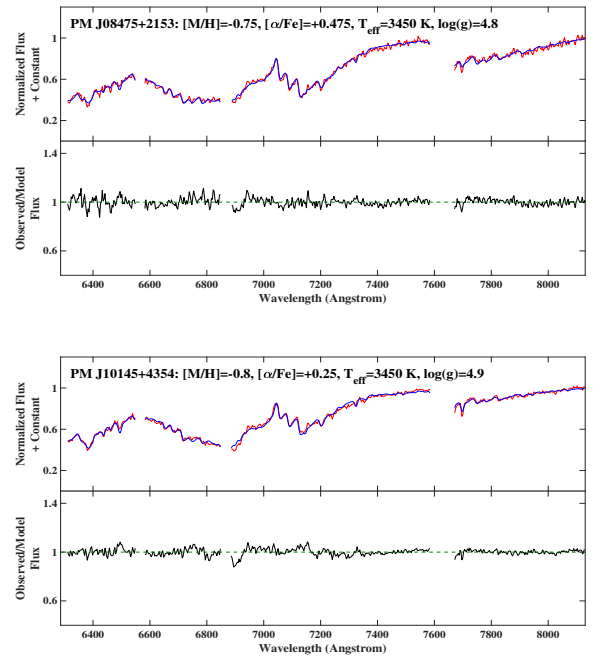

**Figure 19.** Comparison between the flux-corrected spectrum (red) and their respective best-fit model (blue), and the observed-to-model spectrum ratio (black) of two near-solar metallicity M dwarfs: PM J12139+2934 and PM J15192+2304.

**Figure 21.** Comparison between the flux-corrected spectrum (red) and their respective best-fit model (blue), and the observed-to-model spectrum ratio (black) of two moderately metal-poor M subdwarfs: PM J08475+2153 and PM J10145+4354.

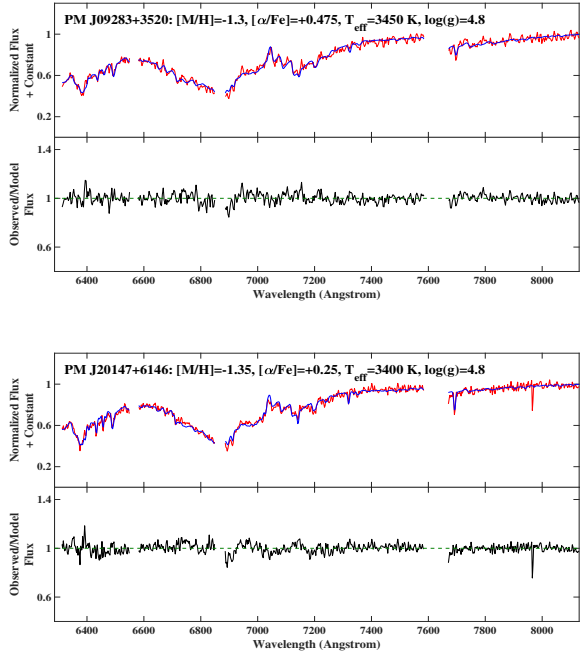

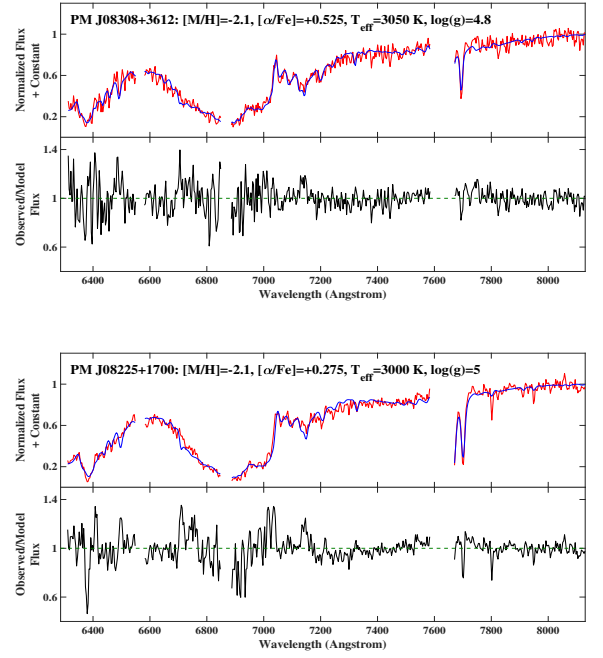

**Figure 22.** Comparison between the flux-corrected spectrum (red) and their respective best-fit model (blue), and the observed-to-model spectrum ratio (black) of two significantly metal-poor M subdwarfs: PM J09283+3520 and PM J20147+6146.

**Figure 24.** Comparison between the flux-corrected spectrum (red) and their respective best-fit model (blue), and the observed-to-model spectrum ratio (black) of two very extreme metal-poor M subdwarfs: PM J08308+3612 and PM J08225+1700.

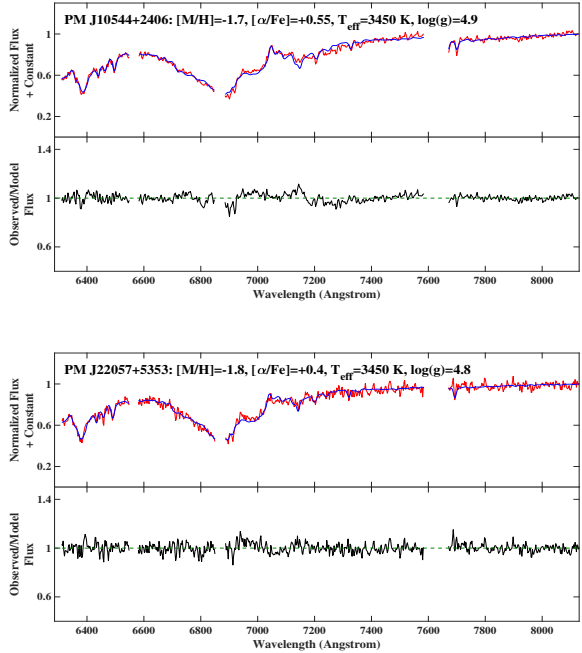

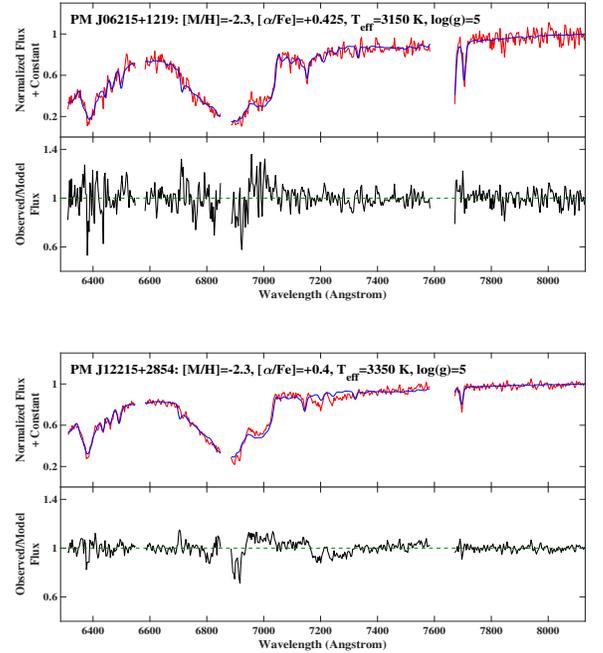

**Figure 23.** Comparison between the flux-corrected spectrum (red) and their respective best-fit model (blue), and the observed-to-model spectrum ratio (black) of two extreme metal-poor M subdwarfs: PM J10544+2406 and PM J22057+5353.

**Figure 25.** Comparison between the flux-corrected spectrum (red) and their respective best-fit model (blue), and the observed-to-model spectrum ratio (black) of two ultra metal-poor M subdwarfs: PM J06215+1219 and PM J12215+2854.

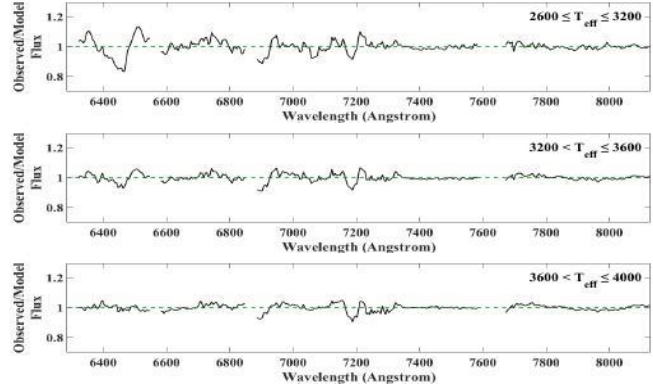

**Figure 26.** Average of the observed-to-model spectra ratios for the subsamples with -0.5 < [M/H] < +0.5 dex, and 2600 ≤ $T_{eff}$ ≤ 3200 K (252 stars), 3200 < $T_{eff}$ ≤ 3600 K (750 stars), and 3600 < $T_{eff}$ ≤ 4000 K (253 stars).

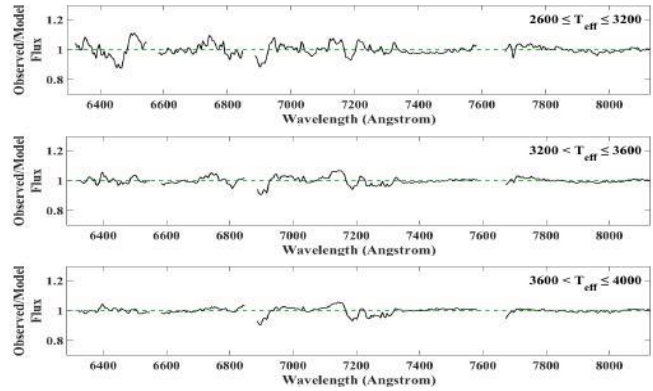

**Figure 27.** Average of the observed-to-model spectra ratios for the subsamples with -1.5 < [M/H] ≤ -0.5 dex, and 2600 ≤ $T_{eff}$ ≤ 3200 K (27 stars), 3200 < $T_{eff}$ ≤ 3600 K (123 stars), and 3600 < $T_{eff}$ ≤ 4000 K (35).

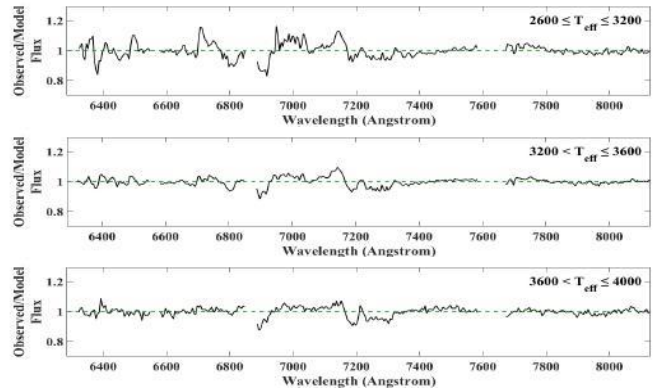

**Figure 28.** Average of the observed-to-model spectra ratios for subsamples with -2.5 < [M/H] ≤ -1.5 dex, and 2600 ≤ $T_{eff}$ ≤ 3200 K (16 stars), 3200 < $T_{eff}$ ≤ 3600 K (43 stars), and 3600 < $T_{eff}$ ≤ 4000 K (3 stars).

5.4), the best-fit convolution factors are only accurate within ±1 Å.

Figure 18 displays the flux-corrected and best-fit model spectra of two super metal-rich M dwarfs. Most spectral lines and features are well reproduced by BT-Settl models. However, the observed/model flux ratio of these two stars reveals significant disagreement be-tween the observed and synthetic spectra over the five regions 1, 2, 3, 4, and 5. Figure 19 presents the spectra of two near-solar metallicity M dwarfs. Compared with the more metal-rich M dwarfs shown in Figure 18, there is significantly more agreement between model and observations, particularly in the spectral regions 1, 2 and 4. However, the disagreement in the region 6 remains significant. The fact that the metal-rich stars show larger discrepancies at blue wavelengths suggest that this is due to deficiencies in the BT-Settl model itself, rather that improper telluric/sky line corrections, which would affect all stars equally. Figure 20 shows the spectra of two slightly metal-poor M dwarfs. The consistency between the observed and model spectra for these two stars clearly improves over most spectral regions, although the discrepancies in regions 3 and 5 remain significant. The compatibility between observed and model spectra even more significantly improves for the two moderately metal-poor M subdwarfs, as shown in Figure 21. There are some minor differences between the observed spectra and synthetic models, but the most important discrepancies occur only in region 3.

**Table 6.** Spectral Regions for Observed vs. Model Spectrum Analysis

| Region Number | Wavelength Range (Å) |
|---|---|
| 1 | 6370-6530 |
| 2 | 6690-6820 |
| 3 | 6885-6930 |
| 4 | 6950-7100 |
| 5 | 7120-7240 |
| 6 | 7240-7320 |

Figure 22 shows the spectra of two significantly metal-poor M subdwarfs. The rather good consistency between the observed and best-fit model spectrum of both stars over many spectral regions, except for some lines and features in the regions 1 and 3, is evident. However, we note a significant difference around 7965 Å for the second star (the lower panel), which in this specific case is due to an artifact absorption in the observed spectrum. The spectra of two very metal-poor M subdwarfs are shown in Figure 23. Although the spectrum of the second star (the lower panel) is noisier than the first one (the upper panel), the observed/model flux ratios of both stars show good consistency between the observed and model spectra. However, the discrepancies in regions 3 and 5, especially for the second star is still notable. The apparent difference around 7680 Å, close to the KI atomic line, in the spectrum of the second star is due to what looks like instrumental noise. In Figures 24 and 25, we present four of the most metal-poor stars in our sample. Since our very metal-poor M subdwarfs are typically dim, their spectra appear to be noisier than more metal-rich stars. The overall shape of the observed spectra of these low-metallicity stars are, however, good matches to their corresponding best-fit models. The lower panel in Figure 24 shows a noticeable difference near the wavelengths 7800 Å and 7950 Å, which we interpret to be instrumental artifacts. Among these four stars, PM J12215+2854 (the lower panel in Figure 25) has the lowest instrumental noise, and provides the most accurate information on the level of consistency between observed and model spectra. This star demonstrates clear discrepancies over some part of all six spectral regions defined in Table 6.

Although the agreement between the observed spectra and the best-fit models obtained from our pipeline is generally acceptable, we remove those stars with [M/H] and [α/Fe] values at the edge of the [M/H]-[α/Fe] grid shown in Figure 14. A best-fit model at the border of the grid may not imply the real minimum $\chi^2$ as the parameter values beyond the grid have not examined in the pipeline. In addition, due to possible unsolved problems in the flux calibration or the insufficient resolving power of some observed spectra as well as unknown issues in model spectra, the fitting process may not converge at a model within the grid. As a result, such best-fit models are not reliable, and are excluded from our following analysis. Nevertheless, owing to the small number of grid points in log $g$ and the well-constrained values of $T_{eff}$ determined by our method (Section 5.6), stars with effective temperature and surface gravity at the upper or lower limit of $T_{eff}$ and log $g$ grids are included in our sample. After trimming, our revised sample consists of 1,544 stars, including 1,227 dM, 138 sdM, 139 esdM, and 40 usdM stars. The physical parameters of these stars are listed in Table 8.

The above comparison of observed spectra with their best-fit synthetic models suggests a trend between BT-Settl model deficiencies and the metallicity of stars. Model deficiencies are also expected to depend on effective temperature. To improve this analysis, we combine the observation-to-model spectral ratios of similar stars together. We divide our sample of 1,544 stars into nine subsamples with different ranges of metallicity and temperature, and take the average of the observed/model flux ratios from all the stars in each subsample. Since the wavelengths of observed spectra are not the same, the direct calculation of average over data points at a single wavelength is not possible. For this reason, we divide the entire wavelength range of spectra into bins of 5 Å. We then compute the mean value of the ratios over each bin of each spectrum separately, and finally find the average of these mean values at every single bin through all stars in each subsample. Our initial sample included 42 stars whose starting wavelength is in the range 3200-3300 Å; we exclude these stars from this analysis as their two or three first bins have no flux data. The final nine subsamples are tabulated in Table 7.

The combined, average ratio of observed to model spectra for these subsamples are shown in Figures 26-28. In each of the three metallicity subgroups, the agreement between observed and model spectra increases as one goes from low to high temperatures. In Figure 26, we compare the ratio plots of three subsamples with high [M/H], and low, moderate and high $T_{eff}$. The most prominent discrepancies in the low-$T_{eff}$ plot are found in the regions 1, 3 and 5, with a maximum difference of $\sim$ 15%, 10%, and 8%, respectively. These differences, particularly those in the region 1, significantly decrease for higher temperatures as seen in the middle and bottom panels of Figure 26. Figure 27 shows the ratio plots of three subsamples with moderate [M/H], and low, moderate and high $T_{eff}$. The most important disagreements in the low-$T_{eff}$ plot occur in the regions 1, 3, and 5, with a maximum difference of $\sim$ 10%, 10%, and 5%, respectively. These discrepancies, especially those in the regions 1 and 5, are considerably reduced for subsamples with higher temperature ranges. Figure 28 compares the ratio plots of three subsamples with low [M/H], and low, moderate and high $T_{eff}$. In the low-$T_{eff}$ panel, there are obvious discrepancies in the regions 1, 2, 3, and 5, where even best-fit synthetic spectra sometimes fail to reproduce some observed features, with a maximum difference of $\sim$ 15%, 15%, 18%, and 12%, respectively. Similar to the two previous figures, these differences are notably mitigated for subsamples with higher temperatures.

### 5.5.2. *Effective Temperature and Metallicity Distribution*

In Figure 29, we plot the distribution of effective temperatures for our sample of observed M dwarfs and subdwarfs, as inferred from the present model-fit pipeline. Values range from $T_{eff}$=2850 K to 4000 K. The general trend of this distribution is similar to that of the subtype distribution in Figure 4, which is expected from the near-linear relationship between temperature and spectral subtype (Section 7.1). The histogram peaks at $T_{eff}$ $\sim$ 3400 K, which is corresponds to subtype M4.0 where the subtype distribution also shows a maximum.

Figure 30 plots the metallicity distribution of the sample, and shows a wide range of values from very metal-rich M dwarfs with [M/H] = +0.45 dex to very metal-poor M subdwarfs with [M/H] = -2.45 dex. The overall metallicity distribution is consistent with the distribution of metallicity classes in Figure 5, suggesting a tight correlation between these two independent measurements of metal content (Section 7.2). The majority of stars have metallicities between [M/H] = -0.5 and

+0.5 dex, and thus mostly belong to the Galactic disk. The relative numbers of metal-rich and metal-poor stars is of course largely determined by our target selection; our high proper-motion sample overrepresents nearby stars mainly from the thin disk, but also tends to overrepresent high velocity stars from the local Galactic halo population. In the end, owing to the rarity of halo objects in the Solar Neighborhood, the sample is dominated by the thin disk stars. The abrupt decrease at [M/H] = -0.5 dex, however, suggests that the sample underselects metal-poor thick disk stars (-1.0 $\lesssim$ [M/H] $\lesssim$ -0.5 dex) which are not common in the Solar vicinity, and also do not have extremely high velocities relative to the Sun. Interestingly, the distribution of metal-poor stars shows a peak at [M/H] $\sim$ -1.4 dex, which may be a result of the overselection of halo stars relative to thick disk objects.

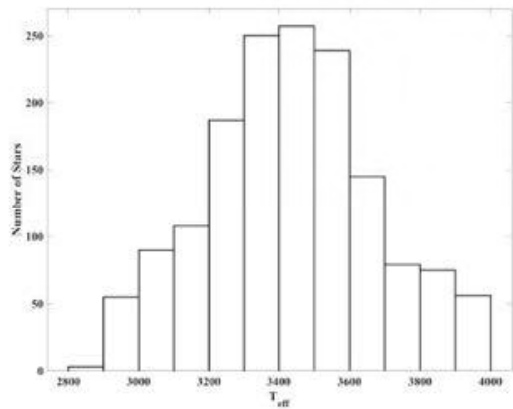

**Figure 29.** The effective temperature distribution of 1,544 stars from the coolest stars with $T_{eff}$ = 2850 K to the hottest stars with $T_{eff}$ = 4000 K.

### 5.6. *Estimated Model-Fit Uncertainties*

The uncertainty in the estimated values of the physical parameters is derived based on a 10% deviation of the $\chi^2$ from from its minimum value, as estimated in the fourth or fifth pass in the minimization algorithm (Section 5.4 and Appendix B). This is a conservative choice when using low/medium-resolution spectra, as compared to the 5% deviation chosen in the previous high-resolution analysis of Rajpurohit et al. 2014.

To infer the uncertainty of [$\alpha$/Fe], we allow $T_{eff}$, [M/H], and [$\alpha$/Fe] to jointly vary, but keep log $g$ fixed, and assemble a subset of all the models in the grid having $\chi^2$ values equal to or less than 1.1 times the minimum value obtained from the fourth pass, $(\chi^2_{min})_{fourth}$.

| Subsample | Number of Stars | Effective Temperature (K) | Metallicity |
|---|---|---|---|
| High $T_{eff}$ - High [M/H] | 253 | $3600 < T_{eff} \leqslant 4000$ | $-0.5 < $ [M/H] $ < +0.5$ dex |
| High $T_{eff}$ - Moderate [M/H] | 35 | $3600 < T_{eff} \leqslant 4000$ | $-1.5 < $ [M/H] $ \leqslant -0.5$ dex |
| High $T_{eff}$ - Low [M/H] | 3 | $3600 < T_{eff} \leqslant 4000$ | $-2.5 < $ [M/H] $ \leqslant -1.5$ dex |
| Moderate $T_{eff}$ - High [M/H] | 750 | $3200 < T_{eff} \leqslant 3600$ | $-0.5 < $ [M/H] $ < +0.5$ dex |
| Moderate $T_{eff}$ - Moderate [M/H] | 123 | $3200 < T_{eff} \leqslant 3600$ | $-1.5 < $ [M/H] $ \leqslant -0.5$ dex |
| Moderate $T_{eff}$ - Low [M/H] | 43 | $3200 < T_{eff} \leqslant 3600$ | $-2.5 < $ [M/H] $ \leqslant -1.5$ dex |
| Low $T_{eff}$ - High [M/H] | 252 | $2600 \leqslant T_{eff} \leqslant 3200$ | $-0.5 < $ [M/H] $ < +0.5$ dex |
| Low $T_{eff}$ - Moderate [M/H] | 27 | $2600 \leqslant T_{eff} \leqslant 3200$ | $-1.5 < $ [M/H] $ \leqslant -0.5$ dex |
| Low $T_{eff}$ - Low [M/H] | 16 | $2600 \leqslant T_{eff} \leqslant 3200$ | $-2.5 < $ [M/H] $ \leqslant -1.5$ dex |

**Table 7.** Subsamples of Stars for Combined Observed/Model Flux Ratios

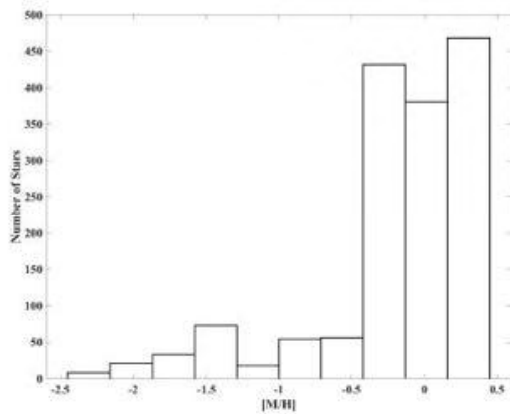

**Figure 30.** The metallicity distribution of 1,544 stars from very metal-rich M dwarfs with [M/H] = +0.45 dex to very metal-poor M subdwarfs with [M/H] = -2.45 dex.

We then calculate the uncertainty of [α/Fe] using the following equation:

$$\delta X = \sqrt{\frac{\sum_{i=1}^{N}(X_i - \overline{X})^2}{N-1}} \qquad (4)$$

where $X$ denotes the parameter [α/Fe], $X_i$ indicates the [α/Fe] value of the $i$th component in the subset of N models that satisfy the condition above, and $\overline{X}$ is the average of all $X_i$ values. The subset of N acceptable models defines a limited volume in the three-dimensional space of $T_{eff}$, [M/H], and [α/Fe], roughly centered around the fourth-pass best-fit values. The projection of this volume on the [M/H]-[α/Fe] plane defines a closed boundary. Some examples are shown in Figure 31, which represents the [M/H]-[α/Fe] parameter space for 8 stars of different metallicities, from very metal-rich M dwarfs to very metal-poor M subdwarfs. Model points within the boundary (black lines) in each panel have $\chi^2$ values deviating by a maximum of 10% from the minimum value corresponding to the fourth-pass best-fit model, whose projection on the [M/H]-[α/Fe] plane is presented by a filled red circle. The average values $\overline{[M/H]}$ and $\overline{[\alpha/Fe]}$ of the acceptable models also define a point on this plane, which is shown by an open blue circle; this point does not generally coincide with the best fit value (filled red circle).

The extent of each closed boundary is a measure of the uncertainty for both parameters [M/H] and [α/Fe]. Depending on the observed noise levels and the consistency between the observed and model spectra, the stretch of the boundary varies from one star to another. Most importantly, once notices that the boundaries are elliptical, which shows that [M/H] and [α/Fe] are not independent parameters to the fits, and that their errors are correlated. The orientation of the stretch in [M/H]-[α/Fe] space however suggests that the sum of the two parameters, i.e. [M/H] + [α/Fe], is best constrained by this model-fit procedure and likely has the most accurate value. This indicates that spectral features in M dwarfs and subdwarfs are mostly correlated with [α/H], assuming [Fe/H] is largely correlated with [M/H] (Eq. 5).

In order to determine the uncertainty of $T_{eff}$, [M/H] and $\log g$, we vary these parameter together, while fixing [α/Fe], and collect a subset of models having $\chi^2$ values equal or less than 1.1 times the minimum value which is obtained from the fifth pass, $(\chi^2_{min})_{fifth}$. We next compute the uncertainties using Eq.4, where $X$ here stands for either $T_{eff}$, [M/H], or $\log g$, $X_i$ is the parameter value of the $i$th component in the above subset, $\overline{X}$ is the average of $X$, and N is the total number of acceptable models. The two parameters $T_{eff}$ and $\log g$ show no obvious correlation with each other, or with one of the chemical abundance parameters [M/H] and [$\alpha$/Fe], which suggest they act like independent parameters.

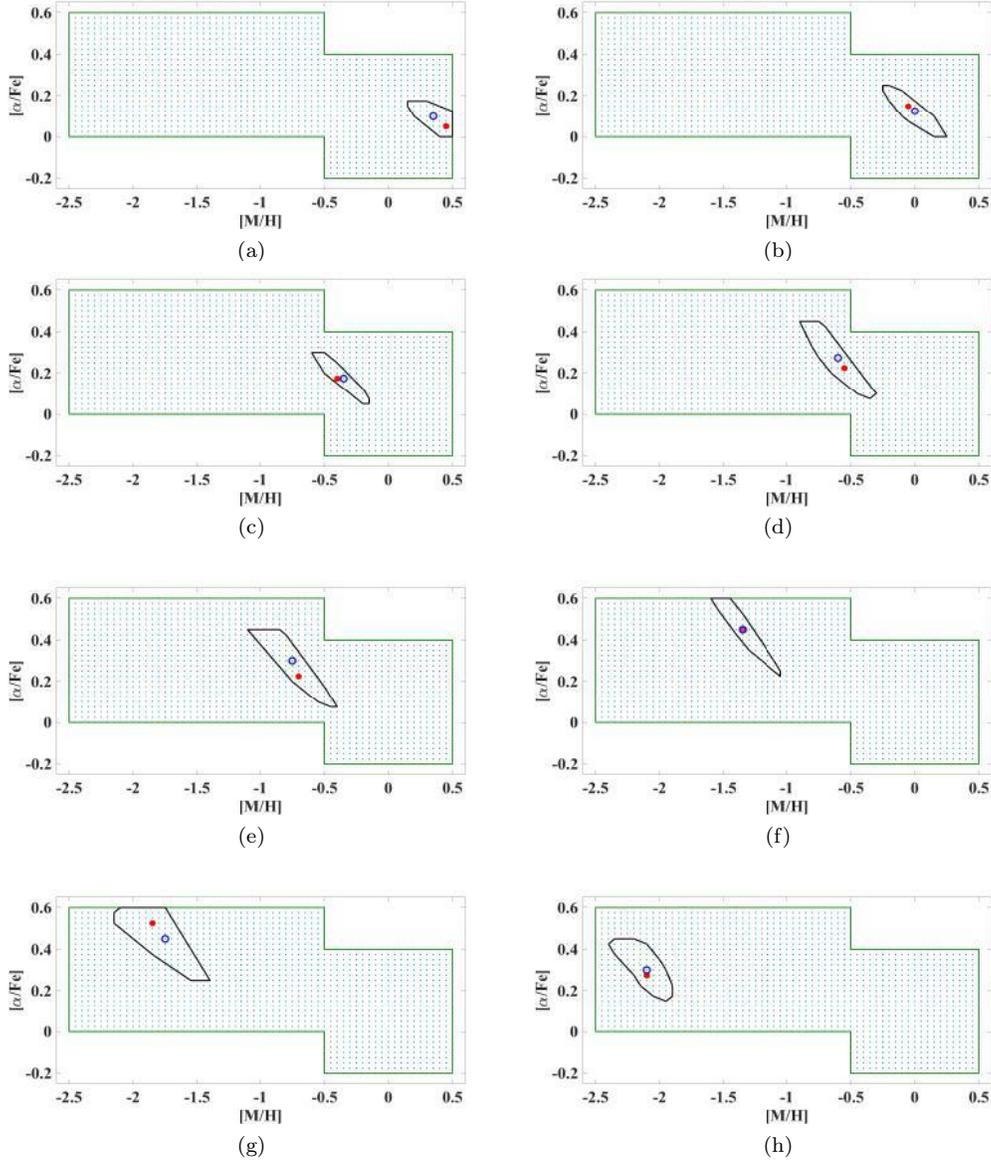

**Figure 31.** Results of the model-fit procedure represented in the [M/H]-[α/Fe] parameter plane. The closed black shape is the projected boundary of the subset of acceptable models with < 10% departure from the minimum $\chi^2$ associated with the fourth-pass best-fit results. The location of the best-fit model is shows by a filled red circle, while the average of [M/H] and [α/Fe] of all acceptable models within the boundary is shown by a open blue circle. The eight stars represented in this figure, and their adopted physical parameters (from the fifth pass) along with their final uncertainty in [α/Fe] derived from Eq.4, are as follows: **(a)** PM J10438+2648, $T_{\mathrm{eff}}$ = 3150 K, [M/H] = +0.45 dex, [α/Fe] = 0.05 dex, log $g$ = 4.9 dex, and δ[α/Fe] = 0.052 dex. **(b)** PM J08236+6413, $T_{\mathrm{eff}}$ = 3050, [M/H] = -0.05 dex, [α/Fe] = 0.15 dex, log $g$ = 4.9 dex, and δ[α/Fe] = 0.073 dex. **(c)** PM J00194+0450, $T_{\mathrm{eff}}$ = 2950, [M/H] = -0.40 dex, [α/Fe] = 0.175 dex, log $g$ = 5.0 dex, and δ[α/Fe] = 0.074 dex. **(d)** PM J07063-0151, $T_{\mathrm{eff}}$ = 3200, [M/H] = -0.55 dex, [α/Fe] = 0.225 dex, log $g$ = 4.8 dex, and δ[α/Fe] = 0.121 dex. **(e)** PM J03369+0250, $T_{\mathrm{eff}}$ = 3350, [M/H] = -0.70 dex, [α/Fe] = 0.225 dex, log $g$ = 5.0 dex, δ[α/Fe] = 0.122 dex. **(f)** PM J02342+1745, $T_{\mathrm{eff}}$ = 3300, [M/H] = -1.35 dex, [α/Fe] = 0.45 dex, log $g$ = 4.8 dex, and δ[α/Fe] = 0.102 dex. **(g)** PM J23370+5027, $T_{\mathrm{eff}}$ = 3300, [M/H] = -1.85 dex, [α/Fe] = 0.525 dex, log $g$ = 4.8 dex, and δ[α/Fe] = 0.100 dex. **(h)** PM J08225+1700, $T_{\mathrm{eff}}$ = 3000, [M/H] = -2.10 dex, [α/Fe] = 0.275 dex, log $g$ = 5.0 dex, and δ[α/Fe] = 0.084 dex.

We calculate the uncertainty of these physical parameters for 1,544 stars in our revised sample using the method described above. The average uncertainty for each parameter is determined as follows: $\overline{\delta T_{eff}} \simeq 49$ K, $\overline{\delta[M/H]} \simeq 0.1$ dex, $\overline{\delta[\alpha/Fe]} \simeq 0.09$ dex, and $\overline{\delta \log g} \simeq 0.1$ dex. The inferred uncertainty values for all stars in our sample are recorded in Table 8. In some cases, we calculate effective temperature errors that are smaller than our parameter grid size, which seems unreasonable, and is likely due to the limited resolution of the grid. For those stars, we assign an uncertainty of 25 K, as an adopted minimum value. Of course these values do not represent the total uncertainties, as systematic errors due to the insufficiency and/or interpolation of models as well as undetected issues with the observed spectra have not been taken into account.

## 6. COMMON PROPER-MOTION PAIRS

Our spectroscopic sample includes 48 common proper motion pairs, for which we have collected spectra from both components. The close similarity in the proper motions along with similar values of the Gaia parallaxes confirms that these pairs are wide binary systems. The primary and secondary components of these 48 pairs along with their inferred classification and chemical parameters [M/H] and [$\alpha$/Fe] are listed in Table 9. Each binary comprises two M dwarfs, or in a few cases two M subdwarfs, whose metallicities are expected to be the same, assuming the two components were formed from the same parent molecular cloud. This suggests that any difference in the metallicity estimates from the two stars should be a measure of the imprecision in our procedure, which may be a reflection of instrumental noise, calibration errors, or issues in the model-fit.

We therefore use our parameter estimates for these binaries to evaluate the magnitude of the precision. In Figure 32 (a), we plot the metallicity of the primary versus that of its companion for all our common proper motion binaries. We find the [M/H] values to be in relatively good agreement, as they show a strong correlation. We calculate a Spearman coefficient of 0.78 for this relationship. We also calculate the average of the differences between the two estimates, $\Delta[M/H]$ = -0.017 dex, and the standard deviation of the differences around that average, $\sigma(\Delta[M/H]) = 0.22$ dex, which is a measure of the precsion. The [$\alpha$/Fe] values of the two components in these binaries are also relatively consistent, as shown in Figure 32 (b), though the smaller scale in the two axes causes the error bars to appear larger in this figure. We find a Spearman coefficient of 0.497, an average difference of $\Delta[\alpha/Fe] = 0$, and a standard deviation of $\sigma(\Delta[\alpha/Fe])$ =0.08 dex. Figure 32 (c) compares the values of the combined chemical parameter [M/H]+[$\alpha$/Fe] between the primaries and their companions, showing a stronger correlation between these values, as compared to panels (a) and (b). We determine a Spearman coefficient of 0.836, an average difference of $\Delta([M/H]+[\alpha/Fe])$ = -0.017, and a standard deviation of $\sigma(\Delta([M/H]+[\alpha/Fe])) = 0.16$ dex. The relatively small standard deviation of [$\alpha$/Fe] is consistent with the narrower range of values constrained by our model-fitting procedure, as illustrated in Figure 31, compared with the larger range of values for the other two parameters [M/H] and [M/H]+[$\alpha$/Fe], in our selected model grid. The comparison between the standard deviations of [M/H] and [M/H]+[$\alpha$/Fe] shows our pipeline is generally more precise in measuring the combined parameter [M/H]+[$\alpha$/Fe] compared with [M/H] alone.

In Appendix A.4, we list the four most prominent outliers for which the difference between the [M/H] value of the primary and its companion is more than twice the estimated precision. Although there is a tighter correlation between the [M/H]+[$\alpha$/Fe] values of the primaries and the corresponding secondaries, as compared to their [M/H] values, the described four binaries also appear to be the outliers for this correlation. Despite these few outliers, we find that the majority of the wide binaries have estimated values of [M/H] and [M/H]+[$\alpha$/Fe] that are in agreement within the estimated precision of the model-fit method. This makes us confident that the method does provide consistent estimates of these two parameters for M dwarfs and subdwarfs. Our future high-resolution spectroscopy will reveal the reason of the inconsistencies in the these outliers.

## 7. COMPARISON BETWEEN CLASSIFICATION AND PHYSICAL PARAMETERS

### 7.1. *Relationship between Spectral Subtype and Effective Temperature*

In order to examine the consistency between the spectral types and effective temperatures determined from the model-fit pipline, we plot the median value of $T_{eff}$ for each spectral type in the sample of 1,544 stars, which are divided into the four original metallicity classes dM, sdM, esdM, and usdM, as shown in panels (a), (b), (c), and (d) of Figure 33, respectively, and described in the caption. To better compare between M dwarfs and M subdwarfs, we overplot dM stars in panels (b), (c) and (d) as well. The vertical error bars indicate the interquartile ranges, in which the difference between the median and the 25th percentile (the first quartile) determines the length below the data points, and the differ-

**Table 8.** Spectroscopic Catalog of the 1,544 Stars in Our Survey: Stellar Parameters Derived from BT-Settl Model Fit.

| Name | $T_{eff}$ (K) | [M/H] | [$\alpha$/Fe] | $\log(g)$ |
|---|---|---|---|---|
| PM J00012+0659 | $2900.00 \pm 25.00$ | $+0.00 \pm 0.13$ | $+0.1750 \pm 0.0663$ | $4.90 \pm 0.11$ |
| PM J00031+0616 | $3200.00 \pm 25.00$ | $+0.30 \pm 0.11$ | $+0.1250 \pm 0.0758$ | $5.10 \pm 0.11$ |
| PM J00051+4547 | $3750.00 \pm 143.61$ | $-0.20 \pm 0.17$ | $+0.1750 \pm 0.1141$ | $4.80 \pm 0.08$ |
| PM J00101+1327 | $3250.00 \pm 33.00$ | $-0.40 \pm 0.10$ | $+0.2250 \pm 0.1298$ | $5.00 \pm 0.14$ |
| PM J00110+0420 | $3250.00 \pm 44.11$ | $-1.90 \pm 0.16$ | $+0.4500 \pm 0.0821$ | $5.00 \pm 0.13$ |
| PM J00119+3303 | $3350.00 \pm 38.85$ | $+0.30 \pm 0.08$ | $+0.0750 \pm 0.0586$ | $5.10 \pm 0.11$ |
| PM J00132+6919N | $3400.00 \pm 29.41$ | $+0.30 \pm 0.09$ | $+0.0750 \pm 0.0673$ | $5.20 \pm 0.08$ |
| PM J00132+6919S | $3400.00 \pm 90.80$ | $+0.40 \pm 0.09$ | $+0.0750 \pm 0.0673$ | $5.20 \pm 0.08$ |
| PM J00133+3908 | $3650.00 \pm 106.90$ | $+0.15 \pm 0.11$ | $+0.0500 \pm 0.0839$ | $4.90 \pm 0.11$ |
| PM J00146+6546 | $3250.00 \pm 25.00$ | $-1.70 \pm 0.06$ | $+0.5750 \pm 0.0686$ | $4.80 \pm 0.09$ |
| PM J00147-2038 | $3300.00 \pm 42.65$ | $+0.10 \pm 0.11$ | $+0.2000 \pm 0.0939$ | $5.10 \pm 0.11$ |
| PM J00162+1951E | $3250.00 \pm 30.11$ | $+0.40 \pm 0.06$ | $+0.0750 \pm 0.0649$ | $4.80 \pm 0.08$ |
| PM J00162+1951W | $3200.00 \pm 25.00$ | $+0.45 \pm 0.06$ | $+0.0750 \pm 0.0655$ | $4.90 \pm 0.10$ |
| PM J00178+0006 | $3300.00 \pm 41.17$ | $-0.05 \pm 0.11$ | $+0.1750 \pm 0.0935$ | $5.00 \pm 0.14$ |
| PM J00179+2057W | $3700.00 \pm 128.42$ | $-0.35 \pm 0.14$ | $+0.2250 \pm 0.1312$ | $4.80 \pm 0.08$ |
| PM J00183+4401 | $3600.00 \pm 115.31$ | $-0.25 \pm 0.10$ | $+0.1250 \pm 0.1152$ | $4.80 \pm 0.08$ |
| PM J00184+4401 | $3250.00 \pm 25.00$ | $-0.20 \pm 0.08$ | $+0.1750 \pm 0.1154$ | $5.20 \pm 0.08$ |
| PM J00188+2748 | $3250.00 \pm 25.00$ | $+0.10 \pm 0.07$ | $+0.1750 \pm 0.0632$ | $5.20 \pm 0.08$ |
| PM J00190+0420 | $3300.00 \pm 34.31$ | $+0.10 \pm 0.09$ | $+0.1250 \pm 0.0875$ | $5.10 \pm 0.11$ |
| PM J00194+0450 | $2950.00 \pm 25.00$ | $-0.40 \pm 0.10$ | $+0.1750 \pm 0.0744$ | $5.00 \pm 0.13$ |

NOTE—This table is available in its entirety in a machine-readable form in the online journal. A portion is shown here for guidance regarding its form and content.

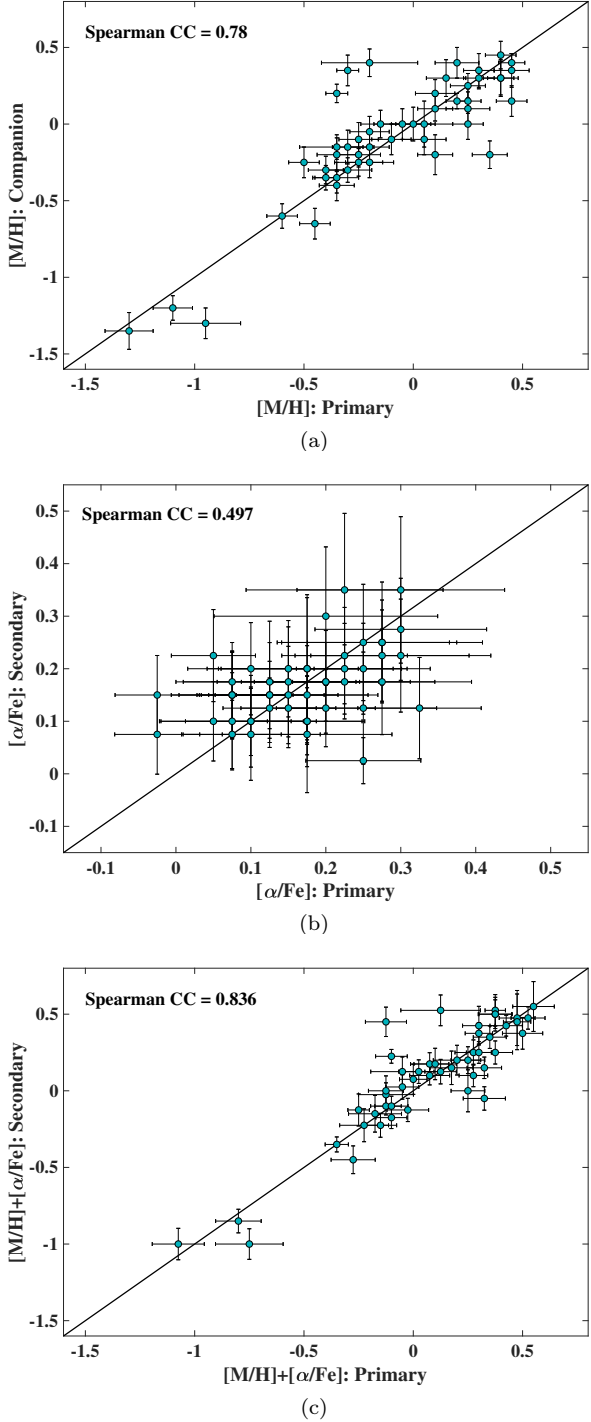

**Figure 32.** Comparison between the metallicity estimates for the primary and secondary components in 48 common proper motion pairs in our sample. **(a)** Comparison between the parameters [M/H] from both components. **(b)** Comparison between the parameters [α/Fe]. **(c)** Comparison between the combined parameters [M/H]+[α/Fe].

ence between the median and 75th percentile (the third quartile) determines the length above the data points.

Apart from slight deviations at early and late subtypes, the median of effective temperature generally decreases as a function of spectral subtype. The median temperature of early-type, metal-poor stars is also typically lower than that of early-type, metal-rich stars, so that an M0.0 dwarf, for example, has a higher effective temperature than an esdM0.0 subdwarf. The slope of the temperature-subtype trend however varies with the metallicity class, and becomes shallower as one goes from metal-rich dM stars to the most metal poor usdM stars. This yields higher values of median temperature at later subtypes for metal-poor stars, as compared to metal-rich stars. As already pointed out in Section 3.5, the color $G_{BP}$ - $G_{RP}$ is less sensitive to subtype for metal-poor M subdwarfs, in comparison to metal-rich M dwarfs. Similarly, the slope change from panel (a) to (d) indicates that spectral subtype becomes less sensitive to effective temperature as one moves from metal-rich to metal-poor stars.

Using their spectroscopic sample of 1564 brightest M dwarfs ($J < 9$) in the northern sky, which is dominated by dM stars, Lépine et al. (2013) inferred the effective temperature of these stars by comparing their observed spectra with BT-Settl synthetic models (the AGSS2009 project[12], Allard et al. 2011). The models include $T_{eff}$ from 3000 K to 5000 K in steps of 100K, [M/H]= -1.5, -1, -0.5, 0.0, 0.0, and +0.5 dex, and log $g$ = 4.0, 4.5, and 5.0, while [α/Fe] is kept fixed as a function of [M/H], the same as described in Section 5.2 for the CIFIST project. The authors presented the median temperature versus spectral subtype from K7.0 to M5.5, and found a plateau in the subtype range M1-M3 (Figure 13 in Lépine et al. 2013). Such a plateau can also be observed for our dM stars in panel (a) near subtype dM2.0, which indicates some level of consistency between the two studies.

### 7.2. *Relationship between Metallicity Class and Chemical Parameters*

In Figure 34 (a), we show the median of metallicity values, estimated using our model-fit pipeline, for each subsample of 12 metallicity classes obtained from the present spectral classification. The vertical error bars are as described in Section 7.1, and set by the interquartile range. Aside from small apparent deviations at MC=5 and MC=8, the median value of [M/H] decreases with increasing metallicity class, from very metal-rich M dwarfs with median metallicity = +0.35 dex at MC=1 to very metal-poor M subdwarfs having median metallicity = -1.95 dex at MC=12. This denotes a strong agreement between two independent measurements of metal content; one using empirical classification templates, assembled based on the shapes of the TiO and CaH molecular bands near 7500 Å, and one using BT-Settl synthetic models calculated from theoretical physics along with various assumptions about stellar atmospheres.

**Table 9.** Classification and Chemical Parameters of the 48 Common Proper-Motion Pairs

| Companion | | | | | Primary | | | | |
|---|---|---|---|---|---|---|---|---|---|
| Name | Spectral Type | Metallicity Class | [M/H] | [α/Fe] | Name | Spectral Type | Metallicity Class | [M/H] | [α/Fe] |
| PM J00162+1951E | M4.0 | 1 | +0.40 ± 0.06 | +0.0750 ± 0.0649 | PM J00162+1951W | M4.5 | 2 | +0.45 ± 0.06 | +0.0750 ± 0.0655 |
| PM J00184+4401 | M4.0 | 2 | −0.20 ± 0.08 | +0.1750 ± 0.1154 | PM J00183+4401 | M1.5 | 2 | −0.25 ± 0.10 | +0.1250 ± 0.1152 |
| PM J00235+7711 | M5.0 | 3 | +0.45 ± 0.09 | +0.0750 ± 0.0757 | PM J00234+7711 | M3.0 | 3 | +0.40 ± 0.07 | −0.0250 ± 0.0567 |
| PM J00358+5241N | M4.5 | 2 | +0.00 ± 0.09 | +0.1750 ± 0.0705 | PM J00358+5241S | M3.0 | 2 | −0.15 ± 0.09 | +0.2250 ± 0.1204 |
| PM J02210+3652 | M4.5 | 3 | −0.25 ± 0.10 | +0.1250 ± 0.1067 | PM J02210+3653 | M4.0 | 3 | −0.50 ± 0.07 | +0.2500 ± 0.0978 |
| PM J02275+0749W | M4.5 | 2 | +0.00 ± 0.11 | +0.2000 ± 0.0865 | PM J02275+0749E | M4.5 | 2 | +0.00 ± 0.10 | +0.2500 ± 0.0893 |
| PM J02456+4457 | M5.5 | 2 | +0.35 ± 0.10 | +0.1000 ± 0.0638 | PM J02456+4456 | M1.0 | 3 | −0.30 ± 0.05 | +0.1750 ± 0.0733 |
| PM J02515+5922W | M4.0 | 2 | −0.10 ± 0.10 | +0.2000 ± 0.0863 | PM J02515+5922E | M4.0 | 2 | +0.05 ± 0.10 | +0.2250 ± 0.0760 |
| PM J02565+5526N | M3.5 | 2 | −0.15 ± 0.08 | +0.2750 ± 0.0971 | PM J02565+5526S | M1.0 | 2 | −0.35 ± 0.17 | −0.3000 ± 0.1145 |
| PM J04059+7116W | M6.0 | 2 | +0.30 ± 0.12 | +0.1750 ± 0.0749 | PM J04059+7116E | M4.5 | 2 | +0.40 ± 0.06 | +0.0750 ± 0.0750 |
| PM J04278+2630S | M4.5 | 2 | +0.35 ± 0.11 | +0.1500 ± 0.0810 | PM J04278+2630N | M3.5 | 2 | +0.30 ± 0.07 | +0.0750 ± 0.0709 |
| PM J04561+2554 | M2.5 | 3 | −0.25 ± 0.10 | +0.0750 ± 0.0874 | PM J04561+2553 | M3.0 | 4 | −0.20 ± 0.11 | −0.1000 ± 0.0925 |
| PM J06007+6808 | M4.0 | 2 | −0.10 ± 0.10 | +0.2250 ± 0.0917 | PM J06008+6809 | M4.0 | 2 | −0.10 ± 0.09 | +0.2250 ± 0.0838 |
| PM J06381+2219E | M4.0 | 2 | −0.15 ± 0.10 | +0.2500 ± 0.0808 | PM J06381+2219W | M3.0 | 2 | −0.20 ± 0.09 | +0.2750 ± 0.0996 |
| PM J07430+5109E | M3.5 | 3 | 0.15 ± 0.05 | −0.1000 ± 0.0925 | PM J07430+5109W | M3.0 | 2 | +0.20 ± 0.05 | +0.0750 ± 0.0950 |
| PM J09070+7226 | M5.5 | 2 | +0.35 ± 0.11 | +0.2000 ± 0.0878 | PM J09069+7226 | M4.5 | 2 | +0.45 ± 0.08 | +0.1000 ± 0.0843 |
| PM J09133+4211N | M3.5 | 3 | −0.35 ± 0.08 | +0.2500 ± 0.1150 | PM J09133+4211S | M2.5 | 3 | −0.40 ± 0.06 | +0.2750 ± 0.1340 |
| PM J09144+5241 | M0.5 | 3 | −0.35 ± 0.10 | +0.2000 ± 0.1353 | PM J09143+5241 | M0.5 | 3 | −0.35 ± 0.10 | +0.1750 ± 0.1334 |
| PM J09587+4309S | M2.0 | 4 | −0.30 ± 0.09 | +0.0750 ± 0.1108 | PM J09587+4309N | M1.5 | 4 | −0.40 ± 0.08 | +0.1750 ± 0.1132 |
| PM J10260+5027N | M4.0 | 2 | +0.20 ± 0.09 | +0.1750 ± 0.0749 | PM J10260+5027S | M4.0 | 2 | +0.10 ± 0.09 | +0.2000 ± 0.0708 |
| PM J10344+4618 | M5.5 | 2 | +0.30 ± 0.12 | +0.1250 ± 0.0752 | PM J10345+4618 | M3.5 | 2 | +0.15 ± 0.09 | +0.1500 ± 0.0633 |
| PM J11113+4325E | M3.5 | 2 | +0.10 ± 0.08 | +0.1000 ± 0.0873 | PM J11113+4325W | M3.5 | 2 | +0.10 ± 0.08 | +0.1000 ± 0.0873 |
| PM J11285+5643S | M4.5 | 2 | +0.00 ± 0.10 | +0.1500 ± 0.0825 | PM J11285+5643S | M4.5 | 2 | +0.25 ± 0.07 | +0.0750 ± 0.0812 |
| PM J11421+2505 | M3.0 | 8 | −1.30 ± 0.10 | +0.3000 ± 0.1319 | PM J11421+2506 | M2.0 | 7 | −0.95 ± 0.16 | +0.2000 ± 0.1492 |
| PM J11488+1800E | M4.0 | 2 | +0.30 ± 0.07 | +0.1250 ± 0.0574 | PM J11488+1800W | M4.0 | 2 | +0.30 ± 0.07 | +0.1250 ± 0.0622 |
| PM J12091+4735 | M5.0 | 2 | +0.10 ± 0.11 | +0.1500 ± 0.0641 | PM J12091+4736 | M5.0 | 2 | +0.25 ± 0.10 | +0.1250 ± 0.0727 |
| PM J12117+2400W | M5.0 | 2 | +0.40 ± 0.10 | −0.1000 ± 0.0864 | PM J12117+2400E | M4.0 | 2 | +0.20 ± 0.10 | +0.1750 ± 0.0769 |
| PM J12278+0512W | M4.5 | 2 | +0.40 ± 0.09 | +0.1250 ± 0.0964 | PM J12278+0512E | M2.5 | 2 | −0.20 ± 0.22 | +0.3250 ± 0.0823 |

The relatively large error bars, at least in the metallicity class range MC=5-9, suggests that the metallicity class as determined from spectral classification is a relatively crude estimate of [M/H]. However, one observes a significantly better correlation when we instead compare [M/H]+[α/Fe] with the metallicity class, as shown in Figure 34 (b). The median of [M/H]+[α/Fe] again generally decreases as a function of metallicity class from very metal-rich stars with median [M/H]+[α/Fe] = +0.4 dex at MC =1 to very metal-poor stars with [M/H]+[α/Fe] = -1.575 dex at MC=12. However, the deviation of the trend at MC=5 and 8 is smaller compared to that seen for [M/H] in panel (a). Moreover, there is a clear decrease in scatter around the median values as represented by error bars, when one uses [M/H]+[α/Fe] rather than [M/H].

We perform a linear regression over the data points, represented by a red line in each of panels (a) and (b). The RMSE of each linear fit is also shown in these panels, with RMSE = 0.16 for panel (a) and RMSE = 0.11 for panel (b). The smaller RMSE in panel (b) indicates that the metallicity class from spectral classification is more tightly correlated with the combined parameter [M/H]+[α/Fe] rather than with [M/H]. If we consider the iron abundance [Fe/H] to be correlated with the overall metallicity [M/H] (i.e., [Fe/H] ∼ [M/H]), then this means:

$$[\alpha/\text{Fe}] + [\text{M/H}] \sim [\alpha/\text{Fe}] + [\text{Fe/H}] = [\alpha/\text{H}]. \quad (5)$$

This suggests that it is the general abundance of alpha elements, [α/Fe] that plays the primary role in the empirical spectral classification system of M-type dwarfs and subdwarfs, and there must be a significant relationship between molecular band indices and this parameter which needs to be addressed in future high-resolution analyses.

## 8. PHOTOMETRIC AND KINEMATIC VARIATIONS WITH CHEMICAL PARAMETERS

### 8.1. *Distribution in Color-Color Diagrams*

Figure 35 (a) shows the *J-H* vs. *H-K* color-color diagram (randomized within 0.004) of most of the stars in our sample. We exclude 34 stars that are potentially too faint ($J > 15$) to have reliable 2MASS colors. We also exclude 43 stars with obviously inaccurate 2MASS photometry, with colors beyond the ranges associated with M-type dwarfs, i.e., $0.38 \lesssim J-H \lesssim 0.74$ and $0.11 \lesssim H-K \lesssim 0.45$, which are most likely due to large instrumental errors. The final subset shown in this panel then includes 1468 stars, which are divided into four groups, as described in detail in the caption of Figure 35. Panel (b) in Figure 35 presents the same *J-H* vs. *H-K* diagram, but the stars in Group 1 are further subdivided into three subgroups, and the color code is also explained in the caption. We choose these metallicity groups/subgroups because the extent of their overlaps in the color-color diagram is minimum. In addition, the distributions of these groups/subgroups resemble those of different metallicity classes shown in Figure 8 (a) and (b).

Panel (c) in Figure 35 displays the *G-K* vs. *J-K* diagram of the same subset as above, but excluding 29 more stars (all from Group 1) with inaccurate or missing Gaia magnitudes, or with extreme colors beyond the normal range of M dwarfs/subdwarf, i.e., $2.2 \lesssim G-K \lesssim 5.1$. The final subsample shown in this panel thus comprises 1439 stars, which are divided into the same four metallicity groups as shown in panel (a), and the color code is depicted in the caption. Figure 35 (d) presents the same color-color diagram, in which the stars in Group 1 are again subdivided in the same three subgroups as demonstrated in panel (b), and are described in the caption.

Clearly, all four colors *J-H*, *H-K*, *J-K*, and *G-K* are sensitive to metallicity to some extent; with the more metal-rich stars typically appearing redder that the more metal-poor ones. These color-color diagrams possibly enable one to separate stars by their [M/H] values. We identify a number of clear outliers, whose optical-infrared colors seem to be inconsistent with their metallicity values. The most prominent outliers are described in detail in Appendix A.5.

We again divide the above samples into different groups/subgroups, but this time using different ranges of the combined parameter [M/H]+[α/Fe]. Figure 36 (a) and (b) show the *J-H* vs. *H-K* diagram of the same sample presented in Figure 35 (a) and (b). Similarly, Figure 36 (c) and (d) show the *G-K* vs. *J-K* diagram of the same sample displayed in Figure 35 (c) and (d). The detailed description of the color code of these new groups/subgroups, as defined based on the values of [M/H]+[α/Fe], is presented in the caption of Figure 36. We again choose these groups/subgroups to minimize the extent of their overlaps in the color-color diagrams. Moreover, the number of stars in each above

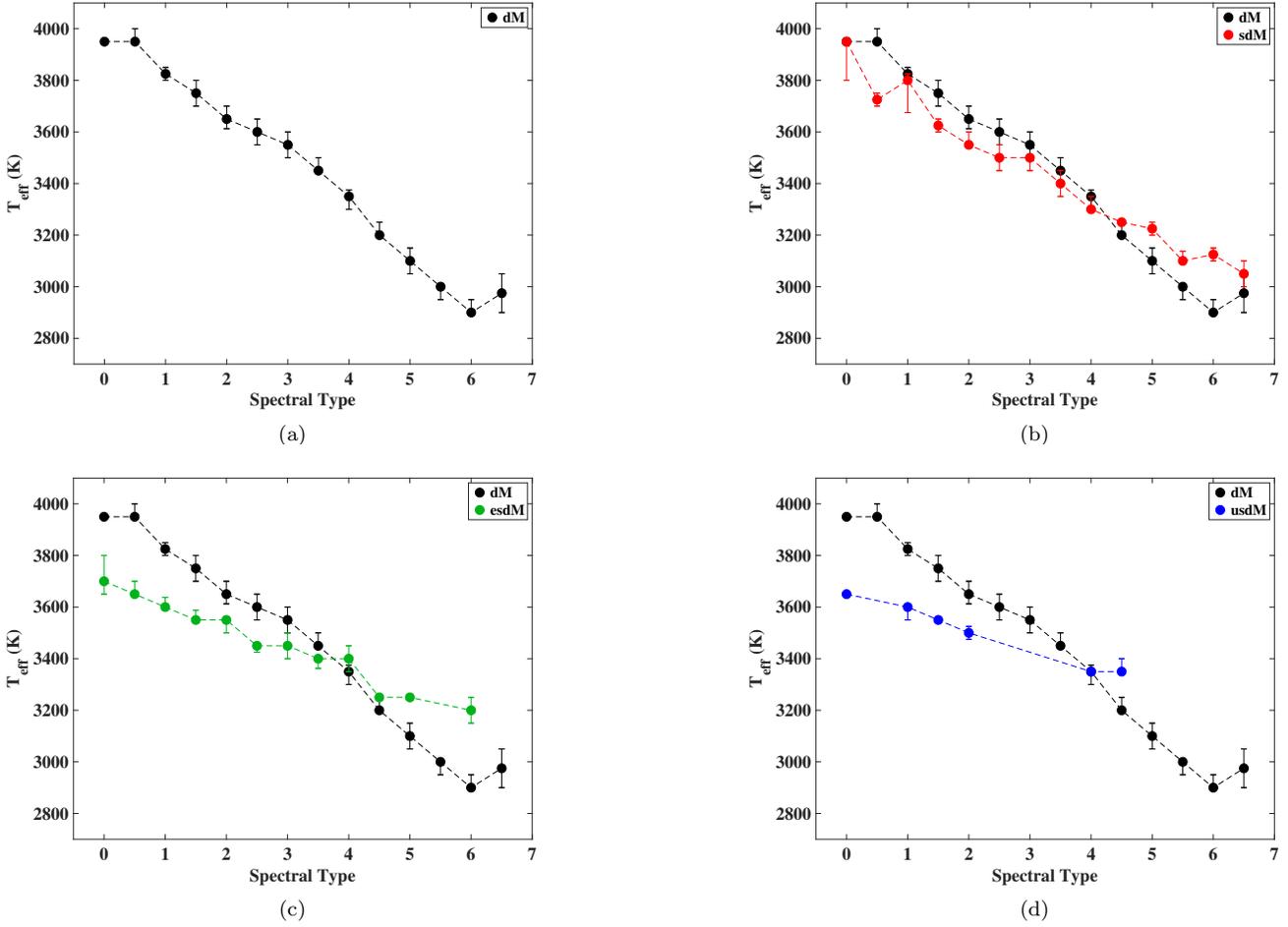

(a)

(b)

(c)

(d)

**Figure 33.** Median values of $T_{eff}$ for each spectral type subsample, calculated separately for all four metallicity classes; **(a)** 1,227 dM stars, overplotted in the other three panels along with **(b)** 138 sdM stars, **(c)** 139 esdM stars, and **(d)** 40 usdM stars. The errors bars are the interquartile ranges, in which the separation of the median values from the first and third quartile determines the length below and above the data points, respectively.

group/subgroup is comparable to that of the equivalent group/subgroup in Figure 35.

To determine which of the metallicity indices, i.e. either [M/H] alone or [M/H]+[α/Fe], produces a better separation of stars in the color-color diagrams, we calculate the histogram intersection of the two normalized color distributions associated with each color and any two groups or subgroups as defined above[13]. Our histogram intersection analysis finds that stellar populations show a better separation when split by [M/H]+[α/Fe] compared to a split by [M/H]. For instance, [M/H+[α/Fe] is marginally better at separating the metal-poor stars from the metal-rich stars by color.

In particular, we notice that the four groups defined based on [M/H]+[α/Fe], presented in Figure 36 panels (a) and (c), show the best separation in the $J$-$K$ color term. Panel (c) in Figure 36 indeed shows that this color term displays the sharpest stratification of these four groups. For example, the group of metal-rich stars (black circles in Figures 35-36, panels (a) and (c)) has an overlap of 0.469 with the moderately metal poor stars (red circles in Figures 35 and 36) in $J$-$K$ when [M/H] is used to define the two groups, whereas the overlap is just 0.456 when [M/H]+[α/Fe] is used to define the groups. The histogram overlaps for the other color terms are larger even for the groups defined by [M/H]+[α/Fe]. On the other hand, results are mixed for the subgroups of metal-rich stars illustrated in panels (b) and (d) in Figures 35-36. The super metal-rich M dwarfs (magenta circles) in particular, are no better separated from the more metal-poor M dwarfs if one uses either [M/H] or [M/H]+[α/Fe]. However, It is the *G-K* color term that best separates these metal-rich subgroups.

---

[13] Histogram intersection is a measure of the similarity between two distinct probability distributions, i.e. normalized histograms, with possible values of the intersection lying between 0 (no overlap) and 1 (identical distributions).

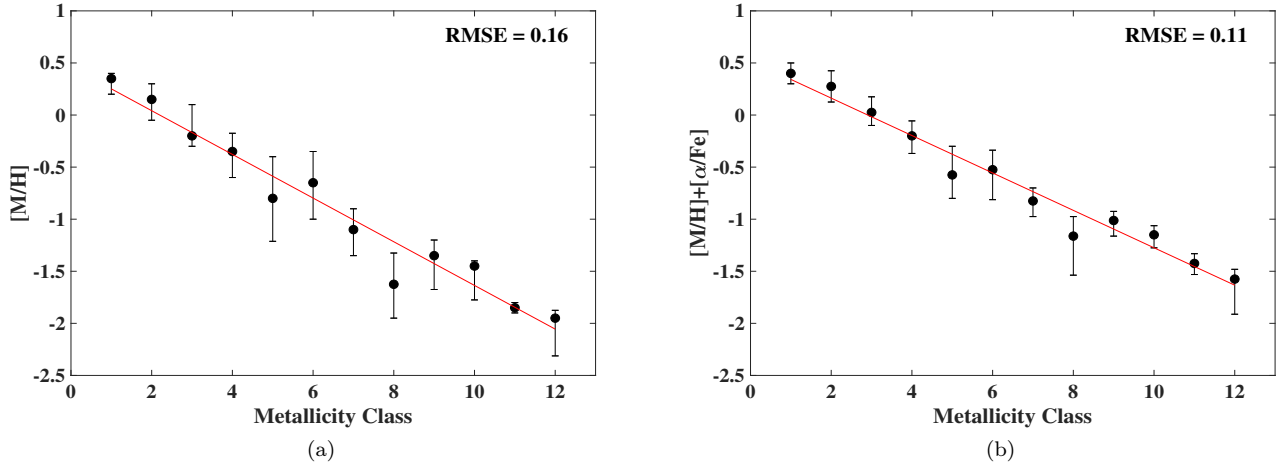

**Figure 34.** **(a)** Median values of [M/H] for each metallicity class subsample, and **(b)** median values of [M/H]+[α/Fe] for each metallicity class subsample of the revised sample including 1,544 M dwarfs/subdwarfs. The errors bars are the interquartile ranges, in which the separation of the median values from the first and third quartile determines the length below and above the data points, respectively. The red lines show the linear best fits over the data points.

### 8.2. *Distance-Transverse Velocity Diagram*

In Figure 37, we plot the transverse velocity, calculated from Eq. 3, versus the distance of 1483 stars in our sample which are cross-matched with the Gaia data. The color coding is based on the same metallicity groups defined in Section 8.1, and described in the caption. The oblique cut in the plot indicates the lower limit of proper motion in our sample ($\mu > 400$ mas/yr), which selects only stars with $V_{trans} \gtrsim 1.8D$, for $V_{trans}$ in km/s and D in pc.

The overall trend for different metallicity ranges in this plot is comparable to that for different metallicity classes as shown in Figure 9; the metal-rich and near-solar metallicity M dwarf, on average, are closer and move more slowly than metal-poor M subdwarfs, which are generally farther away and move faster. This disparity is even clearer in the distance and transverse velocity histograms, as demonstrated in Figures 38 and 39. The maximum of the distribution shifts from lower $V_{trans}$ to higher $V_{trans}$ values, as one plots stars of decreasing metallicity (Group 1 to Group 4).

In spite of the similarity between the separation of spectroscopic metallicity classes (Figures 9-11) and the separation by model fit analysis into metallicity groups (Figures 37-39), there are more extensive overlaps between metallicity classes, particularly in transverse velocity. This suggests the metallicity groups to be a more accurate definition of different stellar populations rather than metallicity classes.

### 8.3. *Hertzsprung-Russell Diagram*

Figure 40 shows the Gaia H-R diagram of 1474 stars in our sample which are cross-matched with the Gaia DR2 - after rejecting 9 stars with inaccurate magnitude/parallax values. The color map is based on our model-fit estimates of metallicity, [M/H], with the most metal-rich stars in the subset shown in dark red, and the most metal poor stars shown in dark blue. There is a clear stratification in this H-R diagram from high to low metallicities, diagonally shifting from the upper right to lower left of the distribution.

In Figure 41 (a), we present the same H-R diagram, in which the stars are divided into the four groups with separate ranges of [M/H]+[α/Fe] as defined in Section 8.1. The color code is described in the caption of Figure 41. Similar to the H-R diagram in panel (a) of Figure 12, we see a clean stratification between these groups, from very metal-rich and near-solar metallicity M dwarfs to very metal-poor M subdwarfs. The metal-rich (in black) are cleanly separated from the metal-poor stars except in one place where a mixture of metal-rich/poor stars is observed: the upper end of the metal-poor sequence in the color-magnitude range $1.80 \lesssim G_{BP}$ - $G_{RP}$ $\lesssim 1.96$ and $9.0 \lesssim M_G \lesssim 10.0$. We believe this is due to the larger uncertainty in estimating metallicity in early-type M dwarfs/subdwarfs, which shows weaker molecular bands.

In panel (b) of Figure 41, we further subdivide the stars in Group $1\alpha$ into three subgroups as defined in Section 8.1, and described in the caption. We notice a clear stratification between these three metal-rich subgroups as well, however, there is a notable overlap between ad-jacent subgroups, in particular between Group1Bα and Group 1Cα for early-type stars ($G_{BP}$ - $G_{RP}$ ≲ 2.3). More importantly, nearly all metal rich stars from Group 1Aα have absolute magnitudes $M_G < 9$, and indeed also have spectral subtypes later than M3.0. This anomaly would appear to suggest that early-type M dwarfs somehow cannot be very metal-rich, which is hard to believe. A more likely explanation is that this may be due to a selection effect: our proper motion selected subset has both a proper motion and a magnitude limit, the combination of which tends to overselects high-velocity (and therefore metal-poor) stars of brighter absolute magnitudes, which are sampled over a much large volume.

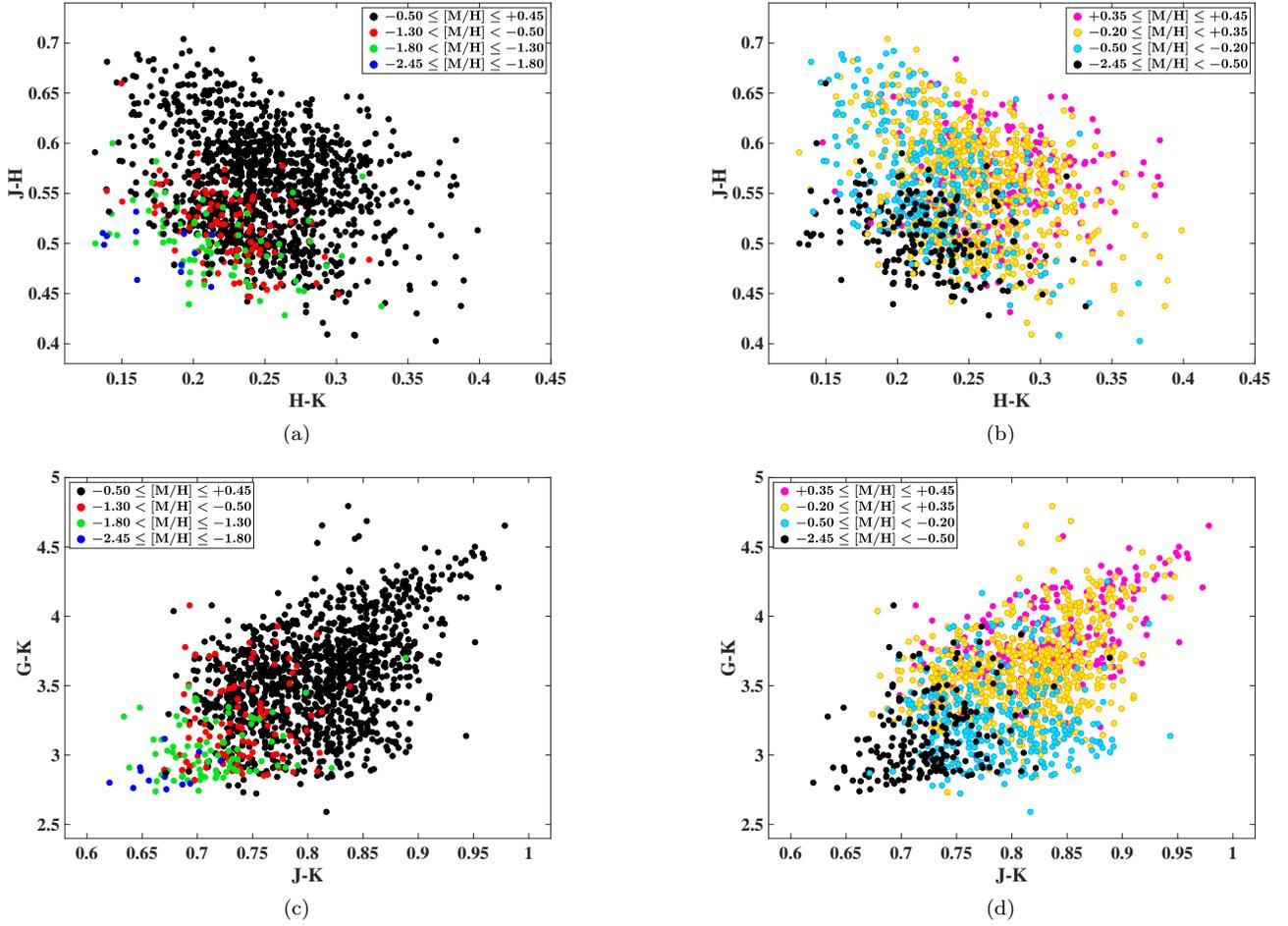

(a)

(b)

(c)

(d)

**Figure 35.** Distribution of ($J$-$H$) vs. ($H$-$K$) for 1,468 stars in our sample, with color-coding to emphasize the metal-poor (left panels) and metal-rich (right panels) stars. **(a)** 1,279 stars with -0.5 ⩽ [M/H] ⩽ +0.45 (black, Group 1), 99 stars with -1.3 <[M/H] < -0.5 (red, Group 2), 78 stars with -1.8 < [M/H] ⩽ -1.3 (green, Group 3), and 12 stars with -2.45 ⩽ [M/H] ⩽ -1.8 (blue, Group 4), and for the same subsample, including **(b)** 228 stars with +0.35 ⩽ [M/H] ⩽ +0.45 (magenta, Group 1A), 732 stars with -0.2 ⩽[M/H] < +0.35 (yellow, Group 1B), 319 stars with -0.5 ⩽ [M/H] < -0.2 (cyan, Group 1C), and 189 stars with -2.45 ⩽ [M/H] < -0.5 (black). Distribution of ($G$-$K$) vs. ($J$-$K$) for 1439 stars, including **(c)** 1,250 stars in Group 1 (black), 99 stars in Group 2 (red), 78 stars in Group 3 (green), and 12 stars in Group 4 (blue), and for the same subset, including **(d)** 226 stars in Group 1A (magenta), 712 stars in Group 1B (yellow), 312 stars in Group 1C (cyan), and 189 stars with -2.45 ⩽ [M/H] < -0.5 (black).

However, another possible cause for these "missing early-type, metal-rich M dwarfs" might be shortcomings of the BT-Settl models. We recall that M dwarfs with the spectral subtype of M3 or later are entirely convective. However, convection in the superadiabatic layer of stars is affected by metallicity. Theoretical simulations have shown that if metallicity increases, the location of the transition region (from efficient to inefficient convection) is forced to lower densities and pressures, yielding larger mean and turbulent velocities over the superadiabatic region (Tanner et al. 2013). Unfortunately, the effect of metallicity on convection is not correctly taken into account in current MLT models. Tanner et al. 2013 have demonstrated that applying a solar-calibrated mixing length parameter to all stellar models cannot be accurate, and any improvement on the treatment of convection has to account for chemical composition. A consequence of this may result in metallicity being systematically overestimated in fully convective stars, based on a comparison of their observed spectra with the BT-Settl model spectra. Therefore, the problem may not be that super metal-rich stars are missing at earlier subtypes, but rather that the "super metal-rich" stars found at later subtypes simply have overestimated metallicity values.

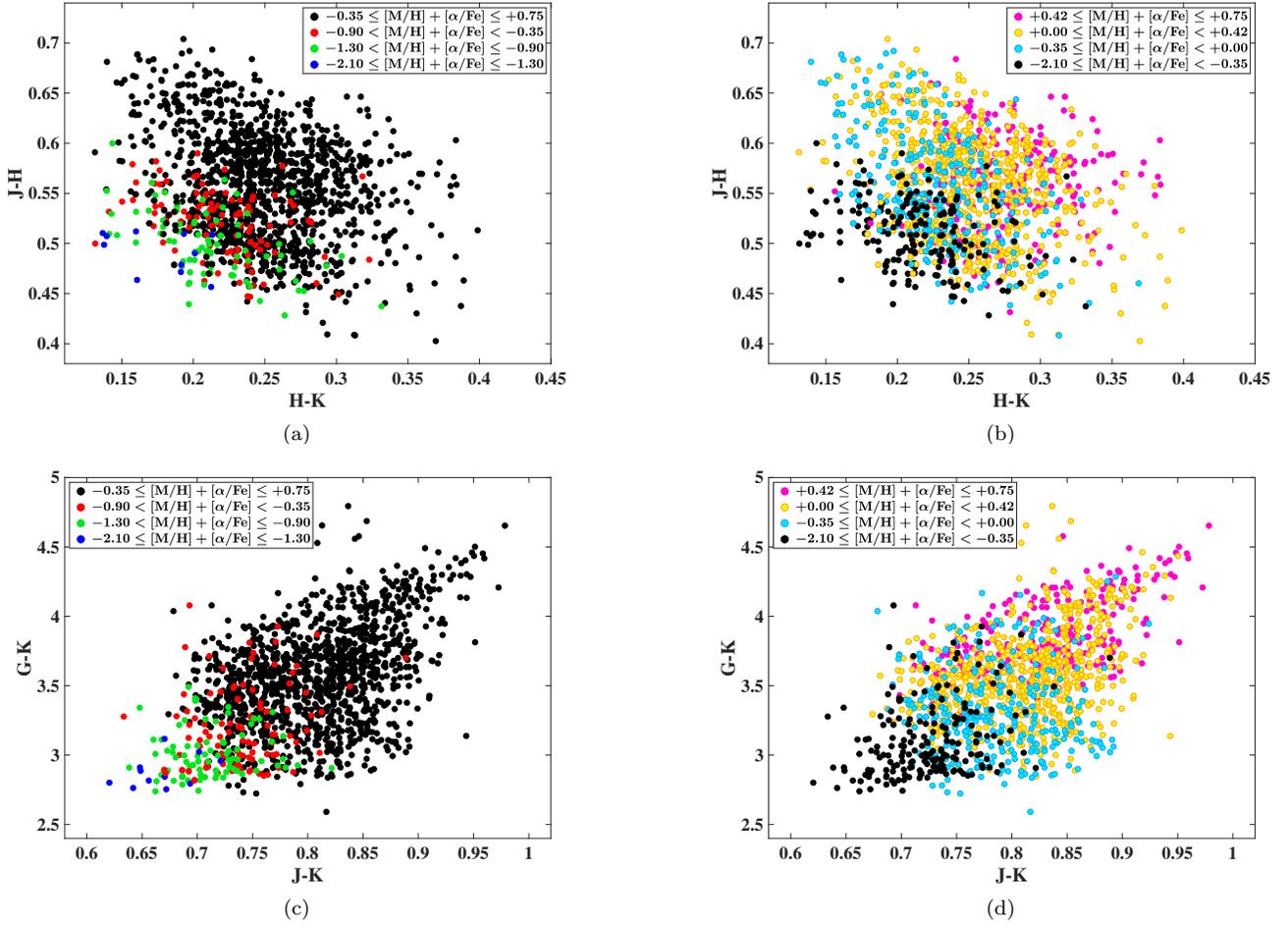

**Figure 36.** Distribution of (*J-H*) vs. (*H-K*) for 1,468 stars in our sample, with color-coding to emphasize the metal-poor (left panels) and metal-rich (right panels) stars. **(a)** 1,278 stars with -0.35 ⩽ [M/H]+[α/Fe] ⩽ +0.75 (black, Group 1α), 100 stars with -0.9 <[M/H]+[α/Fe] < -0.35 (red, Group 2α), 79 stars with -1.3 < [M/H]+[α/Fe] ⩽ -0.9 (green, Group 3α), and 11 stars with -2.1 ⩽ [M/H]+[α/Fe] ⩽ -1.3 (blue, Group 4α), and for the same subsample, including **(b)** 243 stars with +0.42 ⩽ [M/H]+[α/Fe] ⩽ +0.75 (magenta, Group 1Aα), 731 stars with 0 ⩽ [M/H]+[α/Fe] < +0.42 (yellow, Group 1Bα), 304 stars with -0.35 ⩽ [M/H]+[α/Fe] < 0 (cyan, Group 1Cα), and 190 stars with -2.1 ⩽ [M/H]+[α/Fe] < -0.35 (black). Distribution of (*G-K*) vs. (*J-K*) for 1439 stars, including **(c)** 1,249 stars in Group 1α (black), 100 stars in Group 2α (red), 79 stars in Group 3α (green), and 11 stars in Group 4α (blue), and for the same subsample, including **(d)** 239 stars in Group 1Aα (magenta), 713 stars in Group 1Bα (yellow), 297 stars in Group 1Cα (cyan), and 190 stars with -2.1 ⩽ [M/H]+[α/Fe] < -0.35 (black).

Although the above chemically defined groups/subgroups are stratified in the Gaia H-R diagrams, we note several outliers that are found quite far from their expected sequence. The most noticeable of these are explained in Appendix A.6. Despite these outliers and the overlaps between different groups, the clear separation of stars with different ranges of [M/H]+[α/Fe] shows the high precision of our estimates of chemical parameters, and accordingly the high efficiency of our pipeline for future use.

## 9. LOCAL CHEMICAL POPULATIONS OF LOW-MASS STARS: [α/Fe] VERSUS [M/H]

**Table 9.** Continued

| | Companion | | | | | Primary | | | |
|---|---|---|---|---|---|---|---|---|---|
| Name | Spectral Type | Metallicity Class | [M/H] | [α/Fe] | Name | Spectral Type | Metallicity Class | [M/H] | [α/Fe] |
| PM J13195−3506E | M3.5 | 2 | −0.15 ± 0.11 | +0.1500 ± 0.0939 | PM J13195+3506W | M1.5 | 3 | −0.30 ± 0.07 | +0.1750 ± 0.0944 |
| PM J14518+5147W | M5.0 | 3 | −0.60 ± 0.08 | +0.2500 ± 0.1107 | PM J14518+5147E | M3.0 | 3 | −0.60 ± 0.07 | +0.2500 ± 0.1149 |
| PM J15308+5608N | M6.0 | 7 | −1.35 ± 0.12 | +0.3500 ± 0.1457 | PM J15308+5608S | M6.0 | 7 | −1.30 ± 0.11 | +0.2250 ± 0.1314 |
| PM J15353−1743 | M5.0 | 2 | −0.40 ± 0.10 | +0.1750 ± 0.0977 | PM J15353−1742 | M3.5 | 3 | −0.35 ± 0.08 | +0.2000 ± 0.1127 |
| PM J15400+4329S | M4.5 | 2 | −0.30 ± 0.08 | +0.1750 ± 0.1004 | PM J15400+4329N | M3.5 | 2 | −0.30 ± 0.11 | +0.2750 ± 0.1194 |
| PM J15413+1349S | M4.0 | 4 | −0.65 ± 0.10 | +0.2000 ± 0.1409 | PM J15413+1349N | M1.5 | 4 | −0.45 ± 0.07 | +0.1750 ± 0.1148 |
| PM J15531+3444 | M3.5 | 2 | +0.15 ± 0.08 | +0.1000 ± 0.0756 | PM J15531+3445W | M3.5 | 2 | +0.25 ± 0.06 | −0.0500 ± 0.0720 |
| PM J17193−2949N | M5.0 | 2 | +0.00 ± 0.15 | +0.1500 ± 0.0998 | PM J17193−2949S | M4.5 | 2 | +0.05 ± 0.13 | +0.1250 ± 0.0912 |
| PM J18180−3846W | M4.5 | 2 | +0.00 ± 0.10 | +0.1750 ± 0.1047 | PM J18180−3846E | M3.5 | 2 | −0.05 ± 0.13 | +0.1500 ± 0.0950 |
| PM J18427+5937S | M4.0 | 2 | −0.10 ± 0.11 | +0.2250 ± 0.0869 | PM J18427+5937N | M3.5 | 2 | −0.25 ± 0.09 | +0.2750 ± 0.1161 |
| PM J19113−5224E | M3.5 | 2 | +0.25 ± 0.08 | +0.1000 ± 0.0652 | PM J19113+5224W | M3.5 | 1 | +0.25 ± 0.08 | +0.1000 ± 0.0537 |
| PM J19141+2825 | M2.0 | 6 | −1.20 ± 0.08 | +0.3500 ± 0.1394 | PM J19140+2825 | M1.0 | 6 | −1.10 ± 0.09 | +0.3000 ± 0.1385 |
| PM J19282−0200S | M6.5 | 2 | −0.20 ± 0.13 | +0.2000 ± 0.0917 | PM J19282−0200N | M3.5 | 2 | +0.10 ± 0.08 | +0.1500 ± 0.0933 |
| PM J19388+3512E | M5.0 | 3 | −0.20 ± 0.09 | +0.1500 ± 0.0752 | PM J19388+3512W | M4.0 | 3 | +0.35 ± 0.08 | −0.0250 ± 0.0562 |
| PM J20231+1844E | M4.0 | 3 | −0.20 ± 0.11 | +0.2250 ± 0.1075 | PM J20231+1844W | M3.5 | 2 | −0.35 ± 0.09 | +0.3000 ± 0.1201 |
| PM J21000+4004W | M3.0 | 3 | +0.20 ± 0.06 | +0.0250 ± 0.0438 | PM J21000+4004E | M1.0 | 3 | −0.35 ± 0.05 | +0.2500 ± 0.0767 |
| PM J22173−0848S | M5.0 | 2 | −0.05 ± 0.10 | +0.1250 ± 0.0734 | PM J22173−0848N | M5.0 | 3 | −0.20 ± 0.09 | +0.2000 ± 0.0660 |
| PM J22262+0300S | M6.5 | 2 | +0.30 ± 0.11 | +0.1500 ± 0.0838 | PM J22262+0300N | M4.0 | 2 | +0.40 ± 0.08 | +0.0750 ± 0.0469 |
| PM J23293−4127 | M4.5 | 2 | +0.15 ± 0.10 | +0.2250 ± 0.0876 | PM J23293+4128 | M4.0 | 2 | +0.45 ± 0.07 | +0.0500 ± 0.0561 |
| PM J23578+7836 | M3.5 | 3 | −0.25 ± 0.09 | +0.1500 ± 0.0926 | PM J23580+7836 | M3.0 | 3 | −0.25 ± 0.11 | +0.1500 ± 0.1063 |

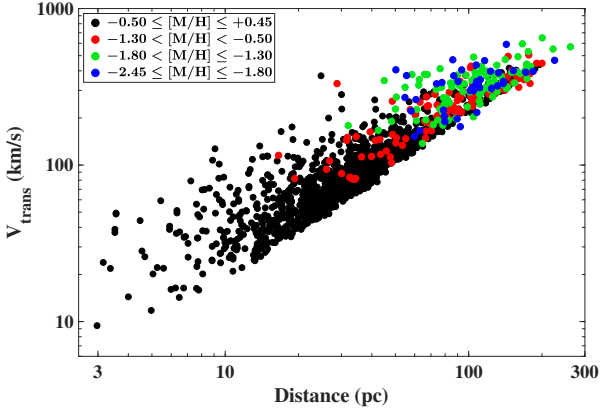

**Figure 37.** Transverse velocity (km/s) vs. distance (pc) of 1,483 stars in our study, color coded based on the same metallicity groups as defined in Figure 35 (a) and (c), including 1,243 stars in Group 1 (black), 106 stars in Group 2 (red), 94 stars in Group3 (green), and 40 stars in Group 4 (blue).

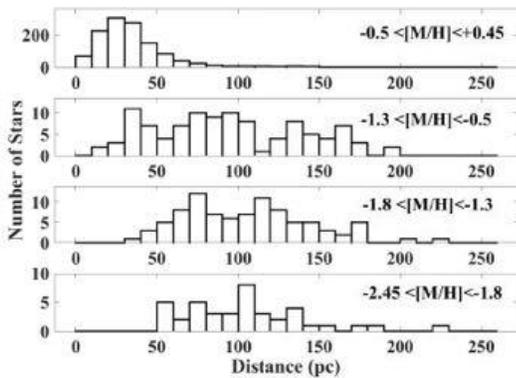

**Figure 38.** Distance distribution of four metallicity groups as described in Figure 35 (a) and (c).

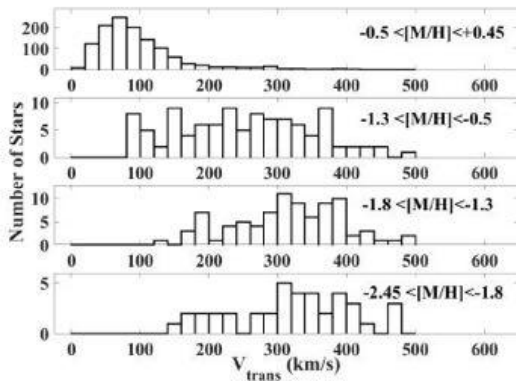

**Figure 39.** Transverse velocity distribution of four metallicity groups as described in Figure 35 (a) and (c).

Figure 42 shows the distribution of [α/Fe] (randomized within ±0.025 dex) and [M/H] (randomized within ±0.05 dex) for the sample of 1,544 stars, color-mapped according to the transverse velocity of each star. The horizontal and vertical error bars are the mean uncertainties of [M/H] and [α/Fe], respectively.

The general trend indicates that metal-rich stars, on average, have lower values of [α/Fe], while metal-poor stars, on average, have higher values of [α/Fe]. This trend is compatible with current views about the enrichment history of the Galaxy; in the early times of the Galactic evolution, it is assumed that type II supernovae were the primary source of heavy elements. The material produced by this type of supernovae is always rich in α elements, as compared to iron, regardless of the metallicity of the progenitor star. However, after a few billion years, when the most important iron producers, type Ia supernovae, had reached their maturity (Heringer et al. 2019), the enrichment of iron overtook that of α elements. This is the time when the ratio of α-element to iron abundance started to drop at [M/H] ∼ -1.1 dex (Croswell 1995; Pagel 1997; Chiappini et al. 1999, 2001; Chiappini 2001; Matteucci 2001). It should be pointed out that the [α/Fe]-[M/H] diagram of more massive stars and red giants has been used to model the Galactic chemical evolution. For instance, the two-infall model (and its revised version) is capable to properly reproduce the relation between [α/Fe] and [M/H] (Chiappini et al. 2001; Chiappini 2001; Romano et al. 2010; Spitoni et al. 2009, 2019).

While our own [α/Fe]-[M/H] diagram is in general agreement with the above narrative, there are many details in our distribution that need to be carefully addressed. As compared to similar diagrams for more massive stars, the metal-rich and near-solar metallicity M dwarfs in our sample, which belong to the Galactic disk, have a higher mean [α/Fe] at any given [M/H]. We believe this may be due to either systematic errors in the pipeline or issues with the BT-Settl models which may have a tendency to overestimate the metal content in convective metal-rich objects (see section 8 above). Another possible reason may be that our high proper-motion sample overselects stars of higher transverse motions, which tend to be on average older, introducing an age bias in the distribution. Haywood et al. (2013) compared the spectroscopically inferred stellar parameters of a sample from Adibekyan et al. (2012) with theoretical isochrones to derive the stellar ages. They found that the high-[α/Fe] disk stars are, in general, older than low-[α/Fe] disk stars. Although the majority of our disk stars have lower velocities than the metal-poor halo stars, our disk subset still comprises a major-ity of stars with relatively high motions ($> 50$ km/s). A large spread of ages could cause both a systematic offset in the metallicity values, and a large spread in [α/Fe].

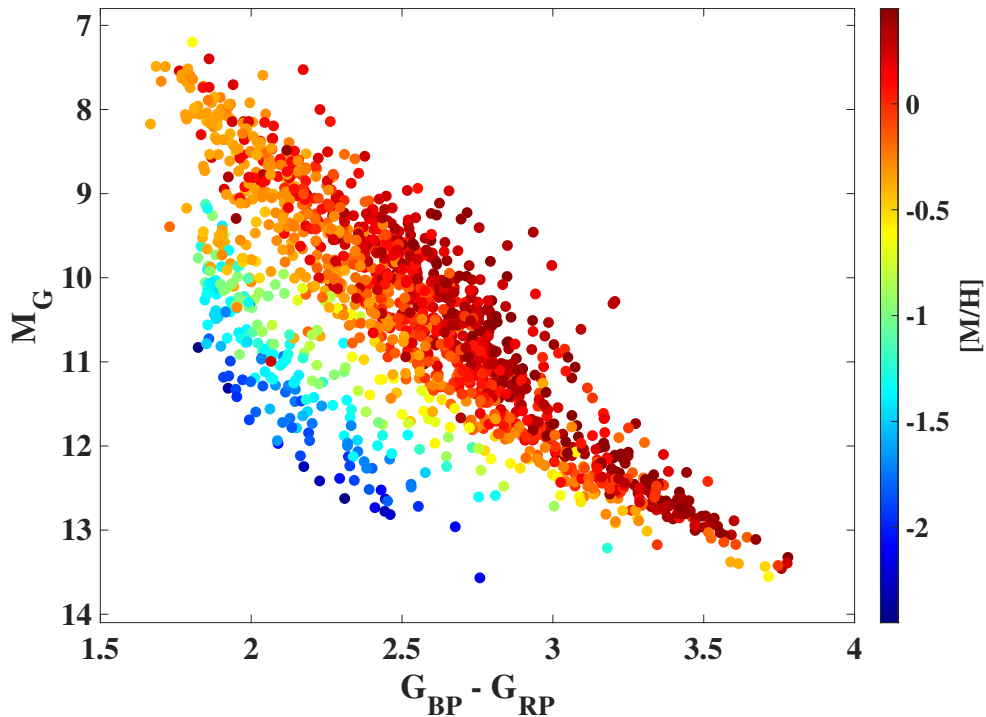

**Figure 40.** H-R diagram of $M_G$ vs. ($G_{BP}$ - $G_{RP}$) color of 1474 M dwarfs/subdwarfs in our sample, color-mapped by [M/H] metallicity index according to the scale shown on the right.

In order to better present the effect of velocity on the distribution of stars in the [α/Fe]-[M/H] diagram, we divide our sample into four groups with different velocity ranges, as plotted in Figure 43 and described in the caption. One notices the spread in [α/Fe] is smaller for stars with lower velocities ($V_{trans} < 50$). All three groups of stars with $V_{trans} < 150$ km/s have metallicity values that associate them with the Galactic disk, but with a few exceptions. We find only a handful of metal-poor stars with $V_{trans} < 150$ km/s, and these stars conceivably have large space motions as well, with the low transverse motion only due to projection effects - in which case we predict these stars should have large radial velocities.

The majority ($\sim 60\%$) of high-velocity stars with $V_{trans} \geqslant 150$ km/s (panel d) show a range of metallicity values clearly consistent with the Galactic halo. On the other hand, it is quite interesting that a significant number of high-velocity stars have metallicity values similar to those of the three lower-velocity groups ([M/H] $\geqslant$ - 0.5). According to the original Geometrically-based definition of the so-called thin disk and thick disk (Gilmore & Reid 1983; Yoshii 1982), thick-disk stars, on aver-age, possess higher velocity dispersion, and therefore higher peculiar velocities in the Solar Neighborhood. High-resolution spectra of kinematically selected nearby stars have shown that thick-disk stars have higher values of [α/Fe], with respect to thin-disk stars with similar metallicities. It has, however, turned out that the chemical definition of stellar populations is more accurate than the geometrical one (e.g., Bovy et al. 2012), and accordingly, several spectroscopic surveys of more massive stars have been employed to define and identify the two disks by revealing a gap in the [α/Fe]-[M/H] diagram (e.g., Fuhrmann 2004; Reddy et al. 2006; Bensby et al. 2007; Lee et al. 2011a, b; Adibekyan et al. 2012, 2013; Haywood et al. 2013; Ramírez et al. 2013; Recio-Blanco et al. 2014). On the other hand, some other studies have not detected a bimodality in the chemical distribution of stars in this diagram, and proposed a single disk without breaking it into two thin and thick substructures (e.g., Bovy et al. 2012). This is consistent with Galactic models under the assumption of radial migration (e.g., Sellwood & Binney 2002; Schönrich & Binney 2009 a,b), and the theoretical analysis of Schönrich & Binney 2009 a, which suggests that the continuous distribution of [α/Fe] can be explained by the standard models of star formation and metal enrichment. As pointed out in Recio-Blanco et al. 2014, sample selection effects can partly be the reason for discrepancies between the chemical distribution of different studies, which have led to different data interpretation. Due to these shortcomings, it has been suggested that stellar age alone is the best way to define these two disk components (Fuhrmann 2011, and references therein). However, the difficulty in measuring accurate stellar ages for large-volumed samples may prevents one to adopt this definition (see Kawata & Chiappini 2016 for a thorough discussions on the thin and thick disks). Although we cannot perceive a clear break or bimodality in the chemical distribution of our disk stars, we conclude that the stars with rather high [α/Fe] (≳ +0.3 dex) around [M/H] ≃ -0.35 dex are most likely to belong to the Galactic thick disk.

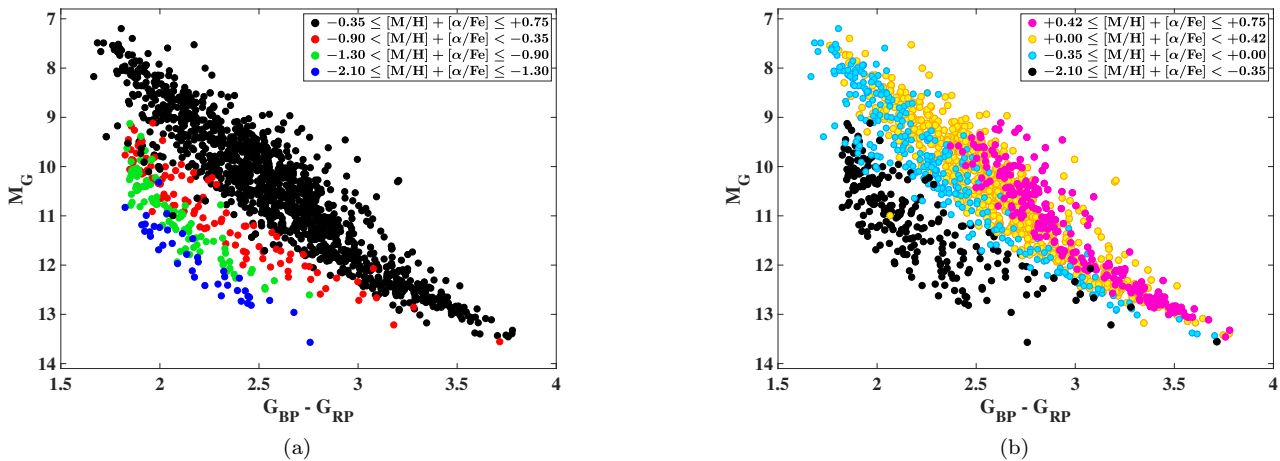

**Figure 41.** H-R diagram of $M_G$ vs. ($G_{BP}$ - $G_{RP}$) color of 1474 stars in our sample, with color-coding to emphasize the metal-poor (left panel) and metal-rich (right panel) stars. **(a)** 1236 stars with -0.35 ⩽ [M/H]+[α/Fe] ⩽ +0.75 (black, Groupα 1), 106 stars with -0.9 < [M/H]+[α/Fe] < -0.35 (red, Groupα 2), 92 stars with -1.3 < [M/H]+[α/Fe] ⩽ -0.9 (green, Groupα 3), and 40 stars with -2.10 ⩽ [M/H]+[α/Fe] < -1.3 (blue, Groupα 4), **(b)** 234 stars with +0.42 ⩽ [M/H]+[α/Fe] ⩽ +0.75 (magenta, Groupα 1A), 705 stars with 0 ⩽ [M/H]+[α/Fe] < +0.42 (yellow, Groupα 1B), 297 stars with -0.35 ⩽ [M/H]+[α/Fe] < 0 (cyan, Groupα 1C), and 238 stars with -2.10 ⩽ [M/H]+[α/Fe] < -0.35 (black, Groupα 2 - Groupα 4).

Perhaps a more striking feature of our [M/H]-[α/Fe] distribution is the clumpy distribution observed for the metal-poor, high-velocity subset. One can see at least three prominent clumps near [M/H] = -0.9, -1.3, and -1.7, and possibly another one near [M/H] = -2.4. The clump near [M/H] = -1.3 could also be two clumps with different values of [α/Fe]. Each of these clumps has a size comparable to, or just slightly larger than the estimated uncertainty, which suggest that these clumps might be even more concentrated than suggested by the diagram. There are various known halo substructures, such as streams from tidally disrupted dwarf galaxies or globular clusters (e.g., Martinez-Delgado et al. 2001, 2010; Pearson 2018). The clumps observed in our metallicity diagram could be the signature of similar streams that happen to be crossing through the solar neighborhood.

There is also a noticeable high-velocity clump with subsolar α-to-iron ratios, i.e., [α/Fe] < 0, and with -0.45 < [M/H] < -0.2 dex. As one possible scenario for these peculiar stars, we may consider the hot gases, ejected from the disk by supernova explosions, which are now falling back to the disk, known as galactic fountains (for more details, see Kahn 1994). The kinematics, structure and composition of the disk are partially influenced through the hydrodynamic and thermodynamic interactions when infalling clouds collide with the Galactic disk (Sung & Kwak 2018). Another possible origin for this odd clump is prograde stellar streams (Necib et al. 2019); massive dwarf galaxies merging with the Milky Way on prograde orbits can be drawn into the plane of the the Galactic disk before being fully disrupted. For example, Necib et al. (2019) have found evidence for such a stream in the Solar Neighborhood, whose stars have a peak metallicity of ∼ -0.5 and ages ∼ 10-13 Gyr. The kinematics of these stars differ from both thin and thick disk. More careful studies are, however, required to better understand the origin of this clump in the [M/H]-[α/Fe] diagram.

We are planning a follow-up study that will analyze the chemical make-up of the local Galactic populations in greater detail. A larger sample can give us a better clue to the understanding of these clumps and their origins. To this end, we are currently applying the present pipeline to an additional subset of lower proper motion stars, as well as to a larger sample of M dwarf/subdwarf spectra collected as part of the SDSS spectroscopic survey, which is available from the SDSS data archive. By comparing the distribution of this extended sample with that of more massive stars, such as F, G, and K dwarfs, from previous studies, we hope to reach a more accurate conclusion on the nature of the Galactic thin-thick disk and halo, as well as a more realistic model of Galactic chemo-dynamical evolution.

## 10. SUMMARY AND DISCUSSION

In this paper, we have presented an extensive set of low-to-medium spectra of 1,544 nearby, high proper motion M dwarf/Msubdwarfs, which were observed at the MDM, Lick, and KPNO observatories. We utilize the spectra of these stars to measure their stellar parameters using two different approaches. One approach uses a set of empirically assembled classification templates, based on the shapes of the TiO and CaH molecular bands near 7000 Å. We develop a template-fit method to determine the spectral subtype and metallicity class of stars in the sample using these templates. The other approach relies on synthetic spectra generated from the BT-Settl stellar atmosphere theoretical models. We also establish a model-fit pipeline to obtain the physical parameters, i.e., $T_{eff}$, [M/H], [$\alpha$/Fe], and log $g$, of the sample. We find an acceptable level of consistency between the observed spectra and their corresponding models, in particular for higher temperatures.

We determine that the relationship between color and spectral subtype depends on the metallicity class, as the color $G_{BP}$ - $G_{RP}$ is more sensitive to subtype for metal-rich M dwarfs, with respect to metal-poor M subdwarfs. This happens because the weaker molecular bands (especially TiO) in the more metal poor stars result in much less extreme color changes with effective temperature. This confirms that photometric colors are not a reliable indicator of effective temperature in M-type dwarfs in general. While the spectral subtype is a more reliable indicator of effective temperature in those stars, the subtype-$T_{eff}$ relationship is also dependent on the metallicity class: $T_{eff}$ as function of subtype has a steeper slope for metal-rich M dwarfs, as compared to metal-poor M subdwarfs. The agreement between metallicity class (from basic classification templates) and metallicity estimates (from model fits) is more reliable, and metallicity class can be regarded as a rough tracer of [M/H]. We determine that metallicity class is somewhat better correlated with the combined parameter [M/H]+[$\alpha$/Fe], rather than [M/H] alone; this is consistent with the fact that the most prominent spectral features in the optical spectra of M-type dwarfs are the TiO bands, with O and Ti both being $\alpha$ elements.

While our low-to-medium resolution spectroscopy may suffer from systematic errors in the classification and/or model fit, due to the homogeneity of our spectroscopic sample and applying the same method to all stars, our results appear to be relatively precise, and correctly distinguish between stars of different metallicity and effective temperature. The level of this precision is revealed by comparing results from stars in common proper motion pairs, which are in reasonably good agreement, and also by examining the distribution of our sample stars in the Gaia H-R diagram. Stars of different spectroscopic metallicity classes, or of different model-fit estimated metallicity values, are greatly stratified in the H-R diagram, with the "layers" of metal-rich stars found to the upper right (redder/overluminous) and "layers" of metal-poor stars to the lower left (bluer/underluminous). We see some level of overlap between different metallicity classes and metallicity ranges at early subtypes, though these overlaps are somewhat reduced when we use the combined parameter [M/H]+[$\alpha$/Fe]. A troubling pattern however emerges in the diagram, which either shows a dearth of super metal-rich stars at earlier subtypes, or an excess of super metal-rich stars at later subtypes. While this could be due to a selection effect, related to the high-proper motion limit of the sample selection, a likely possibility is that late-type M dwarfs have systematically overestimated metallicity values. Theoretical stellar models for partly convective and fully convective stars, which are based on the MLT without the treatment of metallicity-dependent convection, may be systematically underestimating the line strengths in synthetic spectra of metal-rich objects. Comparison with observed spectra would thus tend to overestimates of the metallicity values in fully convective (later-type) stars.

Using a subsample of binary systems, we show that the measured chemical parameters [M/H] and [$\alpha$/Fe] for the two components in these binaries are in a good agreement with each other, which is another indication of the high precision of our method. One more time, this agreement improves if we compare the [M/H]+[$\alpha$/Fe] value of the components, instead of [M/H] alone.

We briefly address the [$\alpha$/Fe]-[M/H] diagram for stars in our sample. The distribution of stars in this diagram includes important information on the structure, kinematics and chemistry of the local Galactic disk and halo. The rather large scatter in [$\alpha$/Fe] at any given [M/H] for disk stars is due to our age-biased sample, as this overselects high-velocity stars, which are typically older than low-velocity stars. We divide the sample into four ranges of transverse velocity, and show that the spread in [$\alpha$/Fe] is smaller for low-velocity stars, as compared to high-velocity stars. There are several distinct clumps in the distribution of high-velocity stars, which may originate from different substructures in the halo.

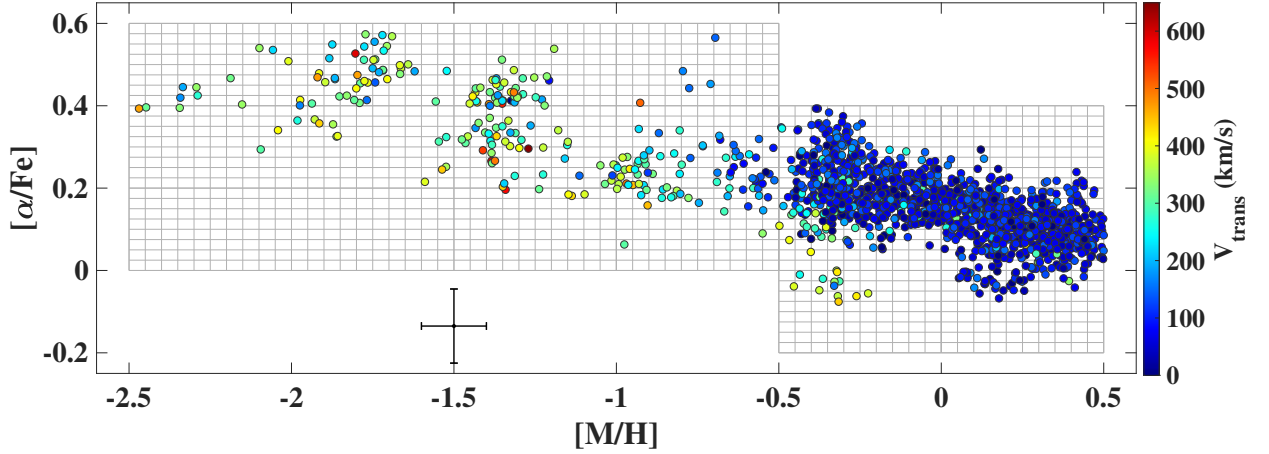

**Figure 42.** The [α/Fe] (randomized within ±0.025 dex) vs. [M/H] (randomized within ±0.05 dex) of 1,544 stars, color-mapped based on the transverse velocities.

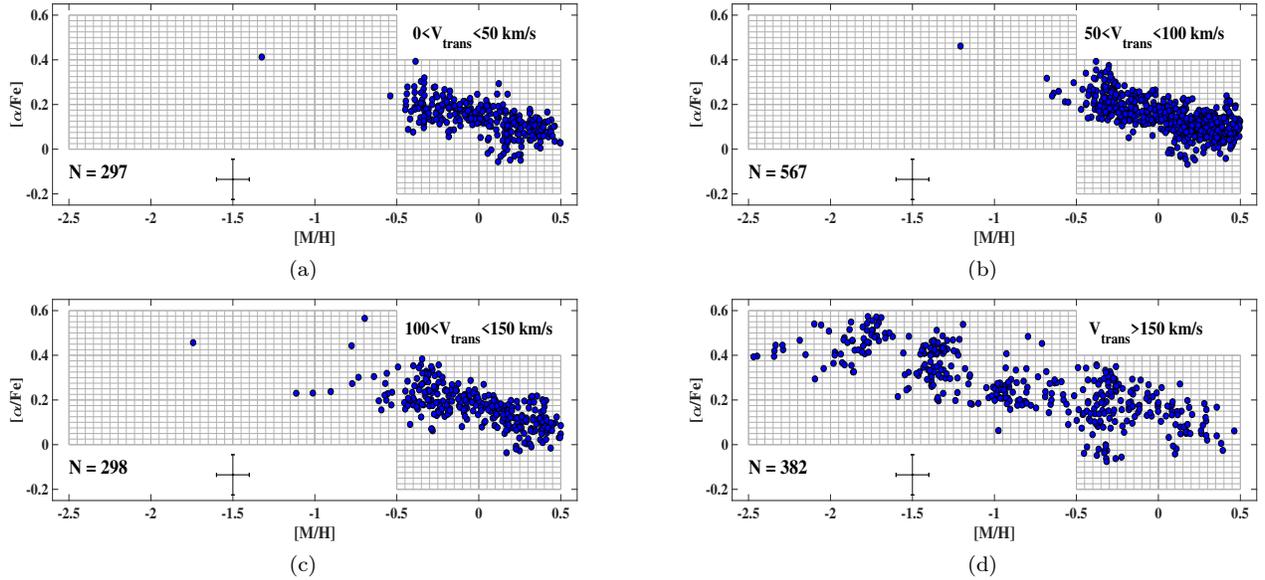

**Figure 43.** The [α/Fe] (randomized within ±0.025 dex) vs. [M/H] (randomized within ±0.05 dex) of **(a)** 297 stars with $0 \leqslant$ $V_{trans} < 50$ km/s, **(b)** 567 stars with $50 \leqslant V_{trans} < 100$ km/s, **(c)** 298 stars with $100 \leqslant V_{trans} < 150$ km/s, and **(d)** 382 stars with $V_{trans} \geqslant 150$ km/s.

In our future work, we aim to expand our spectroscopic sample using a low-proper motion sample, which were mostly observed at the MDM observatory, a large SDSS catalog, as well as a sample from the LAMOST survey. We will apply the pipeline to those spectra with good quality, and examine their distribution in the [α/Fe]-[M/H] diagram. The comparison of this distribution with that of F, G, and K dwarfs from previous studies will enable us to identify the thin-thick disk, if any, and uncover the halo substructure associated with any detected clump of stars in the [M/H]-[α/Fe] plane.

*Acknowledgements.* We gratefully thank the anonymous referee for helpful suggestions that improved our manuscript. We also thank Bob Barr and the staff at MDM observatory for their outstanding support and extremely useful help in conducting the observational program; we are indebted to John Thorstensen for extremely useful suggestions and comments about our MDM program. We wish to thank Elinor Gates and all the staff at Lick observatory for their precious help and support.

We extend our thanks to Philip Muirhead, Mark Veyette, Wei-Chun Jao, and Russel White for helpful suggestions. We would also like to thank Jeffrey Valenti, John Brewer, Ralph Schönrich, and Chao Liu for insightful discussions. The technical support from Justin Cantrell and Jeremy Simmons is greatly appreciated.

This material is based on work supported by the National Science Foundation under grant No. AST 09-08419, the National Aeronautics and Space Administration under grant Nos. NNX15AV65G, NNX16AI63G, and NNX16AI62G issued through the SMD/Astrophysics Division as part of the K2 Guest Observer Program.

This work has made use of data from the European Space Agency (ESA) mission *Gaia* (https://www.cosmos.esa.int/gaia), processed by the *Gaia* Data Processing and Analysis Consortium (DPAC, https://www.cosmos.esa.int/web/gaia/dpac/consortium). Funding for the DPAC has been provided by national institutions, in particular the institutions participating in the *Gaia* Multilateral Agreement.

# APPENDIX

## A. NOTES ON INDIVIDUAL OUTLIERS WITH UNEXPECTED PARAMETERS, PHOTOMETRY, AND PARALLAX

### A.1. *Color-Spectral Types Relationship*

In Figure 7 (b), there are 8 dM stars with a spectral subtype of M6.5 that are unexpectedly found to have relatively blue colors, and are described below:

- PM J02110+7823 ($G_{BP} - G_{RP} = 2.624$), PM J02096+1955 ($G_{BP} - G_{RP} = 2.620$), PM J23557+5613 ($G_{BP} - G_{RP} = 2.418$), and PM J20129+3416S ($G_{BP} - G_{RP} = 2.017$), which have either a high noise level, or at least one artifact emission line in their spectra (or even both). All these spectra are misclassified by our template-fit pipeline, and upon closer examination are better represented by templates of earlier-subtypes.

- PM J13332+6417 ($G_{BP} - G_{RP} = 1.729$) and PM J21115+4634 ($G_{BP} - G_{RP} = 1.95$), which are both misclassified as dM stars because of a few artifact emission lines detected in their spectra. Close examination reveals that they are more likely to be metal-poor M subdwarfs.

- PM J01523+6558 ($G_{BP} - G_{RP} = 2.270$), which is misclassified by our automated pipeline due to a strong emission artifact that was not removed by our rejection algorithm. Upon visual inspection, we find that a template of earlier-subtype and slightly more metal-poor is a better fit for this star.

- PM J09187+2645 ($G_{BP}$ - $G_{RP}$ = 2.253), which has a noisy spectrum and also includes a strong emission artifact, resulting in misclassification by our pipeline. Visual inspection suggests that the stars is unlikely to be an M dwarf/subdwarf, with a spectrum more consistent with a G/K dwarf.

In Figure 7 (c), the star PM J01122+4556 with spectral type M4.0 ($G_{BP}$ - $G_{RP}$ = 1.99) is also located rather below the distribution. The spectrum of this star has a few strong artifact emission lines, and seems to be misclassified (MC=6), as it is similar to a more metal-poor esdM star rather than an sdM star.

### A.2. Photometric Variations with Metallicity Class in Color-Color Diagram

The most important outliers in Figure 8 (c) and (d) are listed as follows:

- The dM star PM J03288+3722 ($J$-$K$ $\simeq$ 0.82, $G$-$K$ $\simeq$ 2.59) which is located well below the main locus. Visual examination of the spectrum indicates that the star is misclassified, as the spectrum looks more like a metal-poor M subdwarf than a metal-rich M dwarf. In any case, the $J$-$K$ color is much too blue, which means either the 2MASS photometry is significantly inaccurate, or it suffers from significant contamination.

- The dM star PM J13332+6417 ($J$-$K$ $\simeq$ 0.74, $G$-$K$ $\simeq$ 2.73) which is below the esdM sequence. The spectrum of this star is relatively noisy and includes a few strong emission artifacts, leading to misclassification by our pipeline. Our careful inspection suggests that the star is a metal-poor M subdwarf, not a metal-rich M dwarf.

- The sdM star PM J14073+3923 ($J$-$K$ $\simeq$ 0.67, $G$-$K$ $\simeq$ 2.87), which is found inside the region of esdM/usdM stars. Visual inspection of the spectrum suggests the star is misclassified, as the spectrum resembles a very metal-poor esdM or usdM star rather than a moderately metal-poor sdM star.

- The sdM star PM J16326+3234 ($J$-$K$ $\simeq$ 0.75, $G$-$K$ $\simeq$ 2.72), which is well below the esdM locus on the plot. Under close scrutiny, we find this star to be likely misclassified, as its spectrum is more similar to a dM star, not an sdM star. This however makes the problem worse, and suggests the 2MASS $J$-$K$ color is significantly in error, or suffers from significant contamination.

- The two sdM stars, PM J08287+0623 ($J$-$K$ $\simeq$ 0.65, $G$-$K$ $\simeq$ 3.34) and PM J18526+2304 ($J$-$K$ $\simeq$ 0.63, $G$-$K$ $\simeq$ 3.28), which are located to the left of the distribution. Our visual inspection however indicates that both stars are correctly classified; this suggests the 2MASS $J$ and $K$ photometry to be significantly in error.

- The sdM star PM J14191+2023 ($J$-$K$ $\simeq$ 0.89, $G$-$K$ $\simeq$ 3.70), which is located well inside the dM sequence. Upon close examination, the star seems to be correctly classified, which suggests possible issues with the 2MASS photometry.

- The sdM star above the diagram, PM J23401+6041 ($J$-$K$ $\simeq$ 0.81, $G$-$K$ $\simeq$ 4.7). Following close visual inspection, this star appears to be misclassified as its spectrum indicates a metal-rich M dwarf rather than a metal-poor sdM star.

### A.3. Photometric Variations with Metallicity Class in H-R Diagram

The most noticeable outliers in Figure 12 are listed below:

- The dM star, PM J13332+6417 ($G_{BP}$ - $G_{RP}$ $\simeq$ 1.73, $M_G$ $\simeq$ 9.39), which is well-below the dM domain, and more closely aligned with the esdM stars. Being to the blue of the esdM locus, however, suggests the star may be a late-K subdwarf. Close inspection of the spectrum reveals significant noise and a number of instrumental artifacts, probably responsible for the misclassification. The underlying spectrum does indeed appear to be more consistent with a metal-poor object, possibly a late-type K subdwarf.

- The two dM stars, PM J15460+5135 ($G_{BP}$ - $G_{RP}$ $\simeq$ 2.07, $M_G$ $\simeq$ 10.99) and PM J10286+3214 ($G_{BP}$ - $G_{RP}$ $\simeq$ 2.19, $M_G$ $\simeq$ 10.64), which are both located well inside the dM distribution, and very far from the dM locus. Visual inspection, however, suggests both stars are correctly classified as dwarfs (dM), which means these stars may have inaccurate Gaia photometry or/and parallaxes.

- Several dM outliers are hovering just above and to the right of the main dM loci, and are most likely overluminous, unresolved binaries. However, those stars which are at the farthest distance from the distribution, i.e. PM J16170+5516 ($G_{BP}$ - $G_{RP}$ $\simeq$ 2.17, $M_G$ $\simeq$ 7.53), PM J15011+0709 ($G_{BP}$ - $G_{RP}$ $\simeq$ 2.66 , $M_G$ $\simeq$ 8.97), PM J10520+0032 ($G_{BP}$ - $G_{RP}$ $\simeq$ 2.93 , $M_G$ $\simeq$ 9.46), PM J10182-2028E ($G_{BP}$ - $G_{RP}$ $\simeq$ 3.20, $M_G$ $\simeq$ 10.31), and PM J22195+6120 ($G_{BP}$ - $G_{RP}$ $\simeq$ 3.21, $M_G$ $\simeq$ 10.28) may suffer from inaccurate Gaia photometry and/or parallaxes.

- The sdM star, PM J06245+4238 ($G_{BP}$ - $G_{RP}$ $\simeq$ 2.34, $M_G$ $\simeq$ 12.13), which is found inside the esdM domain. Close examination of the spectrum shows a rather high level of noise, which has led to the misclassification of this star. The spectrum, in effect, resembles a more metal-poor esdM star than an sdM star.

### A.4. Chemical Parameters in Common Proper-Motion Pairs

The most notable outliers in Figure 32, for which the difference between the [M/H] value of the primary and its companion is more than twice the estimated precision, are listed as follows:

- The wide binary system consisting of two M dwarfs PM J02456+4456 (dM1.0, MC=3, [M/H] = -0.3 dex, and [α/Fe] = +0.175 dex) as the primary and PM J02456+4457 (dM5.5, MC = 2, [M/H] = +0.35 dex, and [α/Fe] = +0.1 dex) as the secondary. Upon close examination, the spectra of both stars appear to be correctly classified, however, the primary does show features that suggest it is more metal-poor than the secondary. The cause of the inconsistency is unclear, but could reflect real compositional differences, or could be caused by contaminated light from an unresolved companion.

- The wide binary system consisting of two M dwarfs PM J12278+0512E (dM2.5, MC=2, [M/H] = -0.2 dex, and [α/Fe] = +0.325 dex) as the primary and PM J12278+0512W (dM4.5, MC=2, [M/H] = +0.4 dex, and [α/Fe] = +0.125 dex) as the secondary. Upon visual inspection, the initial spectral classification of the two stars appears to be correct, with both spectra consistent with the same metallicity class MC=2. The higher metallicity of the secondary from the model fit method therefore appears to be an overestimate. The source of the discrepancy remains unclear.

- The wide binary system consisting of two M dwarfs PM J19388+3512W (dM4.0, MC=3, [M/H] = +0.35 dex, and [α/Fe] = -0.025 dex) as the primary and PM J19388+3512E (dM5.0, MC=3, [M/H] = -0.2 dex, and [α/Fe] = +0.15 dex) as the secondary. Visual inspection shows that the initial spectral classification, suggesting a moderately metal-poor M dwarfs, with metallicity class MC=3, is accurate for both stars. This time it is the estimated [M/H] value of the primary star from the model-fit pipeline that appears to be overvalued, the reason of which is uncertain.

- The wide binary system consisting of two M dwarfs PM J21000+4004E (dM1.0, MC=3, [M/H] = -0.35 dex, and [α/Fe] = +0.25 dex) as the primary and PM J21000+4004W (dM3.0, MC=3, [M/H] = +0.2 dex, and [α/Fe] = +0.025 dex) as the secondary. Careful examination shows that the initial spectral classification of both stars are accurate and the metallicity estimates should be in agreement. However, the metallicity of the secondary from the model-fit appears to be slightly overestimated. We believe this is due to a strong artifact emission in the spectrum.

### A.5. *Photometric Variations with Metallicity in Color-Color Diagram*

The most important outliers in Figure 35 (c) and (d) are described in the following way:

- The two early-type (M0.0-M0.5) stars from Group 1 that are found well in the intermediate region between Group 3 and Group 4; PM J14073+3923 ($J$-$K$ ≃ 0.67, $G$-$K$ ≃ 2.87, [M/H] = -0.35 dex, [α/Fe] = -0.025 dex) and PM J15307+0126 ($J$-$K$ ≃ 0.69, $G$-$K$ ≃ 2.88, [M/H] = -0.35 dex, [α/Fe] = -0.025 dex). Under close examination, both stars seem to erroneous metallicity estimates as their rather noisy spectra are more consistent with metal-poor M subdwarfs rather than near-solar metallicity M dwarfs, and accordingly, these stars do not belong to Group 1.

- The star from Group 1 found at the bottom of the diagram; PM J03288+3722 ($J$-$K$ ≃ 0.82, $G$-$K$ ≃ 2.59, [M/H] = -0.35 dex, [α/Fe] = +0.3 dex). Visual inspection of the spectrum suggests the metallicity estimate is correct, and we therefore suspect an error in the photometry.

- The star from Group 2 that can be seen in the general loci of Group 3 or 4 stars; PM J09003+6646 ($J$-$K$ ≃ 0.67, $G$-$K$ ≃ 2.89, [M/H] = -1.05 dex, [α/Fe] = +0.2 dex). Close scrutiny suggests the metallicity estimate is in error, because the noisy spectrum is similar to that of a very metal-poor star, rather than a moderately metal-poor one in Group 2.

- The star PM J22140+5211 ($J$-$K$ ≃ 0.69, $G$-$K$ ≃ 4.08, [M/H] = -0.6 dex, [α/Fe] = +0.225 dex) from Group 2, which is situated above, and to the right of the distribution. Visual inspection shows a spectrum more consistent with that of a metal rich M dwarf. In any case it appears likely that the photometry is also in error, given the large color offset.

- The two stars from Group 2, which are found well inside the distribution of Group 1; PM J03075+4125 ($J$-$K$ ≃ 0.81, $G$-$K$ ≃ 3.87, [M/H] = -0.55 dex, [α/Fe] = +0.2 dex) and PM J08396+5309 ($J$-$K$ ≃ 0.84, $G$-$K$ ≃ 3.49, [M/H] = -0.75 dex, [α/Fe] = +0.175 dex). Our close visual inspection of the former suggests that the star is more metal-rich than the value from the model-fit, which likely account for the discrepancy. Our close visual inspection of the latter, on the other hand, shows a spectrum consistent with the metallicity estimate from the model fit, which suggests that the problem may be due to the photometry in that case.

- The star from Group 3 which is located well inside the domain of Group 1; PM J14191+2023 ($J$-$K$ ≃ 0.89, $G$-$K$ ≃ 3.70, [M/H] = -1.30 dex, [α/Fe] = +0.45 dex). Upon careful examination, the spectrum appears very consistent with the metallicity estimate from the model fit, and the very red color may be due to a problem with the photometry, either from large instrumental error or from photometric contamination.

### A.6. *Photometric Variations with Chemical Parameters in H-R Diagram*

The most significant outliers in Figure 41 are listed below:

- The M dwarf PM J15460+5135 ($G_{BP}$ - $G_{RP}$ ≃ 2.07, $M_G$ ≃ 10.99, [M/H] = +0.25 dex, and [α/Fe] = +0.05) from Group 1Bα which is located well within the locus of Group 3α stars. Careful inspection shows that the spectrum is completely consistent with that of a metal-rich M dwarf. Therefore we have to assume that the Gaia photometry or/and parallax is erroneous, shifting its location far below and/or to the blue of the Group 1α locus.

- The star from Group 2α, PM J11222+5913 $G_{BP}$ - $G_{RP}$ ≃ 3.08, $M_G$ ≃ 12.07, [M/H] = -0.60 dex , [α/Fe] = +0.2), which is located well inside the locus of Group1α . Our close examination of the spectrum suggests that the star's metallicity may indeed have been underestimated by the pipeline, and thus that the star probably does belong to Group 1.

- The metal-poor M subdwarf from Group 3α, PM J05277+0019 ($G_{BP}$ - $G_{RP}$ ≃ 2.31, $M_G$ ≃ 10.78, [M/H] = -1.35 dex , [α/Fe] = +0.425), which can be seen close to the border of Group 1α and Group 2α. Close examination of the spectrum confirms that the star must be very metal-poor. The odd location in the H-R diagram may be due to inaccurate Gaia data, but it might also suggest that the star is an unresolved binary, which would make the star appear overluminous by up to 1.7 magnitudes, consistent with the observed offset from the locus of Group 3α stars.

## B. MODEL-FIT PROCEDURE

### B.1. *First Pass*

In the first pass, the surface gravity log $g$, α-element enhancement [α/Fe], and convolution factor C are set as fixed parameters. We choose the values of log $g$ and C equal to 5.0 dex and 6.5 Å, respectively, and set [α/Fe] to be a function of [M/H], with [α/Fe] = +0.4 dex for -2.5 ⩽ [M/H] < -1.6 dex, [α/Fe] = -0.25 × [M/H] for -1.6 ⩽ [M/H] < 0 dex, and [α/Fe] = 0 for [M/H] ⩾ 0. This relationship between [α/Fe] and [M/H] is different from the one used in the CIFIST grid, but is in better agreement with results from abundance studies of nearby FGK dwarfs (Adibekyan et al. 2012 and 2013; Recio-Blanco et al. 2014), and it provides a smoother transition of [α/Fe] values from [α/Fe]=0 to [α/Fe]=+0.4 dex.

The initial metallicity values for the first pass ([M/H]₀) is selected based on the pre-determined metallicity classes of stars in our classification pipeline (see section 3.3) as follows: [M/H]₀ = 0 for 1 ⩽ MC ⩽ 3, [M/H]₀ = -0.5 dex for 4 ⩽ MC ⩽ 6, [M/H]₀ = -1.0 dex for $7 \leqslant MC \leqslant 9$, $[M/H]_0$ = -1.5 dex for $10 \leqslant MC \leqslant 12$. The initial values of the polynomial fit order ($n_0$) and of the radial velocity ($v_{r_0}$) are set to 10 and zero, respectively.

Following this, we run four least squares minimization sequences, in each of which only one of the parameters $T_{eff}$, $[M/H]$, $v_r$, or $n$ is allowed to vary. We first perform a minimization through all the effective temperatures from 2600 to 4000 K, in 50 K bins, while $[M/H]$, $v_r$, and $n$ are set equal to their initial values described above. From this run, we obtain the initial best-fit value for the effective temperature $(T_{eff})_1$ with the subscript "1" denoting the first pass. This value together with $[M/H]_0$ and $n_0$ are then used as the fixed parameters in a second minimization sequence where $v_r$ is now allowed to vary from -500 to +500 km/s, with a step size of 10 km/s, yielding an initial best-fit value of $v_r$, $(v_r)_1$. In a third minimization sequence, we utilize $(T_{eff})_1$, $(v_r)_1$, and $n_0$ as the fixed parameters, and vary $[M/H]$ from -2.5 to 0.5 dex, in steps of 0.1 dex, which leads to the initial best-fit metallicity value, $[M/H]_1$. The initial best-fit polynomial order $n_1$ is determined by running a fourth minimization sequence, using $(T_{eff})_1$, $(v_r)_1$, and $[M/H]_1$ as the fixed parameters, and varying $n$ from 6 to 10, in bins of 1.

In the final sequence, we run a least squares minimization by exploring all neighboring grid points around the initial best-fit values, with offsets of $\Delta[M/H] = \pm 0.5$ dex (in steps of 0.1 dex), $\Delta T_{eff} = \pm 50$ K, $\Delta n = \pm 1$, and $\Delta v_r = \pm 10$ km/s to test if these best-fit values indeed minimize the $\chi^2$. These ranges (also similar ranges in the following passes) are selected based on the sensitivity of $\chi^2$ to the primary and secondary parameters, and how these parameters change from one pass to another. During this minimization sequence, the $\chi^2$ values of all possible parameter combinations are compared with each other simultaneously, and none of these parameters (i.e., $[M/H]$, $T_{eff}$, $n$ and $v_r$) is held fixed. Any model that falls outside our selected model grid (see Section 5.2), or any secondary parameter value that is beyond the adopted range, is automatically rejected from the calculation. If all the new best-fit values are the same as the initial ones, the run is stopped and these best fits are adopted as the first-pass, best-fit parameters. Otherwise, if the new best fit of at least one of the parameter has been changed, the sequence is repeated after replacing the initial best fits with the new ones. The process is iterated until convergence is achieved. The best fit parameters at the end of the first pass are denoted $[M/H]_{first}$, $(T_{eff})_{first}$, $(v_r)_{first}$, and $n_{first}$. The parameter $n_{first}$ also happens to be the final value of the polynomial order ($n_{final}$) for the recalibration function, since this value is not allowed to vary in subsequent passes.

### B.2. *Second Pass*

In the second pass, the physical parameters $[\alpha/Fe]$ and $\log g$ are fixed, and the polyfit order $n$ is set as the fixed secondary parameter. We set $[\alpha/Fe]$ as a function of $[M/H]$, as described in the first-pass section, $\log g$ to 5.0 dex, and $n$ to the above determined final value $n_{final}$.

We first perform a least squares minimization sequence by varying the parameter $C$ from 4.5 to 15.5 Å, in steps of 1 Å, while the parameters $T_{eff}$, $[M/H]$, and $v_r$, are fixed to their respective first-pass best-fit values $(T_{eff})_{first}$, $[M/H]_{first}$ and $(v_r)_{first}$. The sequence yields an initial best-fit value for the parameter convolution factor $C_2$, where the subscript "2" denotes the second pass. We then run the final sequence as an iterative minimization process, as described in the previous section, with offsets of $\Delta[M/H] = \pm 0.5$ dex (in steps of 0.1 dex), $\Delta T_{eff} = \pm 50$ K/s, $\Delta v_r = \pm 10$ km/s, and $\Delta C = \pm 1$ Å around $[M/H]_{first}$, $(T_{eff})_{first}$, $(v_r)_{first}$, and $C_2$, respectively. This provides a set of second-pass best-fit values for metallicity and temperature, $[M/H]_{second}$ and $(T_{eff})_{second}$, and also second pass best-fit values for the radial velocity and convolution factor, $(v_r)_{second}$ and $C_{second}$. The latter two also happen to be the adopted final values $(v_r)_{final}$ and $C_{final}$, because they are not allowed to vary in subsequent passes.

### B.3. *Third Pass*

In the third pass, the fixed parameters are the $\alpha$-element abundance $[\alpha/Fe]$, which is set to its value corresponding to $[M/H]$ following the function described above, and the three secondary parameters, $v_r$, $n$, and $C$, which are set to their now adopted final values, i.e., $(v_r)_{final}$, $n_{final}$, and $C_{final}$, respectively.

We first carry out a least squares minimization sequence by varying $\log g$ from 4.8 to 5.2 dex, in steps of 0.1 dex, while fixing the temperature and metallicity to their second-pass best-fit values $(T_{eff})_{second}$ and $[M/H]_{second}$. This provides an initial best-fit value for the gravity parameter $(\log g)_3$, where the subscript "3" denotes the third pass. We then perform an iterative minimization sequence with offsets of $\Delta \log g = \pm 0.1$ dex, $\Delta T_{eff} = \pm 50$ K, and $\Delta[M/H] = \pm 0.5$ dex (in steps of 0.1 dex) around $(\log g)_3$, $(T_{eff})_{second}$, and $[M/H]_{second}$, respectively. This yields the third-pass best fits, $(\log g)_{third}$, $(T_{eff})_{third}$, $[M/H]_{third}$.

### B.4. *Fourth Pass*

In the fourth pass, we fix the primary parameter $\log g$ and all the secondary parameters $v_r$, $n$, and $C$ to their now pre-determined values $(\log g)_{third}$, $(v_r)_{final}$, $n_{final}$, and $C_{final}$. This time, $T_{eff}$, $[M/H]$ and $[\alpha/Fe]$ are all treated as free parameters. In particular, $[\alpha/Fe]$ is no longer set to a function of $[M/H]$ as in all the previous passes. It should be pointed out that if all the four primary parameters vary simultaneously, the resulting best-fit values may not be physically acceptable; more particularly, the variation of $\log g$ and $[\alpha/Fe]$ at the same time may yield unreasonably high values of $[\alpha/Fe]$ for some very metal-rich stars, which is inconsistent with previous studies of FGK dwarfs (Lee et al. 2011a, b; Adibekyan et al. 2012 and 2013; Haywood et al. 2013; Recio-Blanco et al. 2014). This is why $\log g$ is not allowed to vary in this pass.

We first run a least squares minimization sequence by varying $[\alpha/Fe]$, and fixing $T_{eff} = (T_{eff})_{third}$ and $[M/H] = [M/H]_{third}$. This yield an initial best-fit value for the $\alpha$-element abundance $([\alpha/Fe])_4$, where the subscript "4" shows the fourth pass. It should be recalled from Section 4.2 that the range of possible $[\alpha/Fe]$ values depends on $[M/H]_{third}$, and it is $[\alpha/Fe] = [-0.2, +0.4]$ dex for $-0.45 \leqslant [M/H]_{third} \leqslant +0.5$ dex, $[\alpha/Fe] = [-0.2, +0.6]$ dex for $[M/H]_{third} = -0.5$ dex , and $[\alpha/Fe] = [0, +0.6]$ dex for $-2.5 \leqslant [M/H]_{third} < -0.5$ dex. Within two ranges, values of $[\alpha/Fe]$ all vary in steps of 0.025 dex. We then perform a second minimization sequence by varying $[M/H]$ with offsets of $\Delta[M/H] = 1.0$ dex around $[M/H]_{third}$, in steps of 0.05 dex[14], while holding $T_{eff}$ and $[\alpha/Fe]$ fixed to $(T_{eff})_{third}$ and $([\alpha/Fe])_4$, respectively, giving the initial best-fit metallicity ([M/H])$_4$. Since [M/H]$_{third}$ is a good estimate of the final value of [M/H], it is not necessary to examine all values of [M/H], and a variation range of $\Delta$[M/H] = 1.0 dex is sufficient to pinpoint the initial best fit ([M/H])$_4$ near [M/H]$_{third}$.

---

[14] The variation of $[M/H]$ around $[M/H]_{third}$ is symmetric if $\left| [M/H]_{third} - [M/H]_{up/low} \right| \geqslant 0.5$ dex, where $[M/H]_{up/low}$ is either the upper limit $[M/H]$=0.5 dex or the lower limit $[M/H]$=-2.5 dex, and contrarily, the variation range is not symmetric if $\left| [M/H]_{third} - [M/H]_{up/low} \right| < 0.5$ dex. For example, if $[M/H]_{third}$ = -1.4 dex, $[M/H]$ varies over a range of $\Delta[M/H] = \pm 0.5$ dex around -1.4 dex, but if $[M/H]_{third}$ = +0.3 dex, this range is $\Delta[M/H] = +0.2$ dex and -0.8 dex around +0.3 dex.

| Pass | $T_{eff}$ | [M/H] | [$\alpha$/Fe] | log $g$ | $v_r$ | n | C | Output |
|------|-----------|-------|---------------|---------|-------|---|---|--------|
| 1 | V | V | F | F | V | V | F | $(T_{eff})_{first}$, [M/H]$_{first}$, $(v_r)_{first}$, **n$_{final}$** |
| 2 | V | V | F | F | V | F | V | $(T_{eff})_{second}$, [M/H]$_{second}$, **$(v_r)_{final}$**, **C$_{final}$** |
| 3 | V | V | F | V | F | F | F | $(T_{eff})_{third}$, [M/H]$_{third}$, (log $g$)$_{third}$ |
| 4 | V | V | V | F | F | F | F | $(T_{eff})_{fourth}$, [M/H]$_{fourth}$, **[$\alpha$/Fe]$_{final}$** |
| 5 | V | V | F | V | F | F | F | **$(T_{eff})_{final}$, [M/H]$_{final}$, (log $g$)$_{final}$** |

**Table 10.** Summary of which parameters are fixed ("F") and which are allowed to vary ("V") in each of the five passes of our model-fitting pipeline. The last column shows which values are provided as output after each of the passes. Adopted final values for the fit are shown in bold font.

In the next sequence, we carry out an iterative minimization by varying $\Delta T_{eff} = \pm 50$ K, $\Delta$[M/H] = $\pm 0.3$ dex (in steps of 0.05 dex), and $\Delta$[$\alpha$/Fe]= $\pm 0.1$ dex (in steps of 0.025 dex), around $(T_{eff})_{third}$, ([M/H])$_4$, and ([$\alpha$/Fe])$_4$, respectively. The resulting best-fit values are finally used in a single minimization sequence with the variation of $\Delta$[M/H] = $\pm 0.1$ dex (in steps of 0.05 dex), $\Delta T_{eff} = \pm 50$ K, and $\Delta$[$\alpha$/Fe] = $\pm 0.05$ dex (in steps of 0.025 dex)[15]. This provides the fourth-pass best-fit values [M/H]$_{fourth}$, $(T_{eff})_{fourth}$, and [$\alpha$/Fe]$_{fourth}$, the last of which is the final value of the $\alpha$-element enhancement [$\alpha$/Fe]$_{final}$ as well.

### B.5. Fifth Pass

The fixed and varying parameters in this pass are the same as those in the third pass; that is, [$\alpha$/Fe], $v_r$, n, and C, are kept fixed to their adopted final values, [$\alpha$/Fe]$_{final}$, $(v_r)_{final}$, n$_{final}$, and C$_{final}$, respectively, whereas $T_{eff}$, [M/H], and log $g$ are varied as the fitting parameters.

We implement an iterative minimization sequence by varying $\Delta$log $g = \pm 0.1$ dex, $\Delta T_{eff} = \pm 50$ K, and $\Delta$[M/H] = $\pm 0.2$ dex (in steps of 0.05 dex) around (log $g$)$_{third}$, $(T_{eff})_{fourth}$, and [M/H]$_{fourth}$, respectively, yielding the fifth-pass best fits, (log $g$)$_{fifth}$, $(T_{eff})_{fifth}$, [M/H]$_{fifth}$, which are also the final best-fit vales, (log $g$)$_{final}$, $(T_{eff})_{final}$, and [M/H]$_{final}$.

We have summarized the above described five passes in Table 10, identifying the fixed (as shown by "F") and varying (as shown by "V") parameters for each pass. The best-fit parameters generated after each pass are also presented in the last column of this table, and adopted final values for the fit are shown in bold font. The final best fit of the secondary parameters are obtained from the first two passes, as these are not noticeably sensitive to the tuning of best-fit models by adding log $g$ and [$\alpha$/Fe] as free parameters in the subsequent passes. On the other hand, due to their strong correlation with each other as well as with the other atmospheric parameters, $T_{eff}$ and [M/H] vary in all the five passes, refining the best-fit models from one pass to another.

---

[15] Similar to the iterative minimizations, if a model goes beyond the selected grid (Section 4.2) over these parameter ranges, it is automatically removed from the calculation.